\documentclass[aps,prl,twocolumn,superscriptaddress,print]{revtex4-2}
\usepackage[colorlinks=true,linkcolor=blue,citecolor=blue,urlcolor=blue]{hyperref}
\usepackage{amssymb,amsmath}
\usepackage{graphicx}
\usepackage{multirow}
\usepackage{appendix}
\usepackage{mathtools}
\usepackage{amsmath}
\usepackage[table]{xcolor}
\usepackage{verbatim}
\usepackage{gensymb}
\usepackage{subfigure}
\usepackage{bm}
\usepackage{threeparttable}
\usepackage{booktabs}
\usepackage{braket}
\usepackage{xcolor}
\usepackage[normalem]{ulem}
\usepackage{array, makecell}

\begin{document}
\title{Interlayer hybridization in graphene quasicrystal and other bilayer graphene systems}
\author{Guodong Yu}
\affiliation{Key Laboratory of Artificial Micro- and Nano-structures of Ministry of Education and School of Physics and Technology, Wuhan University, Wuhan 430072, China}
\author{Yunhua Wang}
\email{wangyunhua@csrc.ac.cn}
\affiliation{Beijing Computational Science Research Center, Beijing, 100193, China}
\affiliation{Institute for Molecules and Materials, Radboud University, Heijendaalseweg 135, NL-6525 AJ Nijmegen, Netherlands}
\author{Mikhail I. Katsnelson}
\affiliation{Institute for Molecules and Materials, Radboud University, Heijendaalseweg 135, NL-6525 AJ Nijmegen, Netherlands}
\author{Hai-Qing Lin}
\affiliation{Beijing Computational Science Research Center, Beijing, 100193, China}
\author{Shengjun Yuan}
\email{s.yuan@whu.edu.cn}
\affiliation{Key Laboratory of Artificial Micro- and Nano-structures of Ministry of Education and School of Physics and Technology, Wuhan University, Wuhan 430072, China}
\affiliation{Beijing Computational Science Research Center, Beijing, 100193, China}
\affiliation{Institute for Molecules and Materials, Radboud University, Heijendaalseweg 135, NL-6525 AJ Nijmegen, Netherlands}

\begin{abstract}
The incommensurate 30$^{\circ}$ twisted bilayer graphene (BG) possesses both relativistic Dirac fermions and quasiperiodicity with 12-fold rotational symmetry arising from the interlayer interaction [\href{https://science.sciencemag.org/content/361/6404/782}{Ahn et al., Science \textbf{361}, 782 (2018)} and \href{https://www.pnas.org/content/115/27/6928}{Yao et al., Proc. Natl. Acad. Sci. \textbf{115}, 6928 (2018)}]. Understanding how the interlayer states interact with each other is of vital importance for identifying and subsequently engineering the quasicrystalline order in the layered structures. Herein, via symmetry and group representation theory we unravel the interlayer hybridization selection rules governing the interlayer coupling in both untwisted and twisted BG systems. Compared with the only allowed equivalent hybridization in $D_{6h}$ untwisted BG, $D_6$ twisted BG permits equivalent and mixed hybridizations, and $D_{6d}$ graphene quasicrystal allows both equivalent and non-equivalent hybridizations. The energy-dependent hybridization strengths in graphene quasicrystal and $D_6$ twisted BG show two remarkable characteristics: (i) near the Fermi level the weak hybridization owing to the relatively large energy difference between Dirac bands from top and bottom layers, and (ii) in high-energy regions the electron-hole asymmetry of hybridization strength with stronger interlayer coupling for holes, which arises from the non-nearest-neighbor interlayer hoppings and the wave-function phase difference between paring states. These hybridization-generated band structures and their hybridization strength characteristics are verified by the calculated optical conductivity spectra. Our theoretical study paves a way for revealing the interlayer hybridization in van der Waals layered systems.
\end{abstract}

\maketitle
{\textit{Introduction.}}---
Besides the emergent correlated effects\cite{TBG_superconductivity0,TBG_superconductivity1,TBG_superconductivity2,xu2018topological,
TBG_twist_Santos,TBG_flat_band,TBG_flatband1,
TBG_corrected_insulator,kang2019,bernevig2021,TBG_wigner_crystal,FCW_super_TBG_phonon,FCW_super_TBG_Mag_strain,
TBG_supercond_lessthanMag,liu2021} in slightly twisted bilayer graphene (BG), the recently discovered quasicrystal\cite{science_QC,pnas_QC} in 30$^{\circ}$ incommensurately twisted BG has also attracted considerable interests in both experiment \cite{science_QC,30TBG_grown,pnas_QC,cm_QC,30TBG_STM,30TBG_Suzuki,30tBG_onCu_arXiv,30tBG_on_Cu_ACSNano,30TBG_grown_onCu_2dmat} and theory\cite{30TBG_localization,Moon_QC0,30TBG_superlubricity,Moon_QC1,30TBG_quantum_oscillation,Yu_QC_BG,Yu_pressure_QC,Yu_TDBG_QC,QAHE_QC}. Several synthetic methods, including chemical vapor deposition and carbon segregation from the bulk during high temperature annealing, have been used to successfully grow graphene quasicrystal on SiC\cite{science_QC,30TBG_grown}, Pt\cite{pnas_QC}, Cu\cite{30tBG_onCu_arXiv,30tBG_on_Cu_ACSNano,30TBG_grown_onCu_2dmat} and Cu-Ni alloy\cite{cm_QC} substrates. The quasiperiodicity in these samples is experimentally identified by the low-energy electron diffraction\cite{pnas_QC,science_QC,cm_QC,30TBG_grown}, transmission electron microscopy\cite{science_QC,30tBG_on_Cu_ACSNano}, scanning tunneling microscopy\cite{30TBG_STM} and Raman spectroscopy\cite{pnas_QC,30tBG_onCu_arXiv}. The angle-resolved photoemission spectroscopy measurements indicate the multiple Dirac cones together with 12-fold rotational symmetry\cite{pnas_QC,science_QC}. In addition, the low-energy Dirac fermions with unique quasicrystalline order are verified by the magnetotransport measurements\cite{30tBG_onCu_arXiv,30TBG_STM}. Owing to the interlayer scatterings with a constraint of reciprocal lattice vector difference between two layers, i.e., the generalized Umklapp scatterings, replica Dirac cone bands display unbalanced electron distribution features in time- and angle-resolved photoemission spectroscopy measurements (ARPES)\cite{30TBG_Suzuki}. In theoretical aspects, a $\mathbf{k}$-space tight-binding model is constructed to explore the 12-fold symmetric resonant states and the critical characteristic of wave functions as a hallmark of quasicrystalline order is also verified\cite{Moon_QC0}. The quantum oscillations with spiral Fermi surfaces are predicted theoretically due to the quasiperiodicity and weak interlayer coupling\cite{30TBG_quantum_oscillation}. Numerical simulations indicate that a fractal feature happens for the sliding force and the low friction appears as a result of the quasicrystalline structure\cite{30TBG_superlubricity}. The vertical pressure and electric field can be utilized to tune the quasicrystalline electronic states\cite{Yu_pressure_QC}. The numerical calculations show that quasicrystalline electronic states can also exist in 30$^{\circ}$ twisted double BG system\cite{Yu_TDBG_QC}. In doped graphene quasicrystal, a combination of high symmetry and Coulomb interaction possibly enables another superconductivity beyond $d+id$ topological superconductivity in doped graphene\cite{liu2021g}. All of these peculiar physical properties make graphene quasicrystal quite distinctive from graphene monolayer.

\begin{table*}[htbp!]
\caption{{Interlayer hybridization selection rules and classifications in twisted and untwisted BG systems.} }
\centering
\begin{tabular}{c c c c c c c c c c c c}
\hline\hline
& & & \multicolumn{3}{c}{Hybridization classifications}\\
\cline{4-6}
$\theta_t$ & Point groups & Selection rules & Equivalent & Mixed & Non-equivalent\\
\hline

$\theta_t=30^{\circ}$ & $ D_{6d}$ & {{\colorbox{white!10}{\thead{\makecell[c]{ $U_{A_i,A_j}=\delta_{A_i,A_j}U_{A_i,A_j}$\\$U_{E_i,E_j}= \delta_{E_i,E_j}U_{E_i,E_j}$\\
$U_{B_1,ir^{\prime}}=\delta_{B_2,ir^{\prime}}U_{B_1,ir^{\prime}}$\\
$U_{B_2,ir^{\prime}}=\delta_{B_1,ir^{\prime}}U_{B_2,ir^{\prime}}$ }}}}} & \makecell[c]{ $A_1+A_1\Rightarrow A_1+B_2$ \\ $A_2+A_2\Rightarrow A_2+B_1$ \\$E_1+E_1\Rightarrow E_1+E_5$\\$E_2+E_2\Rightarrow E_2+E_4$} & & \makecell[c]{$B_1+B_2\Rightarrow E_3 + E_3$\\$B_2+B_1\Rightarrow E_3 + E_3$} \\

$0^{\circ}<\theta_t<30^{\circ}$ & $ D_{6}$ & \makecell[c]{
$U_{A_i,ir^{\prime}}=$\\ $(\delta_{A_1,ir^{\prime}}+\delta_{A_2,ir^{\prime}})U_{A_i,ir^{\prime}}$\\
$U_{B_i,ir^{\prime}}=$\\ $(\delta_{B_1,ir^{\prime}}+\delta_{B_2,ir^{\prime}})U_{B_i,ir^{\prime}}$\\
$U_{E_i,E_j}= \delta_{E_i,E_j}U_{E_i,E_j}$} & \makecell[c]{ $E_1+E_1\Rightarrow E_1+E_1$\\$E_2+E_2\Rightarrow E_2+E_2$} &\makecell[c]{ $A_{1,2}+A_{1,2}\Rightarrow A_{1,2}+A_{2,1} $\\$B_{1,2}+B_{1,2}\Rightarrow B_{1,2}+B_{2,1} $} &  &  \\

$\theta_t=0^{\circ}$ & $D_{6h}$ & {{\colorbox{white!10}{\thead{\makecell[c]{ $U_{ir,ir^{\prime}}=\delta_{ir,ir^{\prime}}U_{ir,ir^{\prime}}$}}}}} & \makecell[c]{ $A_1+A_1\Rightarrow A_{1g}+A_{2u}$\\$A_2+A_2\Rightarrow A_{2g}+A_{1u}$\\
$B_1+B_1\Rightarrow B_{2g}+B_{1u}$\\$B_2+B_2\Rightarrow B_{1g}+B_{2u}$\\
$E_1+E_1\Rightarrow E_{1g}+E_{1u}$\\$E_2+E_2\Rightarrow E_{2g}+E_{2u}$} & &  \\
\hline \hline
\end{tabular}
\label{table:sl}
\end{table*}
Compared with the conventional quasicrystals where all of the atoms are intrinsically located within a quasiperiodic order\cite{QC_Metallic,QC_symmetry}, graphene quasicrystal is viewed as an extrinsic quasicrystal (i.e., engineered quasicrystals) because its quasiperiodicity arises from the interlayer coupling between two graphene monolayers. Thus, figuring out the origin of quasicrystalline order requires a deep understanding of how the interlayer states interact with each other. However, up to now, there is no theory to describe whether arbitrary two electronic states separately from the two monolayers can be hybridized with each other or not in BG systems. Applying symmetry and point group representation theory we construct the interlayer hybridization selection rules governing which states from two layers are allowed to be hybridized in all BG systems including $D_{6h}$ untwisted BG, $D_6$ twisted BG and $D_{6d}$ graphene quasicrystal. Numerical calculations on interlayer hybridization matrix elements from $p_z$ orbital tight-binding (TB) model, Wannier orbital TB model and density functional theory (DFT) verify the interlayer hybridization selection rules. These allowed hybridizations are classified into three categories: (i) the equivalent hybridization requiring the hybridization paring states with the same irreducible representations (irreps) in all BG systems, (ii) the mixed hybridization with mixed hybridization parings inside $A_i$ as well as parings inside $B_i$ with $i=1,2$ in $D_6$ twisted BG, and (iii) the non-equivalent hybridization with a hybridization paring between $B_1$ and $B_2$ in graphene quasicrystal. Different from an obvious hybridization strength near the Fermi level in untwisted BG, the interlayer hybridization inside low-energy areas is weak in twisted BG because of a relatively large energy difference between the Dirac bands from top and bottom layers. Inside the high-energy areas, an electron-hole asymmetry of hybridization strength exists as a result of non-nearest-neighbor interlayer hoppings and the wave-function phase difference of hybridization paring states. The obtained optical conductivity spectra with remarkably different absorption features at different chemical potentials manifest the hybridization strength characteristics and hybridization-induced band structures in graphene quasicrystal. Our findings significantly explore how the interlayer states couple with each other in both untwisted and twisted BG systems and shed new light on the extrinsic quasicrystal order in 2D layer materials.

\begin{figure*}[!htbp]
\centering
\includegraphics[width=16 cm]{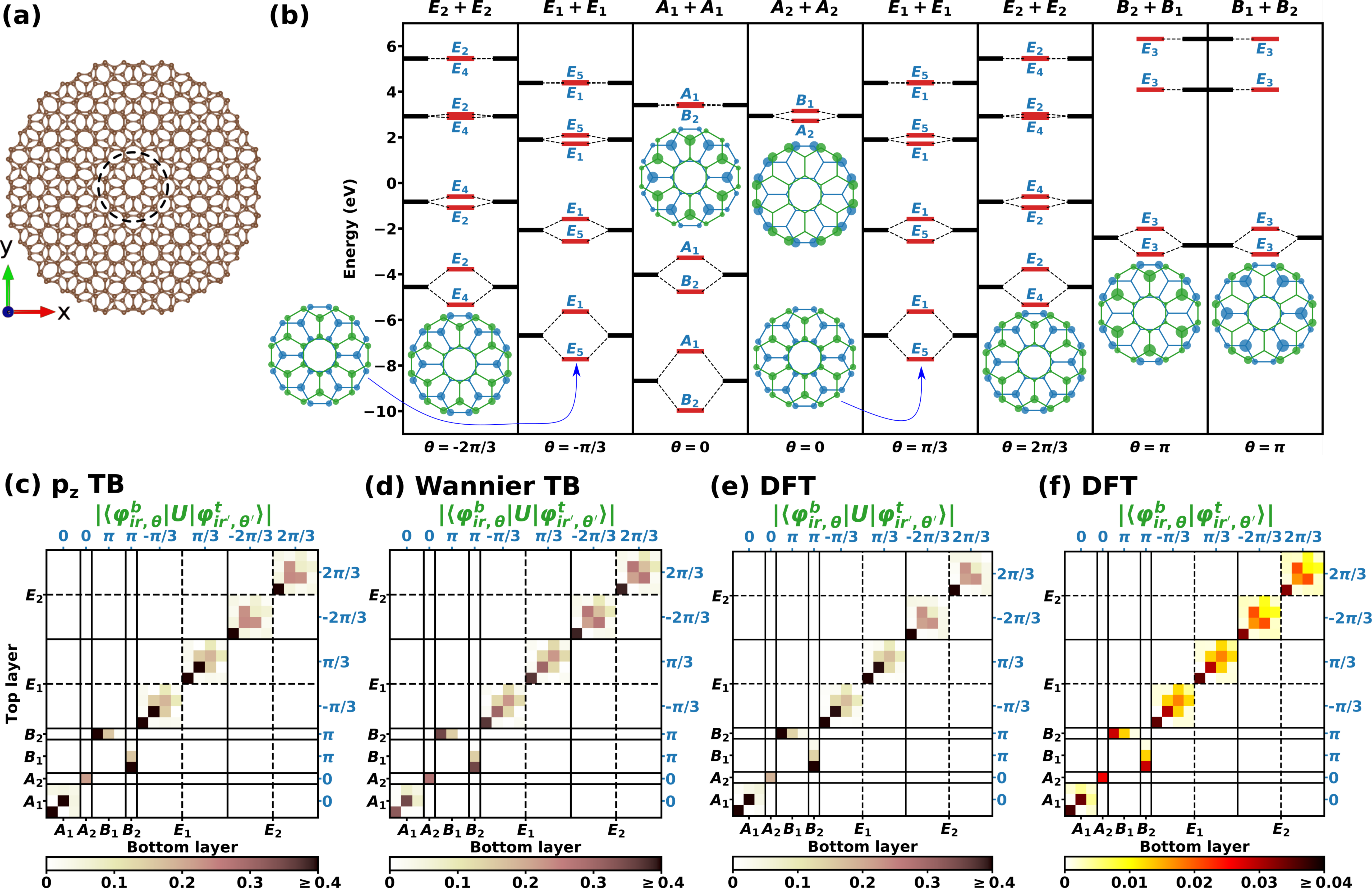}
\caption{(a) A top view on the dodecagonal graphene quasicrystal structure. (b) The eigen energy spectra and their irreps for the size-2 structure [inside the dash circle in (a)] under $C_{6}$ operation. In each subplot of (b), the energy levels (hybridization paring states) of the bottom and top $C_{6v}$ monolayers are correspondingly denoted by the left and right black lines, the energy levels (bonding and antibonding paring states) of the $D_{6d}$ bilayer are denoted by the middle red lines, and the insets show the real-space electron density (denoted by the circle size) for these eigenstates nearest above the insets themselves or signaled by the blue arrows. Interlayer hybridization matrix elements with their absolute values $| \braket{\varphi_{ir,\theta}^b|U|\varphi_{ir^{\prime},\theta^{\prime}}^{t}}|$ in unit of eV under $C_{6}$ operation with their eigenvalues $e^{i\theta}(e^{i\theta^{\prime}})$ and irreps $ir (ir^{\prime})$ from (c) $p_z$ orbital TB, (d) Wannier orbital TB and (e) DFT. (f) The overlap matrix element with its absolute value $| \braket{\varphi_{ir,\theta}^b|\varphi_{ir^{\prime},\theta^{\prime}}^{t}}|$ from DFT.}
\label{fig:hybrid_size2}
\end{figure*}

{\textit{Hybridization selection rule and classification.}}---
For an arbitrary twisted BG consisting of two $C_{6v}$ monolayers with a twist angle $\theta_t$ [for instance, the $D_{6d}$ graphene quasicrystal with $\theta_t=30^{\circ}$ in Fig. \ref{fig:hybrid_size2}(a)], the Hamiltonian includes three terms,
\begin{equation}
H = H_0^b + H_0^t + U,
\label{hamiltonian}
\end{equation}
where $H_0^b$ and $H_0^t$ are the Hamiltonians of the bottom and top layers with corresponding layer indexes $b$ and $t$, respectively, and $U$ is the interlayer coupling. Because of the $\theta_t$-dependent symmetry properties, the twisted BG systems are divided into three categories with corresponding point groups $D_{6h}$ ($\theta_t=0^{\degree}$), $D_6$ ($0^{\circ}<\theta_t<30^{\circ}$) and $D_{6d}$ ($\theta_t=30^{\circ}$), as listed in Table~\ref{table:sl}. The Hamiltonians and reflection operations of the bottom and top layers are connected by the rotation operation $R(\theta_t)$ and the mirror reflection $\sigma_h$ with its mirror plane perpendicular to the $z$ axis. Therefore, we can write the Hamiltonian $H_0^t$ of the top layer as $H_0^t=\left[\sigma_h R(\theta_t)\right] H_0^b \left[\sigma_h R(\theta_t)\right]^{\dagger}$ and the reflection operations between the top and bottom layers as
\begin{equation}
\begin{split}
\sigma_{v,i}^{t} &= R(\theta_t)\sigma_{v,i}^b \left[R(\theta_t)\right]^{\dagger}, \\
\sigma_{d,i}^{t} &= R(\theta_t)\sigma_{d,i}^b \left[R(\theta_t)\right]^{\dagger}, \\
\sigma_{d,i}^{b/t} &= R\left({{\pi}\over{6}}\right)\sigma_{v,i}^{b/t} \left[R\left({{\pi}\over{6}}\right)\right]^{\dagger},
\end{split}
\end{equation}
where $i=0,1,2$, and $\sigma_{v,0}^b=\sigma_x$.
The character projection operator of the irrep $ir$ for a point group $pg$ is defined as
\begin{equation}
P_{ir}^{pg} = {{l_{ir}}\over{h}} \sum_{R\in pg} \chi_{ir}^{*}(R) O_R,
\label{projection_d}
\end{equation}
where $l_{ir}$ and $h$ are the dimension of irrep $ir$ and the order of $pg$, respectively, and $\chi_{ir}(R)$ is the character of matrix representation $O_R$ of the symmetry operation $R$ for irrep $ir$. The projection operator can be used to determine which irrep an arbitrary eigenstate in the point group $pg$ belongs to, according to
\begin{equation}
P_{ir}^{pg}  \left| \varphi_{ir^{\prime}}^{pg} \right>= \delta_{ir,ir^{\prime}} \left| \varphi_{ir^{\prime}}^{pg} \right>.
\label{projection_w}
\end{equation}
Using the projection operator in Eqs.~\eqref{projection_d} and~\eqref{projection_w} and performing some algebraic calculations (see Sec. S2 of \cite{SM}), we obtain the constraint equations of the hybridization matrix element $U_{ir,{ir}^\prime}$, i.e., $\braket{\varphi_{ir}^b|U|\varphi_{ir^\prime}^t}$, for all three BG systems, where $\left| \varphi_{ir}^b \right>$ and $\left| \varphi_{ir^\prime}^t \right>$ are the states from the bottom and top layers with irreps $ir$ and $ir^\prime$, respectively. These constraint equations of $U_{ir,{ir}^\prime}$ listed in Table~\ref{table:sl} indicate which states from the top and bottom layers with corresponding irreps $ir$ and ${ir}^\prime$ are allowed to be hybridized with each other, and hence these constraint equations enable a rule of the interlayer hybridization, namely, the \emph{hybridization selection rule}. Applying these hybridization selection rules we can classify these interlayer hybridizations (see Sec. S3 of \cite{SM}). As shown in Table~\ref{table:sl}, for the $D_{6d}$ graphene quasicrystal, these states with the irreps $A_1, A_2, E_1$ and $E_2$ follow the equivalent hybridization, and these states with irreps $B_1$ and $B_2$ obey the non-equivalent hybridization. Compared with the hybridizations in graphene quasicrystal, the $D_{6h}$ BG with $\theta_t=0^{\circ}$ only allows the equivalent hybridization for all irreps, and the $D_{6}$ twisted BG with $0^{\circ}<\theta_t<30^{\circ}$ permits the equivalent and mixed hybridizations. In graphene monolayer, states with 2D irreps $E_i$ are always degenerate and easily separated by the rotation operation $C_6$. Thus we use $C_6$ to classify the eigenstates from the bottom and top layers by virtue of $C_6|\varphi_{ir,\theta}^{b}\rangle=e^{i\theta} |\varphi_{ir,\theta}^{b}\rangle$ and $C_6|\varphi_{ir^{\prime},\theta^{\prime}}^{t}\rangle=e^{i\theta^{\prime}} |\varphi_{ir^{\prime},\theta^{\prime}}^{t}\rangle$ with $\theta (\theta^{\prime}) = 0, \pm\pi/3, \pi$ and $\pm 2\pi/3$. In the basis functions of $C_6$, we write the hybridization matrix element as
\begin{equation}
\begin{split}
U^{\theta \theta^{\prime}}_{ir,ir^{\prime}}&= \braket{\varphi_{ir,\theta}^b|U|\varphi_{ir^{\prime},\theta^{\prime}}^t} = \braket{\varphi_{ir,\theta}^b|H|\varphi_{ir^{\prime},\theta^{\prime}}^t} \\
&= \braket{\varphi_{ir,\theta}^b|C_6^{\dagger}HC_6|\varphi_{ir^{\prime},\theta^{\prime}}^t} =e^{i(\theta^{\prime}-\theta)}U^{\theta \theta^{\prime}}_{ir,ir^{\prime}}\\
&=\delta_{\theta \theta^{\prime}}U^{\theta \theta^{\prime}}_{ir,ir^{\prime}}.
\end{split}
\label{delta_theta}
\end{equation}
Eq.~\eqref{delta_theta} indicates that the hybridization matrix displays nonzero diagonal elements and zero off-diagonal elements in the basis functions of $C_6$. These hybridization selection rules in Table~\ref{table:sl} are a result of symmetry no matter the system size. Therefore, it is effective to identify the interlayer hybridization selection rules by the numerical calculations on the hybridization matrix elements of a finite-size twisted BG structure keeping the same point group symmetry as that of the infinite system. We consider the graphene quasicrystal system with $D_{6d}$ point group shown in Fig. \ref{fig:hybrid_size2}(a) as an example. Figs. \ref{fig:hybrid_size2}(c)-\ref{fig:hybrid_size2}(e) show the hybridization matrix elements with respect to $C_6$ correspondingly from the $p_z$ orbital TB model, Wannier orbital TB model\cite{fang2016electronic} and density functional theory\cite{soler2002siesta} (see Sec. S4 of \cite{SM}), in a graphene quasicrystal quantum dot with size-2, where the size-2 is the number of zigzag chains from geometrical center to any side (see Sec. S1 of \cite{SM}). The mapped distributions of these nonzero hybridization matrix elements $U^{\theta \theta^{\prime}}_{ir,ir^{\prime}}\neq 0$ from all the three methods manifest the hybridization selection rules of graphene quasicrystal in Table~\ref{table:sl}. The calculations on $U^{\theta \theta^{\prime}}_{ir,ir^{\prime}}$ for other size $D_{6d}$ quasicrystal structures and $D_{6h}$ and $D_6$ BG structures (see Sec. S4 of \cite{SM}) also verify the corresponding hybridization selection rules for three point groups in Table~\ref{table:sl}.

We further explore the characteristics of the two hybridization categories in graphene quasicrystal structures. In the basis functions of $C_6$, for $ir=A_1$, $A_2$, $E_1$ and $E_2$, the antibonding $(+)$ and bonding $(-)$ states can be expressed as $|\phi_{\pm}\rangle =\frac{1}{\sqrt{2}} \left( e^{i\frac{\theta}{2}} |\varphi_{ir}^b\rangle \pm S_{12}|\varphi_{ir}^b\rangle \right)$, and the irreps of $|\phi_{\pm}\rangle$ follow the equivalent hybridization in Table~\ref{table:sl} (see Sec. S3.1 of \cite{SM}). It can be checked that $S_{12}|\phi_{\pm}\rangle=\pm e^{i\frac{\theta}{2}}|\phi_{\pm}\rangle$, which indicates that the states $|\phi_{\pm}\rangle$ are 12-fold symmetric. For the non-equivalent hybridization, both the bonding and antibonding $E_3$ states are 6-fold symmetric (see Sec. S3.1 of \cite{SM}). Fig. \ref{fig:hybrid_size2}(b) shows the eigen energy spectra and their irreps in the quasicrystal quantum dot with size-2. These interlayer hybridizations follow the selection rules of graphene quasicrystal in Table~\ref{table:sl}. In addition, the insets in Fig. \ref{fig:hybrid_size2}(b) also verify the 12-fold rotational symmetry of states generated by equivalent hybridizations and the 6-fold rotational symmetry of states generated by non-equivalent hybridizations.
\begin{figure}[!htbp]
\centering
\includegraphics[width=8.5 cm]{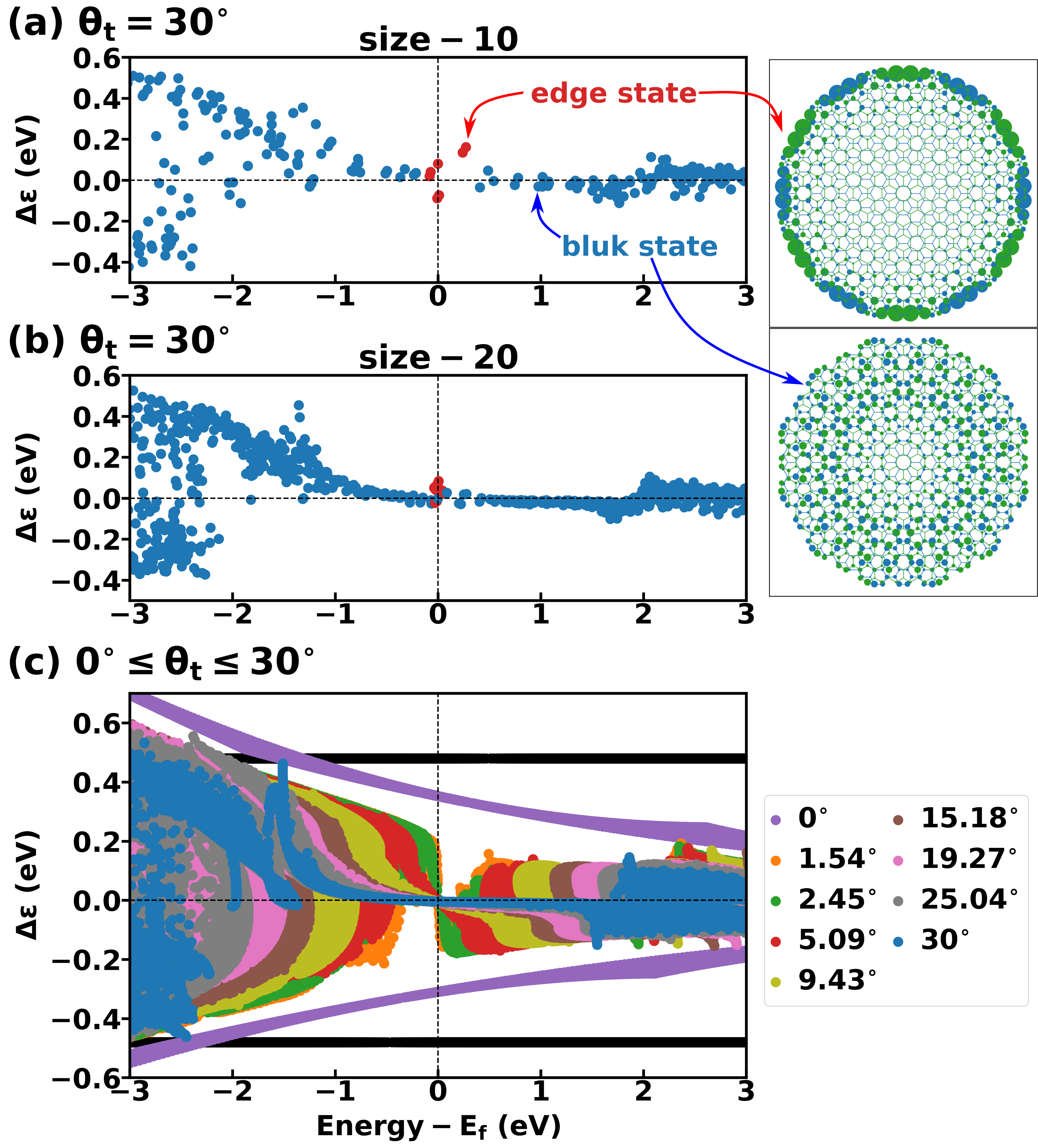}
\caption{The hybridization strengths as a function of energy for dodecagonal graphene quasicrystal structures with size-10 in (a) and size-20 in (b) and for other infinite-size twisted BG systems with various twist angles in (c), where the size-$\infty$ quasicrystal is calculated using a periodic 15/26 approximant\cite{Yu_QC_BG}, and the two black lines represent the hybridization strengths of AA-stacking BG only with the nearest-neighbor interlayer hopping. The right two insets show the real-space electron density for the signaled states indicated by arrows in (a).}
\label{fig:hybrid_strength}
\end{figure}

\begin{figure*}[!htbp]
\centering
\includegraphics[width=16 cm]{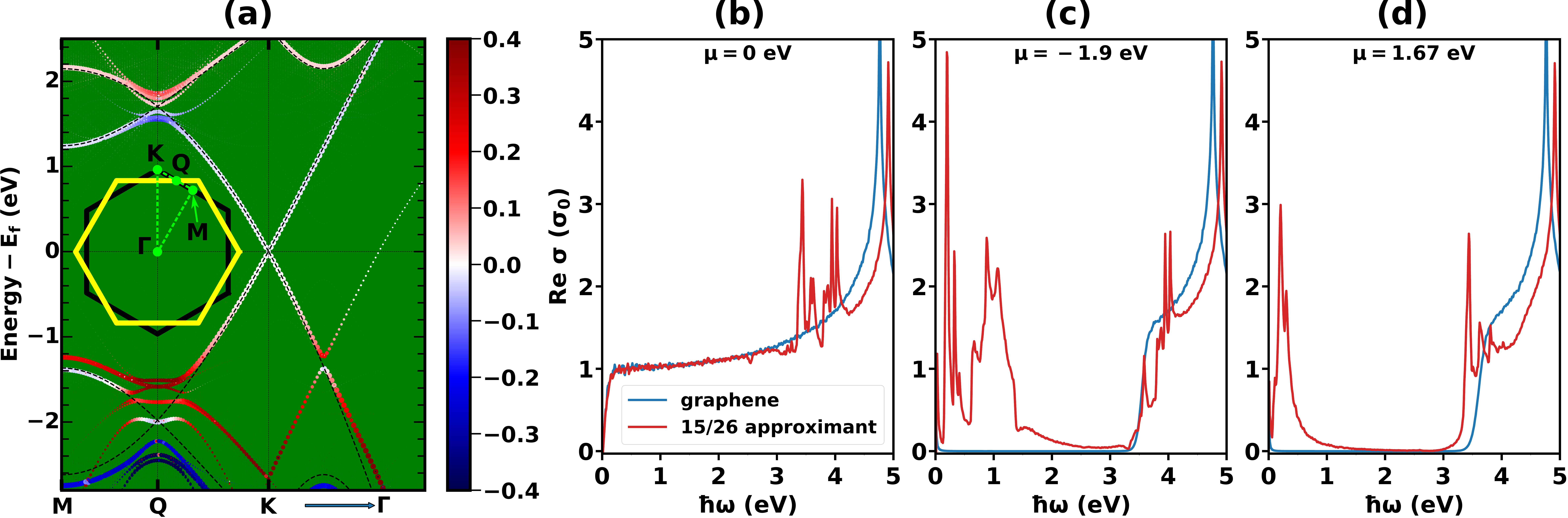}
\caption{(a) The calculated unfolded band structures for dodecagonal graphene quasicrystal within a periodic 15/26 approximant, where the color map denotes the interlayer hybridization strength $\triangle\varepsilon$, and black dashed lines stand for the band structures of the bottom and top graphene monolayers. The real part of optical conductivity as a function of photon energy $\hbar\omega$ at $\mu=0$ eV in (b), $\mu=-1.9$ eV in (c), and $\mu=1.67$ eV in (d), where the red and blue lines are for quasicrystal and graphene monolayer, respectively.}
\label{fig:ops:hyb}
\end{figure*}

{\textit{Hybridization strength.}}---
Following the interlayer hybridization rules, we can determine which states of the two $C_{6v}$ monolayers can be hybridized with each other. For an arbitrary eigenstate $|\phi\rangle$ of twisted BG with energy $\varepsilon$, we can also decouple $|\phi\rangle$ as $c |\varphi^b\rangle + d |\varphi^t\rangle$, where $|\varphi^b\rangle$ and $|\varphi^t\rangle$ are the components on the bottom and the top layers, $|c|^2 +|d |^2=1$, and $\braket{\varphi^b|\varphi^b}=\braket{\varphi^t|\varphi^t}=1$. To measure the energy-dependent interlayer coupling, the hybridization strength is defined as
\begin{equation}
  \Delta \varepsilon =
    \begin{cases}
      \varepsilon- max(\bar{\varepsilon}_{b}, \bar{\varepsilon}_{t}), & \text{if $\varepsilon> max(\bar{\varepsilon}_{b}, \bar{\varepsilon}_{t})$},\\
      \varepsilon- min(\bar{\varepsilon}_{b}, \bar{\varepsilon}_{t}), & \text{if $\varepsilon< min(\bar{\varepsilon}_{b}, \bar{\varepsilon}_{t})$},
    \end{cases}
\label{eq:hybrid_strength_eq}
\end{equation}
where $\bar{\varepsilon}_{b}=\braket{\varphi^b | H_0^b |\varphi^b}$ and $\bar{\varepsilon}_{t}=\braket{\varphi^t | H_0^t |\varphi^t}$ are the energy averages of states $|\varphi^{b}\rangle$ and $|\varphi^{t}\rangle$. Here, the first and second cases with correspondingly positive and negative values reflect the energy-dependent interlayer couplings contributing to the antibonding and bonding states, respectively. To see how the interlayer hybridizations vary on the energy, we use the $p_z$ orbital TB model to numerically compute the hybridization strength as a function of energy for graphene quasicrystal structures with size-10, size-20 and infinite-size within a periodic 15/26 approximant\cite{Yu_QC_BG}, AA-stacking BG and other twisted BG systems with various twist angles, as shown in Fig. \ref{fig:hybrid_strength}. For graphene quasicrystal systems in Figs. \ref{fig:hybrid_strength}(a)-\ref{fig:hybrid_strength}(c), the hybridization strengths inside the low-energy area about from $-$1.0 eV to 1.5 eV are small, except the edge states denoted by red dots in Figs. \ref{fig:hybrid_strength}(a) and \ref{fig:hybrid_strength}(b) with localized electron density at the edge compared with that of bulk states inside quasicrystal quantum dot structures, as demonstrated in the right insets. It means that the interlayer coupling inside the low-energy area is weak such that graphene quasicrystal has a similar low-energy dispersion as that of a decoupled graphene bilayer\cite{TBG_decouple0,TBG_decouple1,30tBG_on_Cu_ACSNano,science_QC,pnas_QC}. The hybridization strengths are also weak near the Fermi level in other $D_6$ twisted BG systems in Fig. \ref{fig:hybrid_strength}(c) except a relatively large strength in $D_{6h}$ AA-stacking BG. Such weak and strong hybridization strengths near the Fermi level respectively in twisted and untwisted systems can be correspondingly well understood by Dirac bands together with the second-order non-degenerate perturbation theory and the first-order degenerate perturbation theory (see Sec. S5.2 of \cite{SM}). Inside high-energy areas, the hybridization strengths in the negative energy region are larger than those inside the positive energy region for all BG systems. Such a electron-hole asymmetrical hybridization arises from the non-nearest-neighbor interlayer hoppings and the phase
differences of hybridization paring states (see Sec. S5.3 of \cite{SM}). Taking AA-stacking BG as an example in Fig. \ref{fig:hybrid_strength}(c), the hybridization strengths (denoted by the two black lines) are independent on energy if only the nearest-neighbor interlayer hopping is included. However, the electron-hole asymmetrical hybridization (denoted by the two purple lines) occurs because the non-nearest-neighbor interlayer hoppings should be taken into account in real systems. From negative to positive energy the phase difference of hybridization paring wave functions varies gradually in a trend from $0$ (parallel) to $\pi$ (anti-parallel), which enables a stronger interlayer coupling of the valence band than that of the conduction band (see Sec. S5.3 of \cite{SM}). In addition, we further reveal that how the additional interlayer potential difference induced by a vertical electric field enhances the hybridization strengths near the Fermi level (see Sec. S5.4 of \cite{SM}) and how the electric field selects the hybridization in $\bm{k}$ space for the resonant quasicrystalline states in graphene quasicyrstalline (see Sec. S6 of \cite{SM}).

{\textit{Proposals identifying hybridization-generated band structures and hybridization strengths experimentally.}}---
Fig. \ref{fig:ops:hyb}(a) shows the unfolded energy band structures of graphene quasicrystal with a supercell of the periodic 15/26 approximant along the same $k$ path of the primitive unit cell of graphene. The hybridization-induced band structures of graphene quasicrystal are represented by the dot lines with the size of dot as the value of spectral weight (i.e., $p_{n^\prime\bm{k}}=\sum_{n,s}\left| \braket{\phi_{n\bm{k}}^s|\varphi_{n^\prime\bm{k}}} \right|^2$\cite{Yu_TDBG_QC}) determining the unfolded band structures, where $\varphi_{n^\prime\bm{k}}$ is the eigenstate of the quasicrystal with the band index $n^\prime$ and wave vector $\bm{k}$, and $\phi_{n\bm{k}}^s$ is the eigenstate of graphene with the band index $n$ and the layer index $s$. The color of the dot line represents the interlayer hybridization strength $\Delta \varepsilon $ in Eq.~\eqref{eq:hybrid_strength_eq}. We see again the two hybridization characteristics in graphene quasicrystal: (i) the weak hybridization strength inside low-energy area and (ii) the electron-hole asymmetrical hybridization inside high-energy areas. Optical conductivity with its real part corresponding to the optical absorption manifests the interband transitions as a result of optical selection rule\cite{wang2021polarization}, and hence is employed to determine the allowed transitions of energy states with their symmetry properties. From the Kubo-Greenwood formula\cite{wang2021polarization}, Figs. \ref{fig:ops:hyb}(b)-\ref{fig:ops:hyb}(d) show the calculated real part of optical conductivity as a function of photon energy $\hbar\omega$ at $\mu=0$ eV, $-1.9$ eV, and $1.67$ eV, respectively, where the chemical potential $\mu$ can be tuned by a gate voltage. In Fig. \ref{fig:ops:hyb}(b) at $\mu=0$ eV, the optical conductivity of quasicrystal is almost the same as the that of graphene for about $\hbar\omega<2.5$ eV, which indicates the weak hybridization inside the low-energy area. In Figs. \ref{fig:ops:hyb}(c) and \ref{fig:ops:hyb}(d) with $\mu=-1.9$ eV and $1.67$ eV, respectively, the optical conductivity spectra show remarkably different energy positions of absorption peaks, which arise from the interband transitions between these hybridization-generated states in negative and positive high-energy areas, respectively (see Sec. S7 of \cite{SM}). Thus, the electron-hole asymmetrical hybridization can be characterized by optical conductivity spectrum experimentally. On the other hand, these hybridization-generated band structures with their hybridization strengths in graphene quasicrystal can also be measured by ARPES\cite{science_QC,pnas_QC}.

{\textit{Conclusion.}}---
By means of the symmetry and point group representation theory, we build the hybridization rules determining which states from two layers are allowed to be hybridized with each other in both untwisted and twisted BG systems. We also perform an hybridization classification according to the hybridization selection rules. The $D_{6h}$ untwisted BG only allows equivalent hybridization; the $D_6$ twisted BG allows equivalent and mixed hybridizations; and the $D_{6d}$ graphene quasicrystalline permits equivalent and non-equivalent hybridizations. These hybridization rules and classifications are identified by the numerical results on interlayer hybridization matrix elements. The energy-dependent hybridization strengths are mapped in graphene quasicrystalline and other BG systems. Except an obvious interlayer coupling in untwisted BG, the hybridization strength near the Fermi level is weak in twisted BG systems. The non-nearest-neighbor interlayer hoppings and the wave-function phase difference between hybridization paring states are responsible for the electron-hole asymmetry of hybridization strength. The calculated optical conductivity spectra in graphene quasicrystalline manifest the hybridization strength characteristics. Our results not only deeply explore how the interlayer states couple with each other in both twisted and untwisted BG systems, but also shed new light on the extrinsic quasicrystals in van der Waals layered structures. At last, in view of the successful experimental synthesis of graphene quasicrystal\cite{science_QC,30TBG_grown,pnas_QC,cm_QC,30TBG_STM,30TBG_Suzuki,30tBG_onCu_arXiv,30tBG_on_Cu_ACSNano,30TBG_grown_onCu_2dmat}  and the state-of-art fabrication technology of graphene-based nanostructures\cite{Ritter2009,Ruffieux2016,Nguyen2017}, we expect that the hybridization selection rules and the electron-hole asymmetrical effect of hybridization strength are verified experimentally by optical absorption spectrum and ARPES.

{\textit{Acknowledgements.}}---
S.Y. acknowledges the support by the National Science Foundation of China (No. 11774269). H.-Q. L. acknowledges the financial support from NSAF (No. U1930402) and NSFC (No. 11734002). G.Y. and Y.W. acknowledge the support from China Postdoctoral Science Foundation (Grant Nos. 2018M632902 and 2019M660433) and NSFC (No. 11832019). MIK acknowledges the support by the JTC-FLAGERA Project GRANSPORT and the ERC Synergy Grant, project 854843 FASTCORR.

\bibliography{main}
\end{document}


\title{Supplementary Material for ``Interlayer hybridization in graphene quasicrystal and other bilayer graphene systems''}
\author{Guodong Yu}
\affiliation{Key Laboratory of Artificial Micro- and Nano-structures of Ministry of Education and School of Physics and Technology, Wuhan University, Wuhan 430072, China}
\author{Yunhua Wang}
\email{wangyunhua@csrc.ac.cn}
\affiliation{Beijing Computational Science Research Center, Beijing, 100193, China}
\affiliation{Institute for Molecules and Materials, Radboud University, Heijendaalseweg 135, NL-6525 AJ Nijmegen, Netherlands}
\author{Mikhail I. Katsnelson}
\affiliation{Institute for Molecules and Materials, Radboud University, Heijendaalseweg 135, NL-6525 AJ Nijmegen, Netherlands}
\author{Hai-Qing Lin}
\affiliation{Beijing Computational Science Research Center, Beijing, 100193, China}
\author{Shengjun Yuan}
\email{s.yuan@whu.edu.cn}
\affiliation{Key Laboratory of Artificial Micro- and Nano-structures of Ministry of Education and School of Physics and Technology, Wuhan University, Wuhan 430072, China}
\affiliation{Beijing Computational Science Research Center, Beijing, 100193, China}
\affiliation{Institute for Molecules and Materials, Radboud University, Heijendaalseweg 135, NL-6525 AJ Nijmegen, Netherlands}

\maketitle
\tableofcontents

\section{S1. Structures}

\begin{figure}[!htbp]
\centering
\includegraphics[width=6 cm]{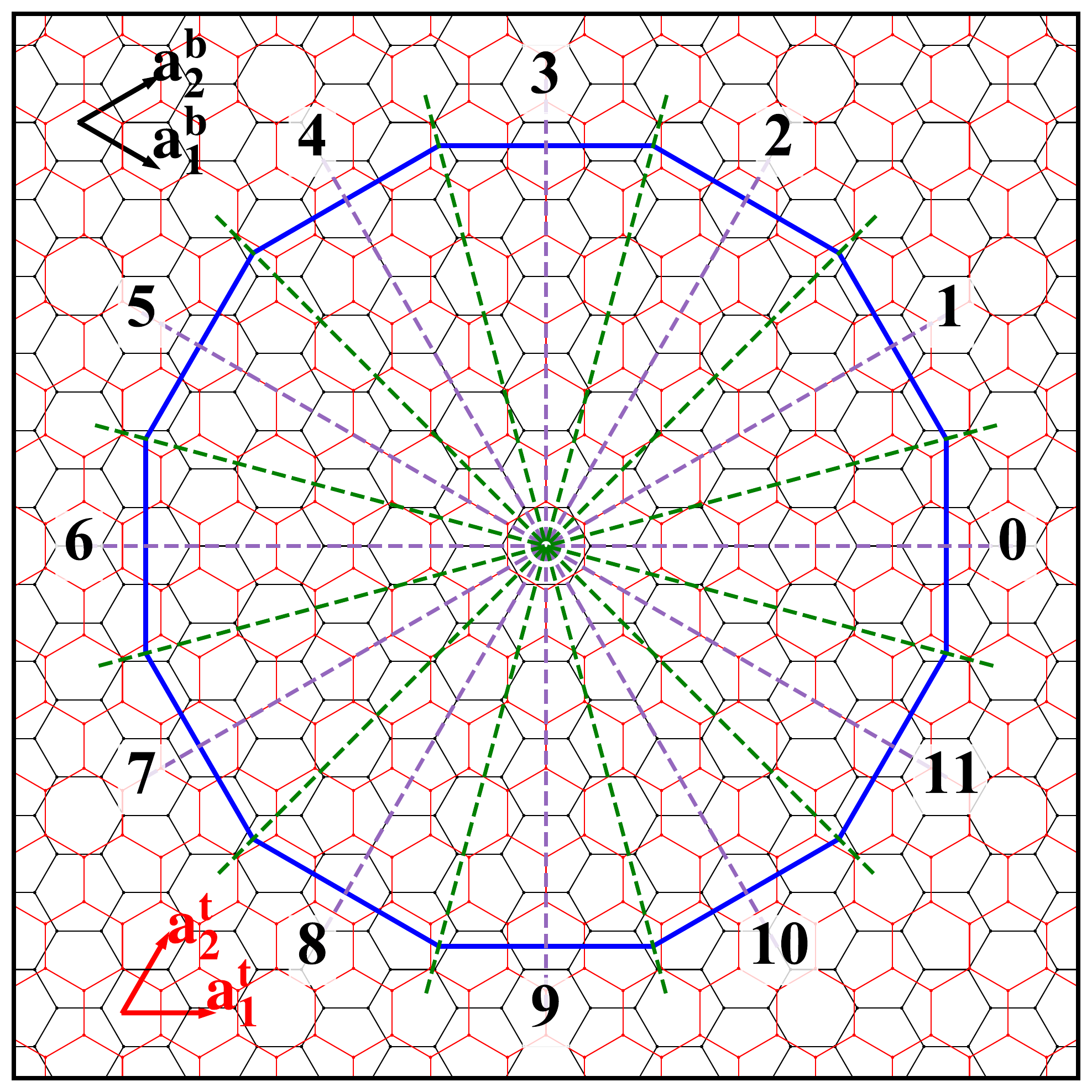}
\caption{The lattice structure and symmetry of a dodecagonal graphene quasicrystal quantum dot with $12$ sides labelled by blue color and $12$ numbers. Because the $12$ edges with even and odd numbers are respectively along the zigzag edges of the black bottom and red top monolayers, the size of the dodecagonal quantum dot can be measured by the number of zigzag chains from origin to any one side. Here, the size of the dodecagonal quantum dot with blue edges is $6$, namely size-$6$. In general, if the number of zigzag chains from center to the $12$ edges is $n$, the size of the quantum dot is labelled by size-$n$. The six $\sigma_d$ reflection planes and six $2$-fold rotational axes are represented respectively by the dashed purple and dashed green lines. For each $C_{6v}$ monolayer, the vertical planes containing the lines (0-6, 2-8, 4-10) and lines (1-7, 3-9, 5-11) are the reflection planes of three $\sigma_v$ and three $\sigma_d$ operations, respectively. The dodecagonal graphene quasicrystal quantum dot with size-$n$ remains $D_{6d}$ symmetry, and both of the bottom and top graphene quantum dots remain $C_{6v}$ symmetry.}
\label{fig:struct_30}
\end{figure}
\begin{figure}[!htbp]
\centering
\includegraphics[width=6 cm]{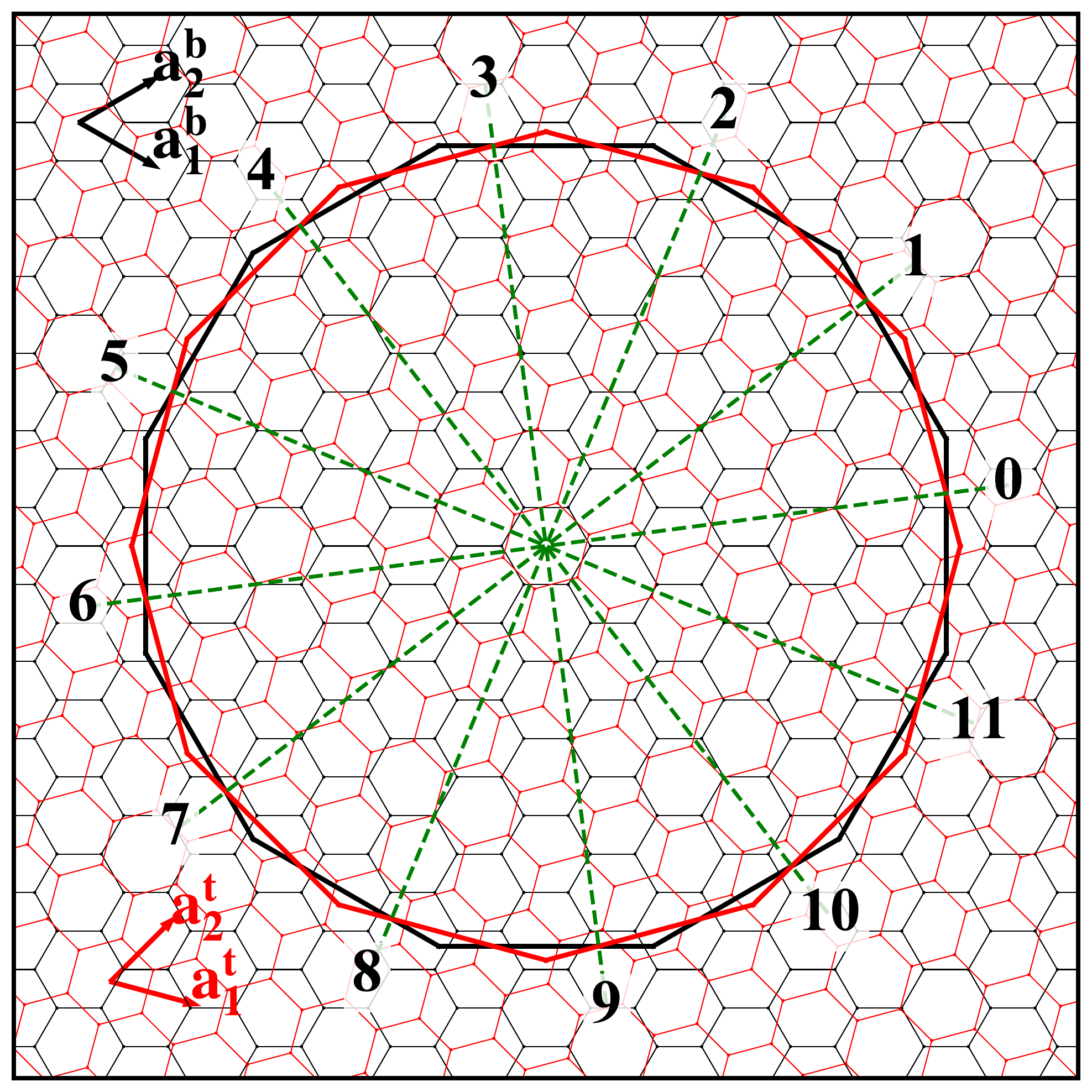}
\caption{The lattice structure and symmetry of a $D_{6}$ twisted BG quantum dot with $\theta_t=15^{\circ}$. The edges of size-$6$ quantum dot structure consist of the bottom black dodecagon and the top red dodecagon. Lines (0-6, 2-8, 4-10) and lines (1-7, 3-9, 5-11) stand for the $2$-fold axes of three $C_{2}^{\prime}$ and three $C_{2}^{\prime\prime}$ rotations, respectively.}
\label{fig:struct_15}
\end{figure}

\begin{figure}[!htbp]
\centering
\includegraphics[width=6 cm]{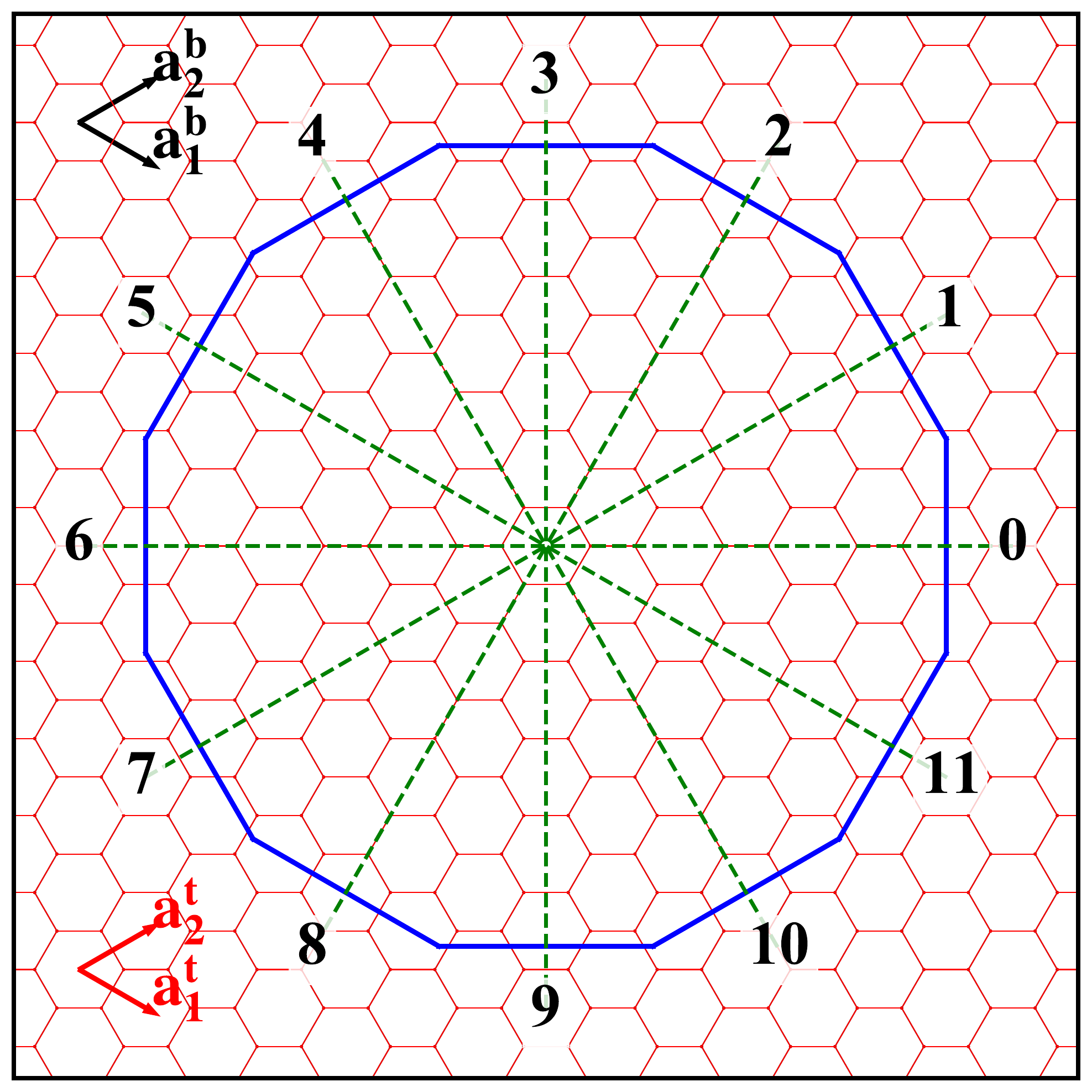}
\caption{The lattice structure and symmetry of a $D_{6h}$ untwisted BG quantum dot. The edges of size-$6$ quantum dot are the blue dodecagons for both two layers from top view. The vertical planes containing the lines (0-6, 2-8, 4-10) and lines (1-7, 3-9, 5-11) are the reflection planes of three $\sigma_v$ and three $\sigma_d$. Lines of (0-6, 2-8, 4-10) and lines of (1-7, 3-9, 5-11) stand for the 2-fold axes of three $C_{2}^{\prime}$ and three $C_{2}^{\prime\prime}$ rotations, respectively.}
\label{fig:struct_0}
\end{figure}

Fig. \ref{fig:struct_30} shows the structure and symmetry operations of graphene quasicrystal ($30^{\circ}$ twisted BG) and its quantum dot. The lattice vectors for bottom and top graphene monolayers are $\bm{a}_1^b=\frac{\sqrt{3}a}{2}\bm{i}-\frac{a}{2}\bm{j},\,\bm{a}_2^b=\frac{\sqrt{3}a}{2}\bm{i}+\frac{a}{2}\bm{j}$ and $\bm{a}_1^t=a\bm{i}+0\bm{j},\,\bm{a}_2^t=\frac{a}{2}\bm{i}+\frac{\sqrt{3}a}{2}\bm{j}$, respectively, where $a=2.46$ {\AA} is the lattice constant of graphene, and $h=3.35$ {\AA} is the interlayer distance. The relative positions of the sublattices in a unit cell for bottom and top graphene monolayers are $\bm{\tau}_{A}^b=\frac{1}{3}\bm{a}_1^b+\frac{1}{3}\bm{a}_2^b, \,\bm{\tau}_{B}^b=\frac{2}{3}\bm{a}_1^b+\frac{2}{3}\bm{a}_2^b$ and $\bm{\tau}_{A}^t=\frac{1}{3}\bm{a}_1^t+\frac{1}{3}\bm{a}_2^t, \, \bm{\tau}_{B}^t=\frac{2}{3}\bm{a}_1^t+\frac{2}{3}\bm{a}_2^t$, respectively. The rotation center is located at the hexagon center of both two layers. The graphene quasicrystal has $D_{6d}$ symmetry, and the graphene monolayers have $C_{6v}$ symmetry. The graphene quasicrystal quantum dot is customized with the same $D_{6d}$ symmetry. The character tables of $C_{6v}$ and $D_{6d}$ are listed in Tables~\ref{table:chi_C6v} and \ref{table:chi_D6d}, respectively.

A twisted BG with $0^{\circ}<\theta_t<30^{\circ}$ is obtained by rotating the top layer of graphene quasicrystal with a angle of $30^{\circ}-\theta_t$ clockwise, as shown in Fig. \ref{fig:struct_15}, where a twisted BG is generated with $\theta_t=15^{\circ}$ as an example. The twisted BG with $0^{\circ}<\theta_t<30^{\circ}$ has $D_6$ symmetry, and the twisted BG quantum dot is customized with the same $D_6$ symmetry. The character table of $D_{6}$ is listed in Tables~\ref{table:chi_D6}.

The untwisted BG is the AA stacking BG with $\theta_t=0^{\circ}$ and $D_{6h}$ symmetry. Fig. \ref{fig:struct_0} shows the structure and symmetry operations of untwisted BG, where the untwisted BG quantum dot is customized with the same $D_{6h}$ symmetry. The character table of $D_{6h}$ is listed in Tables~\ref{table:chi_D6h}.

\section{S2. Derivation of interlayer hybridization selection rules}
\subsection{S2.1. Hybridization selection rules in $D_{6d}$ graphene quasicrystals}

\begin{table*}[htbp!]
\caption{Character table of $C_{6v}$.}
\centering
\setlength{\tabcolsep}{7mm}{
\begin{tabular}{c c c c c c c}
\hline\hline
 $C_{6v}$ & $E$ & 2$C_{6}$ & 2$C_3$ & $C_2$ & 3$\sigma_v$ & 3$\sigma_d$ \\
 \hline
 $A_1$ & 1 &  1 &  1 &  1 &  1 &  1 \\
 $A_2$ & 1 &  1 &  1 &  1 & -1 & -1 \\
 $B_1$ & 1 & -1 &  1 & -1 &  1 & -1 \\
 $B_2$ & 1 & -1 &  1 & -1 & -1 &  1 \\
 $E_1$ & 2 &  1 & -1 & -2 &  0 &  0 \\
 $E_2$ & 2 & -1 & -1 &  2 &  0 &  0 \\
 \hline\hline
\end{tabular}}
\label{table:chi_C6v}
\end{table*}
\begin{table*}[htbp!]
\caption{Character table of $D_{6d}$.}
\centering
\setlength{\tabcolsep}{6mm}{
\begin{tabular}{c c c c c c c c c c c c}
\hline\hline
 $D_{6d}$ & $E$ & 2$S_{12}$ & 2$C_6$ & 2$S_4$ & 2$C_3$ & 2$S_{12}^5$ & $C_2$ & 6$C_2^{\prime}$ & 6$\sigma_d$ \\
 \hline
 $A_1$ & 1 &          1 &  1 &  1 &  1 &           1 &  1 &  1 &  1\\
 $A_2$ & 1 &          1 &  1 &  1 &  1 &           1 &  1 & -1 & -1\\
 $B_1$ & 1 &         -1 &  1 & -1 &  1 &          -1 &  1 &  1 & -1\\
 $B_2$ & 1 &         -1 &  1 & -1 &  1 &          -1 &  1 & -1 &  1 \\
 $E_1$ & 2 & $\sqrt{3}$ &  1 &  0 & -1 & -$\sqrt{3}$ & -2 &  0 &  0 \\
 $E_2$ & 2 &	          1 & -1 & -2 & -1 &           1 &  2 & 	0 &  0\\
 $E_3$ & 2 &          0 & -2 &  0 &  2 &           0 & -2 &  0 &  0 \\
 $E_4$ & 2 &         -1 & -1 &  2 & -1 &          -1 &  2 & 	0 &  0 \\
 $E_5$ & 2 & -$\sqrt{3}$ & 1& 0&	-1& 	$\sqrt{3}$ &	-2 & 0& 	0\\
  \hline\hline
  \end{tabular}}
 \label{table:chi_D6d}
 \end{table*}

The incommensurate 30$^{\circ}$ twisted BG has $D_{6d}$ point group symmetry in Table~\ref{table:chi_D6d}. Due to the 30$^{\degree}$ twist angle, the mirror reflections of the two layers have the relationships of $\sigma_{v,i}^{t}=\sigma_{d,i}^{b}$ and $\sigma_{d,i}^t=\sigma_{v,j}^b$ with $j=(i+1)$ mod $3$, where $i,j=0,1,2$. From the character table of $C_{6v}$ in Table~\ref{table:chi_C6v}, one can find that the projection operators of the two layers have the following properties:

(i) for irrep $ir\in \{A_1, A_2, E_1,E_2\}$, the same characters for $\sigma_{v,i}$ and $\sigma_{d,i}$ enable
\begin{equation}
P_{ir}^{C_{6v}^b} = P_{ir}^{C_{6v}^t};
\label{Proj:pD6dp1}
\end{equation}

(ii) for irreps $B_1$ and $B_2$, the opposite characters for $\sigma_{v,i}$ and $\sigma_{d,i}$ enable
\begin{equation}
P_{B_1}^{C_{6v}^b} = P_{B_2}^{C_{6v}^t},\ P_{B_2}^{C_{6v}^b} = P_{B_1}^{C_{6v}^t};
\label{Proj:pD6dp2}
\end{equation}

(iii) for irrep $ir\in \{A_1, A_2, B_1, B_2, E_1, E_2\}$, because the characters of all operations are real numbers and an arbitrary symmetry operation and its inverse operation are inside the same class, the projection operators satisfy
\begin{equation}
\left[P_{ir}^{C_{6v}^b}\right]^{\dagger}=P_{ir}^{C_{6v}^b},\ \left[P_{ir}^{C_{6v}^t}\right]^{\dagger}=
P_{ir}^{C_{6v}^t};
\label{Proj:pD6dp3}
\end{equation}

(iv) for irrep $ir\in \{A_1, A_2, B_1, B_2, E_1, E_2\}$, because $C_{6v}$ is a subgroup of $D_{6d}$, and hence all symmetry operations in $C_{6v}$ commute with the Hamiltonian $H$, i.e,
\begin{equation}
[P_{ir}^{C_{6v}^b}, H] = [P_{ir}^{C_{6v}^t}, H]=0.
\label{Proj:pD6dp4}
\end{equation}
Using Eq.~\eqref{Proj:pD6dp1}, Eq.~\eqref{Proj:pD6dp3} and Eq.~\eqref{Proj:pD6dp4} for irrep $ir\in \{A_1, A_2, E_1,E_2\}$, we write the hybridization matrix element $U_{ir,{ir}^\prime}$ as
\begin{equation}
U_{ir, ir^{\prime}} = \braket{\varphi_{ir}^{b}|H|\varphi_{ir^{\prime}}^t} = \braket{\varphi_{ir}^{b}|P_{ir}^{C_{6v}^b}H|\varphi_{ir^{\prime}}^t} = \braket{\varphi_{ir}^{b}|HP_{ir}^{C_{6v}^b}|\varphi_{ir^{\prime}}^t} = \braket{\varphi_{ir}^{b}|HP_{ir}^{C_{6v}^t}|\varphi_{ir^{\prime}}^t} = \delta_{ir,ir^{\prime}}U_{ir,ir^{\prime}}.
\label{D6d_ruleAE}
\end{equation}
Using Eq.~\eqref{Proj:pD6dp2}, Eq.~\eqref{Proj:pD6dp3} and Eq.~\eqref{Proj:pD6dp4} for irreps $B_1$ and $B_2$, we write the hybridization matrix elements as
\begin{equation}
\begin{split}
U_{B_1, ir^{\prime}} = \braket{\varphi_{B_1}^{b}|H|\varphi_{ir^{\prime}}^t} = \braket{\varphi_{B_1}^{b}|P_{B_1}^{C_{6v}^b}H|\varphi_{ir^{\prime}}^t} = \braket{\varphi_{ir}^{b}|HP_{B_1}^{C_{6v}^b}|\varphi_{ir^{\prime}}^t} = \braket{\varphi_{ir}^{b}|HP_{B_2}^{C_{6v}^t}|\varphi_{ir^{\prime}}^t} = \delta_{B_2,ir^{\prime}}U_{B_1,ir^{\prime}},\\
U_{B_2, ir^{\prime}} = \braket{\varphi_{B_2}^{b}|H|\varphi_{ir^{\prime}}^t} = \braket{\varphi_{B_2}^{b}|P_{B_2}^{C_{6v}^b}H|\varphi_{ir^{\prime}}^t} = \braket{\varphi_{ir}^{b}|HP_{B_2}^{C_{6v}^b}|\varphi_{ir^{\prime}}^t} = \braket{\varphi_{ir}^{b}|HP_{B_1}^{C_{6v}^t}|\varphi_{ir^{\prime}}^t} = \delta_{B_1,ir^{\prime}}U_{B_2,ir^{\prime}}.
\end{split}
\label{D6d_ruleB}
\end{equation}
The constraint equations of the hybridization matrix elements in Eq.~\eqref{D6d_ruleAE} and Eq.~\eqref{D6d_ruleB} endow hybridization rules for these states from the top and bottom layers, i.e., the hybridization selection rule in $D_{6d}$ graphene quasicrystal.

\subsection{S2.2. Hybridization selection rules in $D_6$ twisted BG with $0^{\circ}<\theta_t<30^{\circ}$}

The twisted BG with $0^{\circ}<\theta_t<30^{\circ}$ has $D_6$ point group symmetry in Table~\ref{table:chi_D6}. The intersection between $C_{6v}$ and $D_6$ is $C_6$. From the character tables of $C_{6v}$ and $C_6$ in corresponding Table~\ref{table:chi_C6v} and Table~\ref{table:chi_C6}, one can find that, for 1D irrep $X_i$ with $i=1,2$ and $X=A$ or $B$,
\begin{equation}
\begin{split}
P_{A}^{C_6} \varphi_{X_i}^{b/t} &= {{1}\over{6}}\sum_{i=0}^{5}C_6^i  \varphi_{X_i}^{b/t} = \delta_{A,X}\varphi_{X_i}^{b/t}=(\delta_{A_1,X_i}+\delta_{A_2,X_i})\varphi_{X_i}^{b/t}, \\
P_{B}^{C_6} \varphi_{X_i}^{b/t} &= {{1}\over{6}}\sum_{i=0}^{5}(-1)^iC_6^i   \varphi_{B_i}^{b/t} = \delta_{B,X}\varphi_{X_i}^{b/t}=(\delta_{B_1,X_i}+\delta_{B_2,X_i})\varphi_{X_i}^{b/t}. \\
\end{split}
\label{Proj:pC6}
\end{equation}
Because $C_6$ is a subgroup of $D_6$, all symmetry operations in $C_6$ commute with the Hamiltonian $H$, and hence the projection operators $P_{A}^{C_6}$ and $P_{B}^{C_6}$ also commute with $H$, i.e,
\begin{equation}
[P_{A}^{C_6}, H] = [P_{B}^{C_6}, H]=0.
\label{Proj:pC6Commu}
\end{equation}
In addition, from the character tables of $C_6$, one can also find that,
\begin{equation}
\left[P_{A}^{C_6}\right]^{\dagger}=P_{A}^{C_6},\ \left[P_{B}^{C_6}\right]^{\dagger}=P_{B}^{C_6}.
\label{Proj:pC6Con}
\end{equation}
Using Eqs.~\eqref{Proj:pC6}-\eqref{Proj:pC6Con}, we write the hybridization matrix elements for 1D irreps $X_i$ with $X=A$ or $B$ as
\begin{equation}
\begin{split}
U_{A_i, X_j} &= \braket{\varphi_{A_i}^{b}|H|\varphi_{X_j}^t} = \braket{\varphi_{A_i}^{b}|P_{A}^{C_{6}}H|\varphi_{X_j}^t} = \braket{\varphi_{A_i}^{b}|HP_{A}^{C_{6}}|\varphi_{X_j}^t} = \delta_{A,X}U_{A_i,X_j}=(\delta_{A_1,X_i}+\delta_{A_2,X_i})U_{A_i,X_j},\\
U_{B_i, X_j} &= \braket{\varphi_{B_i}^{b}|H|\varphi_{X_j}^t} = \braket{\varphi_{B_i}^{b}|P_{B}^{C_{6}}H|\varphi_{X_j}^t} = \braket{\varphi_{B_i}^{b}|HP_{B}^{C_{6}}|\varphi_{X_j}^t} = \delta_{B,X}U_{B_i,X_j}=(\delta_{B_1,X_i}+\delta_{B_2,X_i})U_{B_i,X_j}.
\end{split}
\label{D6_AB_rule}
\end{equation}
For 2D irreps $E_i$ with $i=1,2$, because all the characters of the operation classes $\sigma_{v}$ and $\sigma_{d}$ in $C_{6v}$ and the operation classes $C_2^{\prime}$ and $C_2^{\prime\prime}$ in $D_6$ are 0, the projection operators for irreps $E_i$ in $C_{6v}$ and $D_6$ have following properties
\begin{equation}
\begin{split}
P_{E_i}^{C_{6v}^{b}}&=P_{E_i}^{C_{6v}^{t}}=P_{E_i}^{D_{6}},\\
\left[P_{E_i}^{C_{6v}^b}\right]^{\dagger}&=P_{E_i}^{C_{6v}^b}=\left[P_{E_i}^{C_{6v}^t}\right]^{\dagger}=
P_{E_i}^{C_{6v}^t},\\
[P_{E_i}^{C_{6v}^{b}}, H] &= [P_{E_i}^{C_{6v}^{t}}, H]=0.
\end{split}
\label{Proj:pD6}
\end{equation}
Using Eq.~\eqref{Proj:pD6} we write the hybridization matrix elements for 2D irreps $E_i$ as
\begin{equation}
U_{E_i,E_j} = \braket{\varphi_{E_i}^{b}|H|\varphi_{E_j}^{t}} = \braket{\varphi_{E_i}^{b}|P_{E_i}^{C_{6v}^b}H|\varphi_{E_j}^{t}}= \braket{\varphi_{E_i}^{b}|HP_{E_i}^{C_{6v}^t}|\varphi_{E_j}^{t}} = \delta_{E_i, E_j} U_{E_i, E_j}.
\label{D6_E_rule}
\end{equation}
The constraint equations of the hybridization matrix elements in Eq.~\eqref{D6_AB_rule} and Eq.~\eqref{D6_E_rule} endow hybridization rules for these states from the top and bottom layers, i.e., the hybridization selection rule in $D_6$ twisted BG.

\begin{table*}[htbp!]
\caption{Character tables of $D_6$.}
\centering
\setlength{\tabcolsep}{7mm}{
\begin{tabular}{c c c c c c c}
\hline\hline
 $D_{6}$ & $E$ & 2$C_{6}$ & 2$C_3$ & $C_2$ & 3$C_{2}^{\prime}$ & 3$C_{2}^{\prime\prime}$ \\
 \hline
 $A_1$ & 1 &  1 &  1 &  1 &  1 &  1 \\
 $A_2$ & 1 &  1 &  1 &  1 & -1 & -1 \\
 $B_1$ & 1 & -1 &  1 & -1 &  1 & -1 \\
 $B_2$ & 1 & -1 &  1 & -1 & -1 &  1 \\
 $E_1$ & 2 &  1 & -1 & -2 &  0 &  0 \\
 $E_2$ & 2 & -1 & -1 &  2 &  0 &  0 \\
  \hline\hline
  \end{tabular}}
 \label{table:chi_D6}
 \end{table*}
\begin{table*}[htbp!]
\caption{Character table of $C_{6}$. Here $\varepsilon=e^{{\pi i}\over{3}}$ in this table.}
\centering
\setlength{\tabcolsep}{7mm}{
\begin{tabular}{c c c c c c c}
\hline\hline
 $C_{6}$ & $E$ & $C_{6}$ & $C_3$ & $C_2$ & $C_3^{2}$ & 3$C_{6}^{5}$ \\
 \hline
 $A$ & 1 &  1 &  1 &  1 &  1 &  1 \\
 $B$ & 1 & -1 &  1 & -1 &  1 & -1 \\
 \multicolumn{1}{c}{\multirow{2}{*}{$E_1$}} & 1 &  $\varepsilon$ & -$\varepsilon^*$ & -1 & -$\varepsilon$ & $\varepsilon^*$ \\
 \multicolumn{1}{c}{} & 1 & $\varepsilon^*$ &  -$\varepsilon$ & -1 & -$\varepsilon^*$ & $\varepsilon$ \\
 \multicolumn{1}{c}{\multirow{2}{*}{$E_2$}} & 1 & -$\varepsilon^*$ & -$\varepsilon$ &  1 & -$\varepsilon^*$ &-$\varepsilon$ \\
 \multicolumn{1}{c}{} & 1 & -$\varepsilon$ & -$\varepsilon^*$ &  1 &  -$\varepsilon$ &  -$\varepsilon^*$ \\
  \hline \hline
  \end{tabular}}
 \label{table:chi_C6}
 \end{table*}

\subsection{S2.3. Hybridization selection rules in $D_{6h}$ untwisted BG}

Graphene monolayer has $C_{6v}$ point group symmetry (see Table~\ref{table:chi_C6v}). The untwisted bilayer graphene (BG), i.e, AA stacking BG with $\theta_t=0^\circ$, has $D_{6h}$ point group symmetry in Table~\ref{table:chi_D6h}. The local projection operators $P_{ir}^{C_{6v}^b}$ and $P_{ir}^{C_{6v}^t}$ from the bottom and top layers in untwisted BG have following three properties, where $ir$ is the irreducible representation (irrep), and $b (t)$ denotes the bottom (top) layer.

(i) Because of the same local axes in bottom and top layers, $P_{ir}^{C_{6v}^b}$ and $P_{ir}^{C_{6v}^t}$ should be the same, i.e.,
\begin{equation}
P_{ir}^{C_{6v}^b} = P_{ir}^{C_{6v}^t}.
\label{Proj:p1}
\end{equation}

(ii) Because in $C_{6v}$ the characters of all operations are real numbers and an arbitrary symmetry operation and its inverse operation are inside the same class, the projection operators satisfy
\begin{equation}
\left[P_{ir}^{C_{6v}^b}\right]^{\dagger}=P_{ir}^{C_{6v}^b}=\left[P_{ir}^{C_{6v}^t}\right]^{\dagger}=
P_{ir}^{C_{6v}^t}.
\label{Proj:p2}
\end{equation}

(iii) $C_{6v}$ is a subgroup of $D_{6h}$, and hence all symmetry operations in $C_{6v}$ commute with the Hamiltonian $H$, i.e,
\begin{equation}
[P_{ir}^{C_{6v}^b}, H] = [P_{ir}^{C_{6v}^t}, H]=0.
\label{Proj:p3}
\end{equation}
Using Eqs.~\eqref{Proj:p1}-\eqref{Proj:p3}, we write the hybridization matrix element $U_{ir,{ir}^\prime}$ in untwisted BG as
\begin{equation}
U_{ir, ir^{\prime}} = \braket{\varphi_{ir}^{b}|H|\varphi_{ir^{\prime}}^t} = \braket{\varphi_{ir}^{b}|P_{ir}^{C_{6v}^b}H|\varphi_{ir^{\prime}}^t} = \braket{\varphi_{ir}^{b}|HP_{ir}^{C_{6v}^b}|\varphi_{ir^{\prime}}^t} = \braket{\varphi_{ir}^{b}|HP_{ir}^{C_{6v}^t}|\varphi_{ir^{\prime}}^t} = \delta_{ir,ir^{\prime}}U_{ir,ir^{\prime}}.
\label{rule_D6h}
\end{equation}
The constraint equation of $U_{ir,{ir}^\prime}$ in Eq.~\eqref{rule_D6h} endows a hybridization rule for these states from the top and bottom layers, i.e., the hybridization selection rule in untwisted BG.
\begin{table*}[htbp!]
\caption{Character table of $D_{6h}$.}
\centering
\setlength{\tabcolsep}{4mm}{
\begin{tabular}{c c c c c c c c c c c c c c c c c}
\hline\hline
 $D_{6h}$ & $E$ & 2$C_{6}$ & 2$C_3$ & $C_2$ & 3$C_2^{\prime}$ & 3$C_{2}^{\prime\prime}$ & $i$ & 2$S_3$ & 2$S_6$ & $\sigma_h$ & 3$\sigma_d$ & 3$\sigma_v$ \\
 \hline
 $A_{1g}$ &  1 &  1 &  1 &  1 &  1 &  1 &  1 &  1 &  1 &  1 &  1 &  1 \\
 $A_{2g}$ &  1 &  1 &  1 &  1 & -1 & -1 &  1 &  1 &  1 &  1 & -1 & -1 \\
 $B_{1g}$ &  1 & -1 &  1 & -1 &  1 & -1 &  1 & -1 &  1 & -1 &  1 & -1 \\
 $B_{2g}$ &  1 & -1 &  1 & -1 & -1 &  1 &  1 & -1 &  1 & -1 & -1 &  1 \\
 $E_{1g}$ &  2 &  1 & -1 & -2 &  0 &  0 &  2 &  1 & -1 & -2 &  0 &  0 \\
 $E_{2g}$ &  2 & -1 & -1 &  2 &  0 &  0 &  2 & -1 & -1 &  2 &  0 &  0 \\
 $A_{1u}$ &  1 &  1 &  1 &  1 &  1 &  1 & -1 & -1 & -1 & -1 & -1 & -1 \\
 $A_{2u}$ &  1 &  1 &  1 &  1 & -1 & -1 & -1 & -1 & -1 & -1 &  1 &  1 \\
 $B_{1u}$ &  1 & -1 &  1 & -1 &  1 & -1 & -1 &  1 & -1 &  1 & -1 &  1 \\
 $B_{2u}$ &  1 & -1 &  1 & -1 & -1 &  1 & -1 &  1 & -1 &  1 &  1 & -1 \\
 $E_{1u}$ &  2 &  1 & -1 & -2 &  0 &  0 & -2 & -1 &  1 &  2 &  0 &  0 \\
 $E_{2u}$ &  2 & -1 & -1 &  2 &  0 &  0 & -2 &  1 &  1 & -2 &  0 &  0 \\
  \hline \hline
  \end{tabular}}
 \label{table:chi_D6h}
\end{table*}

\section{S3. Determination of irreps for bonding and antibonding states}
Usually, for a hybridization process $ir_1+ir_2\Rightarrow ir_3+ir_4$, where two + signs denote paring, the hybridization paring orbitals with irreps $ir_1$ and $ir_2$ are given by left two terms, and the bonding and antibonding paring states with irreps $ir_3$ and $ir_4$ are denoted by right two terms. Next, we will list the hybridization categories and determine the irreps of bonding and antibonding states generated by the interlayer hybridization in both untwisted BG and twisted BG.
\subsection{S3.1. Irreps of bonding and antibonding states in $D_{6d}$ graphene quasicrystal}
Following the hybridization selection rule in Eqs.~\eqref{D6d_ruleAE} and ~\eqref{D6d_ruleB} and using the character tables of $C_{6v}$ in Table~\ref{table:chi_C6v} and $D_{6d}$ in Table~\ref{table:chi_D6d}, we write the equivalent hybridizations as
\begin{equation}
\begin{split}
A_1 + A_1 \Rightarrow A_{1} + B_{2}, \\
A_2 + A_2 \Rightarrow A_{2} + B_{1}, \\
E_1 + E_1 \Rightarrow E_{1} + E_{5}, \\
E_2 + E_2 \Rightarrow E_{2} + E_{4}, \\
\end{split}
\label{hyb:D6d_equi}
\end{equation}
and the non-equivalent hybridizations as
\begin{equation}
\begin{split}
B_1 + B_2 \Rightarrow E_{3} + E_{3}, \\
B_2 + B_1 \Rightarrow E_{3} + E_{3}. \\
\end{split}
\label{hyb:D6d_nonequi}
\end{equation}
For the equivalent hybridization paring states, $ir+ir$, with $ir=A_1, A_2, E_1$ and $E_2$ in Eq.~\eqref{hyb:D6d_equi}, their bonding $(+)$ and antibonding $(-)$ states can be written as
\begin{equation}
\left|\phi_{\pm}^{ir}\right\rangle =\frac{1}{\sqrt{2}} \left( e^{i\frac{\theta}{2}}\left|\varphi_{ir}^b\right\rangle \pm S_{12}\left|\varphi_{ir}^b\right\rangle\right),
\end{equation}
where $\left|\varphi_{ir}^b\right\rangle$ is the eigenstate of $C_6$ inside the bottom layer with $C_6 \left|\varphi_{ir}^b\right\rangle = e^{i\theta}\left|\varphi_{ir}^b\right\rangle$. Using $S_{12}S_{12}=C_6$, we have $S_{12}\left|\phi_{\pm}^{ir}\right\rangle=\pm e^{i\frac{\theta}{2}} \left|\phi_{\pm}^{ir}\right\rangle$. For each $C_{6v}$ monolayer, we can also find the irrep $ir$ for each eigenstate of $C_6$ with the corresponding eigenvalue $e^{i\theta}$, as follows: (i) $ir=A_1 (A_2)$, for $\theta=0$; (ii) $ir=E_1$, for $\theta=\pm\pi/3$; and (iii) $ir=E_2$, for $\theta=\pm2\pi/3$. Now we use the projection operator $P_{ir}^{D_{6d}}$ to determine which irreps the two states $\left|\phi_{\pm}^{ir}\right\rangle$ belong to in $D_{6d}$ point group.

(i) For $A_1+A_1\Rightarrow A_1 +B_2$ hybridization with $\theta=0$, the projection operators for irreps $A_1$ and $B_2$ in $D_{6d}$ take the form as
\begin{equation}
\begin{split}
P_{A_1}^{D_{6d}} &= \frac{1}{2} P_{A_1}^{C_{6v}} +\frac{1}{24}\left( \sum_{i=0}^{5} S_{12}^{2i+1} + \sum_{i=0}^{5}C_{2,i}^{\prime} \right), \\
P_{B_2}^{D_{6d}} &= \frac{1}{2} P_{A_1}^{C_{6v}} -\frac{1}{24}\left( \sum_{i=0}^{5} S_{12}^{2i+1} + \sum_{i=0}^{5}C_{2,i}^{\prime} \right), \\
\end{split}
\label{eq:D6d_PA1}
\end{equation}
with
\begin{equation}
S_{12}^{2i+1}\left|\phi_{\pm}^{A_1}\right\rangle =\pm\left|\phi_{\pm}^{A_1}\right\rangle,\,
C_{2,i}^{\prime}\left|\phi_{\pm}^{A_1}\right\rangle=
S_{12}\sigma_{d,i}\left|\phi_{\pm}^{A_1}\right\rangle=\pm\left|\phi_{\pm}^{A_1}\right\rangle.
\label{eq:D6d_PA1f}
\end{equation}
Using Eqs.~\eqref{eq:D6d_PA1} and~\eqref{eq:D6d_PA1f}, we have
\begin{equation}
\begin{split}
P_{A_1}^{D_{6d}} \left|\phi_{\pm}^{A_1}\right\rangle &= \frac{1}{2} \left|\phi_{\pm}^{A_1}\right\rangle \pm \frac{1}{2} \left|\phi_{\pm}\right\rangle, \\
P_{B_2}^{D_{6d}} \left|\phi_{\pm}^{A_1}\right\rangle &= \frac{1}{2} \left|\phi_{\pm}^{A_1}\right\rangle \mp \frac{1}{2} \left|\phi_{\pm}\right\rangle.
\end{split}
\label{eq:D6d_PA1t}
\end{equation}
Eq.~\eqref{eq:D6d_PA1t} indicates that $|\phi_{+}^{A_1}\rangle$ and $|\phi_{-}^{A_1}\rangle$ generated by $A_1+A_1$ hybridization have irreps $A_1$ and $B_2$, respectively, in $D_{6d}$ point group.

(ii) For $A_2+A_2\Rightarrow A_2 +B_1$ hybridization with $\theta=0$, the projection operators for $A_2$ and $B_1$ in $D_{6d}$ read
\begin{equation}
\begin{split}
P_{A_2}^{D_{6d}} &= \frac{1}{2} P_{A_2}^{C_{6v}} +\frac{1}{24}\left( \sum_{i=0}^{5} S_{12}^{2i+1} - \sum_{i=0}^{5}C_{2,i}^{\prime} \right), \\
P_{B_1}^{D_{6d}} &= \frac{1}{2} P_{A_2}^{C_{6v}} -\frac{1}{24}\left( \sum_{i=0}^{5} S_{12}^{2i+1} - \sum_{i=0}^{5}C_{2,i}^{\prime} \right), \\
\end{split}
\label{eq:D6d_PA2}
\end{equation}
with
\begin{equation}
S_{12}^{2i+1}\left|\phi_{\pm}^{A_2}\right\rangle =\pm\left|\phi_{\pm}^{A_2}\right\rangle,\,
C_{2,i}^{\prime}\left|\phi_{\pm}^{A_2}\right\rangle =S_{12}\sigma_{d,i}\left|\phi_{\pm}^{A_2}\right\rangle=\mp\left|\phi_{\pm}^{A_2}\right\rangle.
\label{eq:D6d_PA2f}
\end{equation}
Using Eqs.~\eqref{eq:D6d_PA2} and~\eqref{eq:D6d_PA2f}, we have
\begin{equation}
\begin{split}
P_{A_2}^{D_{6d}} \left|\phi_{\pm}^{A_2}\right\rangle &= \frac{1}{2} \left|\phi_{\pm}^{A_2}\right\rangle \pm \frac{1}{2} \left|\phi_{\pm}^{A_2}\right\rangle, \\
P_{B_1}^{D_{6d}} \left|\phi_{\pm}^{A_2}\right\rangle &= \frac{1}{2} \left|\phi_{\pm}^{A_2}\right\rangle \mp \frac{1}{2} \left|\phi_{\pm}^{A_2}\right\rangle.
\end{split}
\label{eq:D6d_PA2t}
\end{equation}
Eq.~\eqref{eq:D6d_PA2t} indicates that $|\phi_{+}^{A_2}\rangle$ and $|\phi_{-}^{A_2}\rangle$ generated by $A_2+A_2$ hybridization have irreps $A_2$ and $B_1$, respectively, in $D_{6d}$ point group.

(iii) For $E_1+E_1\Rightarrow E_1 +E_5$ hybridization with $\theta=\pm\frac{\pi}{3}$, the projection operators for irrep $E_1$ and $E_5$ in $D_{6d}$ read
\begin{equation}
\begin{split}
P_{E_1}^{D_{6d}} &= \frac{1}{2} P_{E_1}^{C_{6v}} +\frac{2}{24}\left( \sqrt{3}S_{12}+\sqrt{3}S_{12}^{11} - \sqrt{3}S_{12}^{5} - \sqrt{3}S_{12}^{7} \right), \\
P_{E_5}^{D_{6d}} &= \frac{1}{2} P_{E_1}^{C_{6v}} +\frac{2}{24}\left( -\sqrt{3}S_{12}-\sqrt{3}S_{12}^{11} + \sqrt{3}S_{12}^{5} + \sqrt{3}S_{12}^{7} \right), \\
\end{split}
\label{eq:D6d_PE1}
\end{equation}
with
\begin{equation}
\left( S_{12} + S_{12}^{11} \right) \left|\phi_{\pm}^{E_1}\right\rangle =\pm \sqrt{3}\left|\phi_{\pm}^{E_1}\right\rangle,\,
\left( S_{12}^{5} + S_{12}^{7} \right) \left|\phi_{\pm}^{E_1}\right\rangle =\mp \sqrt{3}\left|\phi_{\pm}^{E_1}\right\rangle.
\label{eq:D6d_PE1f}
\end{equation}
Using Eqs.~\eqref{eq:D6d_PE1} and~\eqref{eq:D6d_PE1f}, we have
\begin{equation}
\begin{split}
P_{E_1}^{D_{6d}} \left|\phi_{\pm}^{E_1}\right\rangle &= \frac{1}{2} \left|\phi_{\pm}^{E_1}\right\rangle \pm \frac{1}{2} \left|\phi_{\pm}^{E_1}\right\rangle, \\
P_{E_5}^{D_{6d}} \left|\phi_{\pm}^{E_1}\right\rangle &= \frac{1}{2} \left|\phi_{\pm}^{E_1}\right\rangle \mp \frac{1}{2} \left|\phi_{\pm}^{E_1}\right\rangle.
\end{split}
\label{eq:D6d_PE1t}
\end{equation}
Eq.~\eqref{eq:D6d_PE1t} indicates that $|\phi_{+}^{E_1}\rangle$ and $|\phi_{-}^{E_1}\rangle$ generated by $E_1+E_1$ hybridization have ireeps $E_1$ and $E_5$, respectively, in $D_{6d}$ point group.

(iv) For $E_2+E_2\Rightarrow E_2 +E_4$ hybridization with $\theta=\pm\frac{2\pi}{3}$, the projection operators for irreps $E_2$ and $E_4$ in $D_{6d}$ are expressed as
\begin{equation}
\begin{split}
P_{E_2}^{D_{6d}} &= \frac{1}{2} P_{E_2}^{C_{6v}} +\frac{2}{24}\left( S_{12}+S_{12}^{11} -2S_{12}^{3} -2S_{12}^{9} +S_{12}^{5} + S_{12}^{7} \right), \\
P_{E_4}^{D_{6d}} &= \frac{1}{2} P_{E_2}^{C_{6v}} +\frac{2}{24}\left( -S_{12}-S_{12}^{11} +2S_{12}^{3} +2S_{12}^{9} -S_{12}^{5} - S_{12}^{7} \right), \\
\end{split}
\label{eq:D6d_PE2}
\end{equation}
with
\begin{equation}
\begin{split}
\left( S_{12} + S_{12}^{11} \right) \left|\phi_{\pm}^{E_2}\right\rangle &=\pm \left|\phi_{\pm}^{E_2}\right\rangle, \\
\left( S_{12}^{3} + S_{12}^{9} \right) \left|\phi_{\pm}^{E_2}\right\rangle &=\mp 2 \left|\phi_{\pm}^{E_2}\right\rangle, \\
\left( S_{12}^{5} + S_{12}^{7} \right) \left|\phi_{\pm}^{E_2}\right\rangle &=\pm  \left|\phi_{\pm}^{E_2}\right\rangle. \\
\end{split}
\label{eq:D6d_PE2f}
\end{equation}
Using Eqs.~\eqref{eq:D6d_PE2} and~\eqref{eq:D6d_PE2f}, we have
\begin{equation}
\begin{split}
P_{E_2}^{D_{6d}} \left|\phi_{\pm}^{E_2}\right\rangle &= \frac{1}{2} \left|\phi_{\pm}^{E_2}\right\rangle \pm \frac{1}{2} \left|\phi_{\pm}^{E_2}\right\rangle, \\
P_{E_4}^{D_{6d}} \left|\phi_{\pm}^{E_2}\right\rangle &= \frac{1}{2} \left|\phi_{\pm}^{E_2}\right\rangle \mp \frac{1}{2} \left|\phi_{\pm}^{E_2}\right\rangle.
\end{split}
\label{eq:D6d_PE2t}
\end{equation}
Eq.~\eqref{eq:D6d_PE2t} indicates that $|\phi_{+}^{E_2}\rangle$ and $|\phi_{-}^{E_2}\rangle$ generated by $E_2+E_2$ hybridization have ireeps $E_2$ and $E_4$, respectively, in $D_{6d}$ point group.

(v) For $B_1 + B_2 \Rightarrow E_{3} + E_{3}$ and $B_2 + B_1 \Rightarrow E_{3} + E_{3}$ non-equivalent hybridizations in Eq.~\eqref{hyb:D6d_nonequi}, the projection operator for irrep $E_3$ in $D_{6d}$ reads
\begin{equation}
P_{E_3}^{D_{6d}} = {{2}\over{24}} (2E - 2C_6 -2C_6^5 + 2C_3 + 2C_3^2 - 2C_2).
\label{eq:D6d_PE3}
\end{equation}
Eq.~\eqref{eq:D6d_PE3} indicates $P_{E_3}^{D_{6d}}$ is a combination of rotation operations, and these rotation operations are actually also symmetry operations of $C_{6v}$. Thus, both $|\phi_{B_1}^{b/t}\rangle$ and $|\phi_{B_2}^{b/t}\rangle$ are eigenstates of $P_{E_3}^{D_{6d}}$, i.e.,
\begin{equation}
\begin{split}
P_{E_3}^{D_{6d}} \left|\phi_{B_1}^{b/t}\right\rangle = \left|\phi_{B_1}^{b/t}\right\rangle, \\
P_{E_3}^{D_{6d}} \left|\phi_{B_2}^{b/t}\right\rangle = \left|\phi_{B_2}^{b/t}\right\rangle. \\
\end{split}
\end{equation}
Consequently, the states generated by $B_1 + B_2$ and $B_2 + B_1$ hybridizations have ireep $E_3$ in $D_{6d}$ point group.

\subsection{S3.2. Irreps of bonding and antibonding states in $D_6$ twisted BG with $0^{\circ}<\theta_t<30^{\circ}$}
Following the hybridization selection rule in Eqs.~\eqref{D6_AB_rule} and ~\eqref{D6_E_rule} and using the character tables of $C_{6v}$ in Table~\ref{table:chi_C6v} and $D_{6}$ in Table~\ref{table:chi_D6}, we write the mixed hybridizations as
\begin{equation}
\begin{split}
A_{1,2} + A_{1,2} &\Rightarrow A_{1,2} + A_{2,1},\\
B_{1,2} + B_{1,2} &\Rightarrow B_{1,2} + B_{2,1}, \\
\end{split}
\label{D6_mix}
\end{equation}
and the equivalent hybridizations as
\begin{equation}
\begin{split}
E_1 + E_1 &\Rightarrow E_1 + E_1,\\
E_2 + E_2 &\Rightarrow E_2 + E_2. \\
\end{split}
\label{D6_equiv}
\end{equation}
The mixed hybridization can be viewed as two coupled equivalent hybridizations. For instance, in Eq.~\eqref{D6_mix}, $A_{1,2} + A_{1,2} \Rightarrow A_{1,2} + A_{2,1}$ can be rewritten as
\begin{equation}
\begin{split}
A_1^{C_{6v}^b}+A_1^{C_{6v}^t} &\Rightarrow A_1^{D_6} + A_2^{D_6}, \\
A_2^{C_{6v}^b}+A_2^{C_{6v}^t} &\Rightarrow A_2^{D_6} + A_1^{D_6}.
\end{split}
\label{D6_coupA}
\end{equation}
The two $A_1$ states on the right side in Eq.~\eqref{D6_coupA} are coupled together to form the eigenstates with irrep $A_1^{D_6}$ in $D_{6}$ twisted BG. Two $A_2$ states on the right side are also coupled together. The same procedure is suitable for $B_{1,2} + B_{1,2} \Rightarrow B_{1,2} + B_{2,1}$ in Eq.~\eqref{D6_mix}, i.e.,
\begin{equation}
\begin{split}
B_1^{C_{6v}^b}+B_1^{C_{6v}^t} &\Rightarrow B_1^{D_6} + B_2^{D_6}, \\
B_2^{C_{6v}^b}+B_2^{C_{6v}^t} &\Rightarrow B_2^{D_6} + B_1^{D_6}.
\end{split}
\label{D6_coupB}
\end{equation}
Different from the independent bonding and antibonding states in equivalent hybridization the hybridization-generated states in mixed hybridization are coupled. Due to a twisted angle $\theta_t$, the hybridization-generated bonding $(+)$ and antibonding $(-)$ states for hybridization paring orbitals $ir+ir$ can be expressed as
\begin{equation}
\left|\phi_{\pm}^{ir}\right\rangle = {{1}\over{\sqrt{2}}} \left( \left|\varphi_{ir}^b\right\rangle \pm \sigma_h R(\theta_t) \left|\varphi_{ir}^b\right\rangle \right).
\label{hybrid_states_D6}
\end{equation}
Now we use the projection operator $P_{ir}^{D_{6}}$ to determine which irreps the two states $\left|\phi_{\pm}^{ir}\right\rangle$ belong to in $D_{6}$ point group.

(i) For $A_1 + A_1 \Rightarrow A_{1} + A_{2}$ hybridization, the projection operator for $A_{1}$ and $A_{2}$ in point group $D_{6}$ read
\begin{equation}
\begin{split}
P_{A_1}^{D_6} = {{1}\over{12}} \sum_{i=0}^{5} C_{6}^i + \sum_{i=0}^2 C_{2,i}^{\prime} + \sum_{i=0}^2 C_{2,i}^{\prime\prime}, \\
P_{A_2}^{D_6} = {{1}\over{12}} \sum_{i=0}^{5} C_{6}^i - \sum_{i=0}^2 C_{2,i}^{\prime} - \sum_{i=0}^2 C_{2,i}^{\prime\prime}, \\
\end{split}
\label{D6:pA}
\end{equation}
where $C_{2,i}^{\prime}=\sigma_h R({{\theta_t}\over{2}}) \sigma_{v,i}^b R^{\dagger}({{\theta_t}\over{2}})$ and $C_{2,i}^{\prime\prime}=\sigma_h R({{\theta_t}\over{2}}) \sigma_{d,i}^b R^{\dagger}({{\theta_t}\over{2}})$. Using Eq.~\eqref{D6:pA} and
\begin{equation}
\begin{split}
C_{6}^i \left|\phi_{\pm}^{A_1}\right\rangle &= \left|\phi_{\pm}^{A_1}\right\rangle, \quad \sigma_{v,i} \left|\phi_{\pm}^{A_1}\right\rangle = \left|\phi_{\pm}^{A_1}\right\rangle, \quad \sigma_{d,i} \left|\phi_{\pm}^{A_1}\right\rangle = \left|\phi_{\pm}^{A_1}\right\rangle,\\
C_{2,i}^{\prime} \left|\phi_{\pm}^{A_1}\right\rangle &= \pm \left|\phi_{\pm}^{A_1}\right\rangle, \quad C_{2,i}^{\prime\prime} \left|\phi_{\pm}^{A_1}\right\rangle = \pm \left|\phi_{\pm}^{A_1}\right\rangle,
\end{split}
\label{D6_A1A1_pre}
\end{equation}
where $\sigma_v R(\theta_t/2)=R^{\dagger}(\theta_t/2) \sigma_v$ is used, we have
\begin{equation}
\begin{split}
P_{A_1}^{D_6} \left|\phi_{\pm}^{A_1}\right\rangle = {{1}\over{2}} \left|\phi_{\pm}^{A_1}\right\rangle \pm {{1}\over{2}} \left|\phi_{\pm}^{A_1}\right\rangle, \\
P_{A_2}^{D_6} \left|\phi_{\pm}^{A_1}\right\rangle = {{1}\over{2}} \left|\phi_{\pm}^{A_1}\right\rangle \mp {{1}\over{2}} \left|\phi_{\pm}^{A_1}\right\rangle.
\end{split}
\label{D6:A1}
\end{equation}
Eq.~\eqref{D6:A1} indicates that $|\phi_{+}^{A_1}\rangle$ and $|\phi_{-}^{A_1}\rangle$ generated by $A_1+A_1$ hybridization have irreps $A_{1}$ and $A_{2}$, respectively, in $D_{6}$ point group.

(ii) For $A_2 + A_2 \Rightarrow A_{2} + A_{1}$ hybridization, using Eq.~\eqref{D6:pA} and
\begin{equation}
\begin{split}
C_{6}^i \left|\phi_{\pm}^{A_2}\right\rangle &= \left|\phi_{\pm}^{A_2}\right\rangle, \quad \sigma_{v,i} \left|\phi_{\pm}^{A_2}\right\rangle = -\left|\phi_{\pm}^{A_2}\right\rangle, \quad \sigma_{d,i} \left|\phi_{\pm}^{A_2}\right\rangle = -\left|\phi_{\pm}^{A_2}\right\rangle,\\
C_{2,i}^{\prime} \left|\phi_{\pm}^{A_2}\right\rangle &= \mp \left|\phi_{\pm}^{A_2}\right\rangle, \quad C_{2,i}^{\prime\prime} \left|\phi_{\pm}^{A_2}\right\rangle = \mp \left|\phi_{\pm}^{A_2}\right\rangle,
\end{split}
\label{D6_A2A2_pre}
\end{equation}
we have
\begin{equation}
\begin{split}
P_{A_2}^{D_6} \left|\phi_{\pm}^{A_2}\right\rangle = {{1}\over{2}} \left|\phi_{\pm}^{A_2}\right\rangle \pm {{1}\over{2}} \left|\phi_{\pm}^{A_2}\right\rangle, \\
P_{A_1}^{D_6} \left|\phi_{\pm}^{A_2}\right\rangle = {{1}\over{2}} \left|\phi_{\pm}^{A_2}\right\rangle \mp {{1}\over{2}} \left|\phi_{\pm}^{A_2}\right\rangle.
\end{split}
\label{D6:A2}
\end{equation}
Eq.~\eqref{D6:A2} indicates that $|\phi_{+}^{A_2}\rangle$ and $|\phi_{-}^{A_2}\rangle$ generated by $A_2+A_2$ hybridization have irreps $A_{2}$ and $A_{1}$, respectively, in $D_{6}$ point group.

(iii) For $B_1 + B_1 \Rightarrow B_{1} + B_{2}$ hybridization, the projection operator for $B_{1}$ and $B_{2}$ in point group $D_{6}$ read
\begin{equation}
\begin{split}
P_{B_1}^{D_6} = {{1}\over{12}} \sum_{i=0}^{5} (-1)^i C_{6}^i + \sum_{i=0}^2 C_{2,i}^{\prime} - \sum_{i=0}^2 C_{2,i}^{\prime\prime}, \\
P_{B_2}^{D_6} = {{1}\over{12}} \sum_{i=0}^{5} (-1)^i C_{6}^i - \sum_{i=0}^2 C_{2,i}^{\prime} + \sum_{i=0}^2 C_{2,i}^{\prime\prime}. \\
\end{split}
\label{D6:pB}
\end{equation}
Using Eq.~\eqref{D6:pB} and
\begin{equation}
\begin{split}
C_{6}^i \left|\phi_{\pm}^{B_1}\right\rangle &= (-1)^i \left|\phi_{\pm}^{B_1}\right\rangle, \quad \sigma_{v,i} \left|\phi_{\pm}^{B_1}\right\rangle = \left|\phi_{\pm}^{B_1}\right\rangle, \quad \sigma_{d,i} \left|\phi_{\pm}^{B_1}\right\rangle = -\left|\phi_{\pm}^{B_1}\right\rangle,\\
C_{2,i}^{\prime} \left|\phi_{\pm}^{B_1}\right\rangle &= \pm \left|\phi_{\pm}^{B_1}\right\rangle, \quad C_{2,i}^{\prime\prime} \left|\phi_{\pm}^{B_1}\right\rangle = \mp \left|\phi_{\pm}^{B_1}\right\rangle,
\end{split}
\label{D6_B1B1_pre}
\end{equation}
we have
\begin{equation}
\begin{split}
P_{B_1}^{D_6} \left|\phi_{\pm}^{B_1}\right\rangle = {{1}\over{2}} \left|\phi_{\pm}^{B_1}\right\rangle \pm {{1}\over{2}} \left|\phi_{\pm}^{B_1}\right\rangle, \\
P_{B_2}^{D_6} \left|\phi_{\pm}^{B_1}\right\rangle = {{1}\over{2}} \left|\phi_{\pm}^{B_1}\right\rangle \mp {{1}\over{2}} \left|\phi_{\pm}^{B_1}\right\rangle.
\end{split}
\label{D6:B1}
\end{equation}
Eq.~\eqref{D6:B1} indicates that $|\phi_{+}^{B_1}\rangle$ and $|\phi_{-}^{B_1}\rangle$ generated by $B_1+B_1$ hybridization have irreps $B_{1}$ and $B_{2}$, respectively, in $D_{6}$ point group.

(iv) For $B_2 + B_2 \Rightarrow B_{2} + B_{1}$ hybridization, using Eq.~\eqref{D6:pB} and
\begin{equation}
\begin{split}
C_{6}^i \left|\phi_{\pm}^{B_2}\right\rangle &= (-1)^i \left|\phi_{\pm}^{B_2}\right\rangle, \quad \sigma_{v,i} \left|\phi_{\pm}^{B_2}\right\rangle = -\left|\phi_{\pm}^{B_2}\right\rangle, \quad \sigma_{d,i} \left|\phi_{\pm}^{B_2}\right\rangle = \left|\phi_{\pm}^{B_2}\right\rangle,\\
C_{2,i}^{\prime} \left|\phi_{\pm}^{B_2}\right\rangle &= \mp \left|\phi_{\pm}^{B_2}\right\rangle, \quad C_{2,i}^{\prime\prime} \left|\phi_{\pm}^{B_2}\right\rangle = \pm \left|\phi_{\pm}^{B_2}\right\rangle,
\end{split}
\label{D6_B2B2_pre}
\end{equation}
we have
\begin{equation}
\begin{split}
P_{B_2}^{D_6} \left|\phi_{\pm}^{B_2}\right\rangle = {{1}\over{2}} \left|\phi_{\pm}^{B_2}\right\rangle \pm {{1}\over{2}} \left|\phi_{\pm}^{B_2}\right\rangle, \\
P_{B_1}^{D_6} \left|\phi_{\pm}^{B_2}\right\rangle = {{1}\over{2}} \left|\phi_{\pm}^{B_2}\right\rangle \mp {{1}\over{2}} \left|\phi_{\pm}^{B_2}\right\rangle.
\end{split}
\label{D6:B2}
\end{equation}
Eq.~\eqref{D6:B2} indicates that $|\phi_{+}^{B_2}\rangle$ and $|\phi_{-}^{B_2}\rangle$ generated by $B_2+B_2$ hybridization have irreps $B_{2}$ and $B_{1}$, respectively, in $D_{6}$ point group.

(v) For equivalent hybridizations $E_i + E_i \Rightarrow E_i + E_i$ with $i=1,2$ in Eq.~\eqref{D6_equiv}, because of $P_{E_i}^{C_{6v}^{b}} =P_{E_i}^{C_{6v}^{t}}=P_{E_i}^{D_{6}}$ in Eq.~\eqref{Proj:pD6}, $|\phi_{+}^{E_i}\rangle$ and $|\phi_{-}^{E_i}\rangle$ generated by $E_i+E_i$ hybridization have irrep $E_{i}$ in $D_{6}$ point group.

\subsection{S3.3. Irreps of bonding and antibonding states in $D_{6h}$ untwisted BG}
Following the hybridization selection rule in Eq.~\eqref{rule_D6h} and using the character tables of $C_{6v}$ in Table~\ref{table:chi_C6v} and $D_{6h}$ in Table~\ref{table:chi_D6h}, we write all the equivalent hybridizations in untwisted BG as
\begin{equation}
\begin{split}
A_1 + A_1 \Rightarrow A_{1g} + A_{2u}, \\
A_2 + A_2 \Rightarrow A_{2g} + A_{1u}, \\
B_1 + B_1 \Rightarrow B_{2g} + B_{1u}, \\
B_2 + B_2 \Rightarrow B_{1g} + B_{2u}, \\
E_1 + E_1 \Rightarrow E_{1g} + E_{1u}, \\
E_2 + E_2 \Rightarrow E_{2g} + E_{2u}. \\
\end{split}
\label{D6h_hyb}
\end{equation}
Thanks to the $\sigma_h$ operation in $D_{6h}$, the hybridization-generated bonding $(+)$ and antibonding $(-)$ states inside an equivalent hybridization of $ir+ir\Rightarrow ir_1+ir_2$ can be expressed as
\begin{equation}
\left|\phi_{\pm}^{ir}\right\rangle = {{1}\over{\sqrt{2}}} \left( \left|\varphi_{ir}^b\right\rangle \pm \sigma_h \left|\varphi_{ir}^b\right\rangle \right).
\label{hybrid_states_0}
\end{equation}
Now we use the projection operator $P_{ir}^{D_{6h}}$ to determine which irreps the two states $\left|\phi_{\pm}^{ir}\right\rangle$ belong to in $D_{6h}$ point group.

(i) For $A_1 + A_1 \Rightarrow A_{1g} + A_{2u}$ hybridization, the relationships between the character projection operator for $A_{1g}$ and $A_{2u}$ in $D_{6h}$ and $A_1$ in point group $C_{6v}$ read
\begin{equation}
\begin{split}
P_{A_{1g}}^{D_{6h}} = {{1}\over{2}} P_{A_1}^{C_{6v}} + {{1}\over{24}}\left( \sum_{j=0}^{2} C_{2,j}^{\prime} + \sum_{j=0}^{2} C_{2,j}^{\prime\prime} + i + S_{3}+ S_{3}^2 + S_6 + S_6^5 + \sigma_h  \right), \\
P_{A_{2u}}^{D_{6h}} = {{1}\over{2}} P_{A_1}^{C_{6v}} - {{1}\over{24}}\left( \sum_{j=0}^{2} C_{2,j}^{\prime} + \sum_{j=0}^{2} C_{2,j}^{\prime\prime} + i + S_{3}+ S_{3}^2 + S_6 + S_6^5 + \sigma_h  \right),
\end{split}
\label{D6h:pA1}
\end{equation}
where
\begin{equation}
\begin{split}
C_{2,j}^{\prime} = C_2 \sigma_{v,j} \sigma_h, &\quad C_{2,j}^{\prime\prime} = C_2 \sigma_{d,j} \sigma_h, \quad S_6^j = C_6^j \sigma_h,\\
i=S_6^3, &\quad S_3=S_6^2, \quad S_3^2=S_6^4.
\end{split}
\label{D6h:CS}
\end{equation}
Using Eq.~\eqref{D6h:pA1}, Eq.~\eqref{D6h:CS} and
\begin{equation}
\begin{split}
P_{A_1}^{C_{6v}}\left|\phi_{\pm}^{A_1}\right\rangle &= \left|\phi_{\pm}^{A_1}\right\rangle,\\
\sigma_h \left|\phi_{\pm}^{A_1}\right\rangle  = \pm \left|\phi_{\pm}^{A_1}\right\rangle, &\quad C_6^i \left|\phi_{\pm}^{A_1}\right\rangle = \left|\phi_{\pm}^{A_1}\right\rangle, \\
\sigma_{v,i} \left|\phi_{\pm}^{A_1}\right\rangle = \left|\phi_{\pm}^{A_1}\right\rangle, &\quad \sigma_{d,i} \left|\phi_{\pm}^{A_1}\right\rangle = \left|\phi_{\pm}^{A_1}\right\rangle, \\
\end{split}
\end{equation}
we have
\begin{equation}
\begin{split}
P_{A_{1g}}^{D_{6h}} \left|\phi_{\pm}^{A_1}\right\rangle = {{1}\over{2}} \left|\phi_{\pm}^{A_1}\right\rangle \pm  {{1}\over{2}} \left|\phi_{\pm}^{A_1}\right\rangle, \\
P_{A_{2u}}^{D_{6h}} \left|\phi_{\pm}^{A_1}\right\rangle = {{1}\over{2}} \left|\phi_{\pm}^{A_1}\right\rangle \mp  {{1}\over{2}} \left|\phi_{\pm}^{A_1}\right\rangle.
\end{split}
\label{D6h:A1}
\end{equation}
Eq.~\eqref{D6h:A1} indicates that $|\phi_{+}^{A_1}\rangle$ and $|\phi_{-}^{A_1}\rangle$ generated by $A_1+A_1$ hybridization have irreps $A_{1g}$ and $A_{2u}$, respectively, in $D_{6h}$ point group.

(ii) For $A_2 + A_2 \Rightarrow A_{2g} + A_{1u}$ hybridization, the relationships between the character projection operator for $A_{2g}$ and $A_{1u}$ in $D_{6h}$ and $A_2$ in point group $C_{6v}$ read
\begin{equation}
\begin{split}
P_{A_{2g}}^{D_{6h}} = {{1}\over{2}} P_{A_2}^{C_{6v}} + {{1}\over{24}}\left( -\sum_{j=0}^{2} C_{2,j}^{\prime} - \sum_{j=0}^{2} C_{2,j}^{\prime\prime} + i + S_{3}+ S_{3}^2 + S_6 + S_6^5 + \sigma_h  \right), \\
P_{A_{1u}}^{D_{6h}} = {{1}\over{2}} P_{A_2}^{C_{6v}} - {{1}\over{24}}\left( -\sum_{j=0}^{2} C_{2,j}^{\prime} - \sum_{j=0}^{2} C_{2,j}^{\prime\prime} + i + S_{3}+ S_{3}^2 + S_6 + S_6^5 + \sigma_h  \right).
\end{split}
\label{D6h:pA2}
\end{equation}
Using Eq.~\eqref{D6h:CS}, Eq.~\eqref{D6h:pA2} and
\begin{equation}
\begin{split}
P_{A_2}^{C_{6v}}\left|\phi_{\pm}^{A_2}\right\rangle =& \left|\phi_{\pm}^{A_2}\right\rangle,\\
\sigma_h \left|\phi_{\pm}^{A_2}\right\rangle  = \pm \left|\phi_{\pm}^{A_2}\right\rangle, &\quad C_6^i \left|\phi_{\pm}^{A_2}\right\rangle = \left|\phi_{\pm}^{A_2}\right\rangle, \\
\sigma_{v,i} \left|\phi_{\pm}^{A_2}\right\rangle = -\left|\phi_{\pm}^{A_2}\right\rangle, &\quad \sigma_{d,i} \left|\phi_{\pm}^{A_2}\right\rangle = -\left|\phi_{\pm}^{A_2}\right\rangle.
\end{split}
\end{equation}
we have
\begin{equation}
\begin{split}
P_{A_{2g}}^{D_{6h}} \left|\phi_{\pm}^{A_2}\right\rangle = {{1}\over{2}} \left|\phi_{\pm}^{A_2}\right\rangle \pm  {{1}\over{2}} \left|\phi_{\pm}^{A_2}\right\rangle, \\
P_{A_{1u}}^{D_{6h}} \left|\phi_{\pm}^{A_2}\right\rangle = {{1}\over{2}} \left|\phi_{\pm}^{A_2}\right\rangle \mp  {{1}\over{2}} \left|\phi_{\pm}^{A_2}\right\rangle.
\end{split}
\label{D6h:A2}
\end{equation}
Eq.~\eqref{D6h:A2} indicates that $|\phi_{+}^{A_2}\rangle$ and $|\phi_{-}^{A_2}\rangle$ generated by $A_2+A_2$ hybridization have irreps $A_{2g}$ and $A_{1u}$, respectively, in $D_{6h}$ point group.

(iii) For $B_1 + B_1 \Rightarrow B_{1u} + B_{2g}$ hybridization, the relationships between the character projection operator for $B_{2g}$ and $B_{1u}$ in $D_{6h}$ and $B_1$ in point group $C_{6v}$ read
\begin{equation}
\begin{split}
P_{B_{1u}}^{D_{6h}} = {{1}\over{2}} P_{B_1}^{C_{6v}} - {{1}\over{24}}\left( -\sum_{j=0}^{2} C_{2,j}^{\prime} + \sum_{j=0}^{2} C_{2,j}^{\prime\prime} + i - S_{3}- S_{3}^2 + S_6 + S_6^5 - \sigma_h  \right), \\
P_{B_{2g}}^{D_{6h}} = {{1}\over{2}} P_{B_1}^{C_{6v}} + {{1}\over{24}}\left( -\sum_{j=0}^{2} C_{2,j}^{\prime} + \sum_{j=0}^{2} C_{2,j}^{\prime\prime} + i - S_{3}- S_{3}^2 + S_6 + S_6^5 - \sigma_h  \right).
\end{split}
\label{D6h:pB1}
\end{equation}
Using Eq.~\eqref{D6h:CS}, Eq.~\eqref{D6h:pB1} and
\begin{equation}
\begin{split}
P_{B_1}^{C_{6v}}\left|\phi_{\pm}^{B_1}\right\rangle =& \left|\phi_{\pm}^{B_1}\right\rangle,\\
\sigma_h \left|\phi_{\pm}^{B_1}\right\rangle  = \pm \left|\phi_{\pm}^{B_1}\right\rangle, &\quad C_6^i \left|\phi_{\pm}^{B_1}\right\rangle = (-1)^i \left|\phi_{\pm}^{B_1}\right\rangle, \\
\sigma_{v,i} \left|\phi_{\pm}^{B_1}\right\rangle = \left|\phi_{\pm}^{B_1}\right\rangle, &\quad \sigma_{d,i} \left|\phi_{\pm}^{B_1}\right\rangle = -\left|\phi_{\pm}^{B_1}\right\rangle,
\end{split}
\end{equation}
we have
\begin{equation}
\begin{split}
P_{B_{1u}}^{D_{6h}} \left|\phi_{\pm}^{B_1}\right\rangle = {{1}\over{2}} \left|\phi_{\pm}^{B_1}\right\rangle \pm  {{1}\over{2}} \left|\phi_{\pm}^{B_1}\right\rangle, \\
P_{B_{2g}}^{D_{6h}} \left|\phi_{\pm}^{B_1}\right\rangle = {{1}\over{2}} \left|\phi_{\pm}^{B_1}\right\rangle \mp  {{1}\over{2}} \left|\phi_{\pm}^{B_1}\right\rangle. \\
\end{split}
\label{D6h:B1}
\end{equation}
Eq.~\eqref{D6h:B1} indicates that $|\phi_{+}^{B_1}\rangle$ and $|\phi_{-}^{B_1}\rangle$ generated by $B_1+B_1$ hybridization have irreps $B_{1u}$ and $B_{2g}$, respectively, in $D_{6h}$ point group.

(iv) For $B_2 + B_2 \Rightarrow B_{2u} + B_{1g}$ hybridization, the relationships between the character projection operator for $B_{1g}$ and $B_{2u}$ in $D_{6h}$ and $B_2$ in point group $C_{6v}$ read
\begin{equation}
\begin{split}
P_{B_{2u}}^{D_{6h}} = {{1}\over{2}} P_{B_2}^{C_{6v}} - {{1}\over{24}}\left( \sum_{j=0}^{2} C_{2,j}^{\prime} - \sum_{j=0}^{2} C_{2,j}^{\prime\prime} + i - S_{3}- S_{3}^2 + S_6 + S_6^5 - \sigma_h  \right), \\
P_{B_{1g}}^{D_{6h}} = {{1}\over{2}} P_{B_2}^{C_{6v}} + {{1}\over{24}}\left( \sum_{j=0}^{2} C_{2,j}^{\prime} - \sum_{j=0}^{2} C_{2,j}^{\prime\prime} + i - S_{3}- S_{3}^2 + S_6 + S_6^5 - \sigma_h  \right).
\end{split}
\label{D6h:pB2}
\end{equation}
Using Eq.~\eqref{D6h:CS}, Eq.~\eqref{D6h:pB2} and
\begin{equation}
\begin{split}
P_{B_1}^{C_{6v}}\left|\phi_{\pm}^{B_2}\right\rangle =& \left|\phi_{\pm}^{B_2}\right\rangle,\\
\sigma_h \left|\phi_{\pm}^{B_2}\right\rangle  = \pm \left|\phi_{\pm}^{B_2}\right\rangle, &\quad C_6^i \left|\phi_{\pm}^{B_2}\right\rangle = (-1)^i \left|\phi_{\pm}^{B_2}\right\rangle, \\
\sigma_{v,i} \left|\phi_{\pm}^{B_2}\right\rangle = -\left|\phi_{\pm}^{B_2}\right\rangle, &\quad \sigma_{d,i} \left|\phi_{\pm}^{B_2}\right\rangle = \left|\phi_{\pm}^{B_2}\right\rangle, \\
\end{split}
\end{equation}
we have
\begin{equation}
\begin{split}
P_{B_{2u}}^{D_{6h}} \left|\phi_{\pm}^{B_2}\right\rangle = {{1}\over{2}} \left|\phi_{\pm}^{B_2}\right\rangle \pm  {{1}\over{2}} \left|\phi_{\pm}^{B_2}\right\rangle, \\
P_{B_{1g}}^{D_{6h}} \left|\phi_{\pm}^{B_2}\right\rangle = {{1}\over{2}} \left|\phi_{\pm}^{B_2}\right\rangle \mp  {{1}\over{2}} \left|\phi_{\pm}^{B_2}\right\rangle. \\
\end{split}
\label{D6h:B2}
\end{equation}
Eq.~\eqref{D6h:B2} indicates that $|\phi_{+}^{B_2}\rangle$ and $|\phi_{-}^{B_2}\rangle$ generated by $B_2+B_2$ hybridization have irreps $B_{2u}$ and $B_{1g}$, respectively, in $D_{6h}$ point group.

(v) For $E_1 + E_1 \Rightarrow E_{1u} + E_{1g}$ hybridization, the relationships between the character projection operator for $E_{1u}$ and $E_{1g}$ in $D_{6h}$ and $E_1$ in point group $C_{6v}$ read
\begin{equation}
\begin{split}
P_{E_{1u}}^{D_{6h}} &= {{1}\over{2}} P_{E_1}^{C_{6v}} - {{2}\over{24}}\left(  2i + S_{3}+ S_{3}^2 - S_6 - S_6^5 - 2\sigma_h  \right), \\
P_{E_{1g}}^{D_{6h}} &= {{1}\over{2}} P_{E_1}^{C_{6v}} + {{2}\over{24}}\left(  2i + S_{3}+ S_{3}^2 - S_6 - S_6^5 - 2\sigma_h  \right).
\end{split}
\label{D6h:pE1}
\end{equation}
In $C_{6v}$ monolayer a 2D $E_1$ state is the eigenstate of $C_6$ with the corresponding eigenvalue $e^{\pm i\pi/3}$, i.e., $C_6|\phi_{\pm}^{E_1}\rangle=e^{\pm i \pi/3}|\phi_{\pm}^{E_1}\rangle$. Using Eq.~\eqref{D6h:CS}, Eq.~\eqref{D6h:pE1} and
\begin{equation}
\begin{split}
P_{E_1}^{C_{6v}}\left|\phi_{\pm}^{E_1}\right\rangle =& \left|\phi_{\pm}^{E_1}\right\rangle, \quad \sigma_h \left|\phi_{\pm}^{E_1}\right\rangle  = \pm \left|\phi_{\pm}^{E_1}\right\rangle, \quad C_6^j \left|\phi_{\pm}^{E_1}\right\rangle = \left(e^{\pm i\pi/3}\right)^j \left|\phi_{\pm}^{E_1}\right\rangle,
\end{split}
\end{equation}
we have
\begin{equation}
\begin{split}
P_{E_{1u}}^{D_{6h}} \left|\phi_{\pm}^{E_1}\right\rangle = {{1}\over{2}} \left|\phi_{\pm}^{E_1}\right\rangle \pm  {{1}\over{2}} \left|\phi_{\pm}^{E_1}\right\rangle, \\
P_{E_{1g}}^{D_{6h}} \left|\phi_{\pm}^{E_1}\right\rangle = {{1}\over{2}} \left|\phi_{\pm}^{E_1}\right\rangle \mp  {{1}\over{2}} \left|\phi_{\pm}^{E_1}\right\rangle.
\end{split}
\label{D6h:E1}
\end{equation}
Eq.~\eqref{D6h:E1} indicates that $|\phi_{+}^{E_1}\rangle$ and $|\phi_{-}^{E_1}\rangle$ generated by $E_1+E_1$ hybridization have irreps $E_{1u}$ and $E_{1g}$, respectively, in $D_{6h}$ point group.

\begin{figure}[!htbp]
\centering
\includegraphics[width=14 cm]{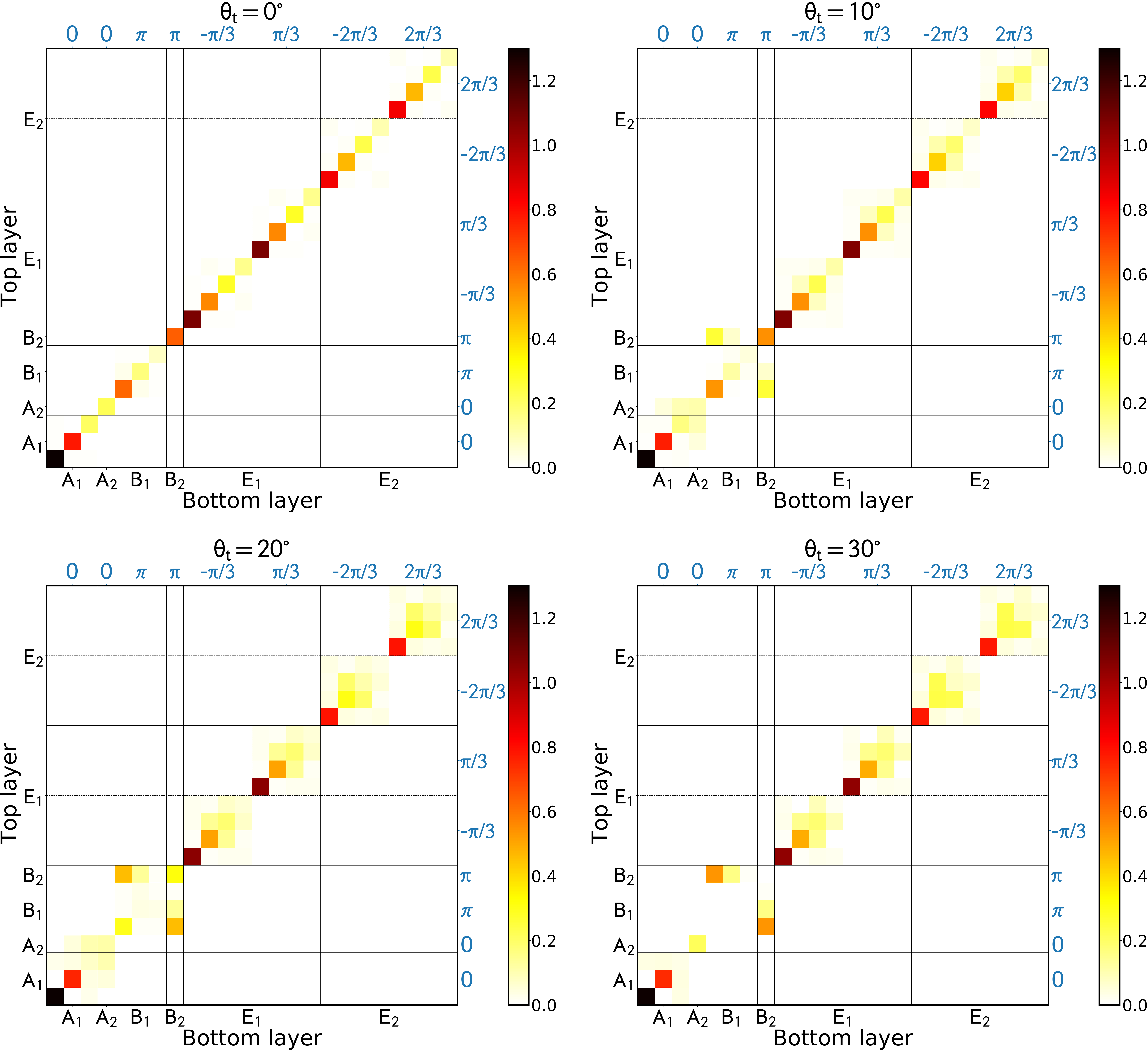}
\caption{Interlayer hybridization matrix elements with their absolute values $| \braket{\varphi_{ir,\theta}^b|U|\varphi_{ir^{\prime},\theta^{\prime}}^{t}}|$ of size-$2$ quantum dot structures in unit of eV from $p_z$ orbital TB model, for untwisted BG with $\theta_t=0^{\circ}$, twisted BG with $\theta_t=10^{\circ}$, twisted BG with $\theta_t=20^{\circ}$, and graphene quasicrystal with $\theta_t=30^{\circ}$. These eigenstates of the two layers $\varphi_{ir,\theta}^{b}$ and $\varphi_{ir^{\prime},\theta^{\prime}}^{t}$ are classified by the irreducible representation $(ir)$ of $C_{6v}$ and $\theta$ in the eigenvalue $e^{i\theta}$ of rotation operation $C_6$.}
\label{fig:size2pz}
\end{figure}

(vi) For $E_2 + E_2 \Rightarrow E_{2g} + E_{2u}$ hybridization, the relationships between the character projection operator for $E_{2g}$ and $E_{2u}$ in $D_{6h}$ and $E_2$ in point group $C_{6v}$ read
\begin{equation}
\begin{split}
P_{E_{2g}}^{D_{6h}} &= {{1}\over{2}} P_{E_2}^{C_{6v}} + {{2}\over{24}}\left(  2i - S_{3}- S_{3}^2 - S_6 - S_6^5 + 2\sigma_h  \right), \\
P_{E_{2u}}^{D_{6h}} &= {{1}\over{2}} P_{E_2}^{C_{6v}} - {{2}\over{24}}\left(  2i - S_{3}- S_{3}^2 - S_6 - S_6^5 + 2\sigma_h  \right).
\end{split}
\label{D6h:pE2}
\end{equation}
In $C_{6v}$ monolayer a 2D $E_2$ state is the eigenstate of $C_6$ with the corresponding eigenvalue $e^{\pm i2\pi/3}$, i.e., $C_6|\phi_{\pm}^{E_2}\rangle=e^{\pm i 2\pi/3}|\phi_{\pm}^{E_2}\rangle$. Using Eq.~\eqref{D6h:CS}, Eq.~\eqref{D6h:pE2} and
\begin{equation}
\begin{split}
P_{E_2}^{C_{6v}}\left|\phi_{\pm}^{E_2}\right\rangle =& \left|\phi_{\pm}^{E_2}\right\rangle, \quad \sigma_h \left|\phi_{\pm}^{E_2}\right\rangle  = \pm \left|\phi_{\pm}^{E_2}\right\rangle, \quad C_6^j \left|\phi_{\pm}^{E_2}\right\rangle = \left(e^{\pm i 2\pi/3}\right)^j \left|\phi_{\pm}^{E_2}\right\rangle,
\end{split}
\end{equation}
we have
\begin{equation}
\begin{split}
P_{E_{2g}}^{D_{6h}} \left|\phi_{\pm}^{E_2}\right\rangle = {{1}\over{2}} \left|\phi_{\pm}^{E_2}\right\rangle \pm  {{1}\over{2}} \left|\phi_{\pm}^{E_2}\right\rangle, \\
P_{E_{2u}}^{D_{6h}} \left|\phi_{\pm}^{E_2}\right\rangle = {{1}\over{2}} \left|\phi_{\pm}^{E_2}\right\rangle \mp  {{1}\over{2}} \left|\phi_{\pm}^{E_2}\right\rangle.
\end{split}
\label{D6h:E2}
\end{equation}
Eq.~\eqref{D6h:E2} indicates that $|\phi_{+}^{E_2}\rangle$ and $|\phi_{-}^{E_2}\rangle$ generated by $E_2+E_2$ hybridization have irreps $E_{2g}$ and $E_{2u}$, respectively, in $D_{6h}$ point group.

\section{S4. Numerical methods}
\subsection{S4.1. $p_z$ orbital TB model}
\begin{figure}[!htbp]
\centering
\includegraphics[width=14 cm]{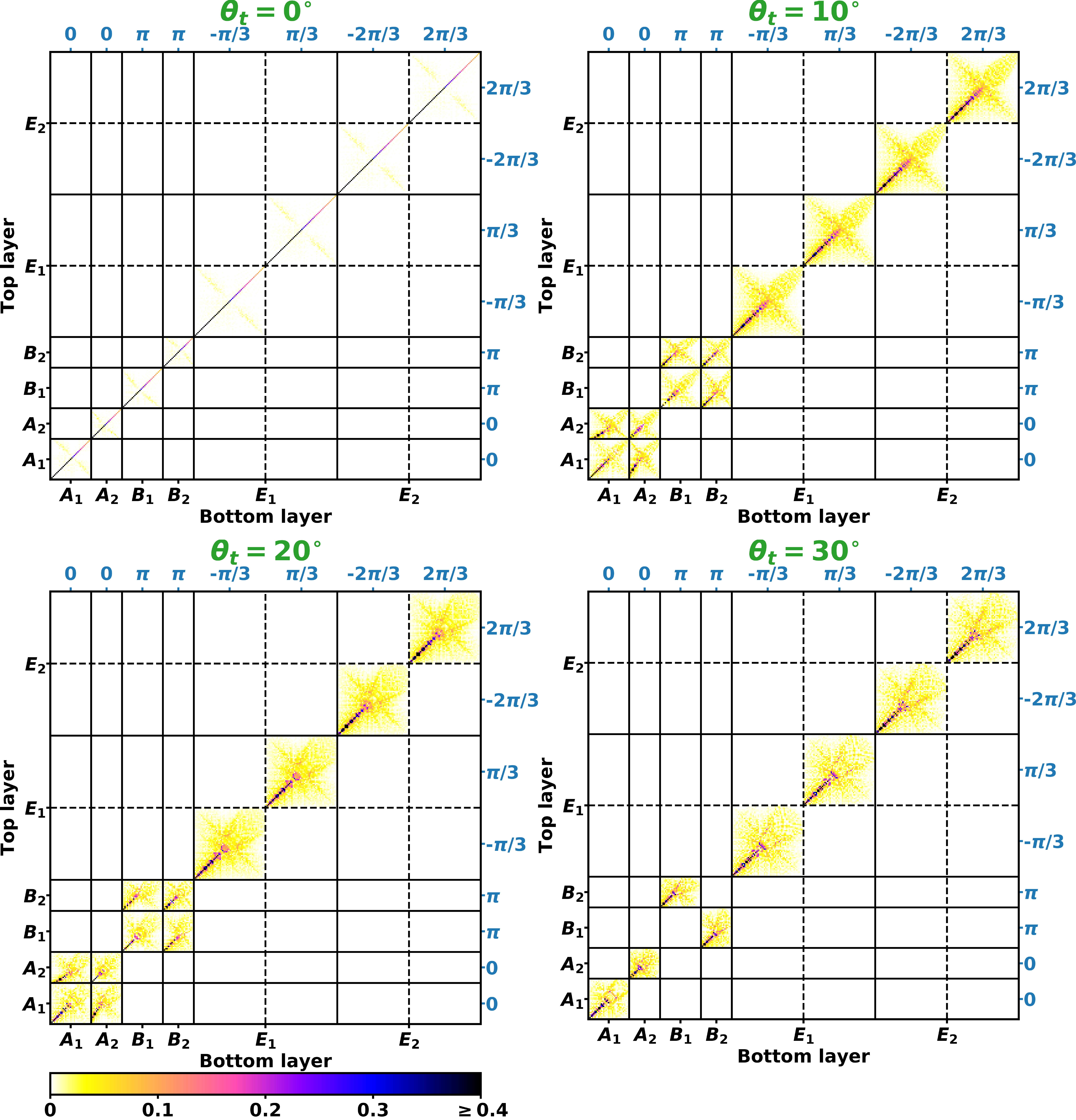}
\caption{Interlayer hybridization matrix elements with their absolute values $| \braket{\varphi_{ir,\theta}^b|U|\varphi_{ir^{\prime},\theta^{\prime}}^{t}}|$ of size-$8$ quantum dot structures in unit of eV from $p_z$ orbital TB model, for untwisted BG with $\theta_t=0^{\circ}$, twisted BG with $\theta_t=10^{\circ}$, twisted BG with $\theta_t=20^{\circ}$, and graphene quasicrystal with $\theta_t=30^{\circ}$. These eigenstates of the two layers $\varphi_{ir,\theta}^{b}$ and $\varphi_{ir^{\prime},\theta^{\prime}}^{t}$ are classified by the irreducible representation $(ir)$ of $C_{6v}$ and $\theta$ in the eigenvalue $e^{i\theta}$ of rotation operation $C_6$.}
\label{fig:size8pz}
\end{figure}

In the $p_z$ orbital based TB model, the hopping energy between site $i$ and $j$ is determined by\cite{Slater_LCAO}
\begin{equation}
t(\bm{r}_{ij}) = n^2 V_{pp\sigma}(\left|\bm{r}_{ij}\right|)+(1-n^2)V_{pp\pi}(\left|\bm{r}_{ij}\right|),
\label{methods:pzTB}
\end{equation}
where $n$ is the direction cosine of relative position vector $\bm{r}_{ij}$ with respect to the $z$ axis. The Slater-Koster parameters $V_{pp\sigma}$ and $V_{pp\pi}$ read
\begin{equation}
\begin{split}
V_{pp\pi}(\left|\bm{r}_{ij}\right|)&=-\gamma_0 e^{2.218(b-\left|\bm{r}_{ij}\right|)}F_c(\left|\bm{r}_{ij}\right|), \\
V_{pp\sigma}(\left|\bm{r}_{ij}\right|)&=\gamma_1 e^{2.218(h-\left|\bm{r}_{ij}\right|)}F_c(\left|\bm{r}_{ij}\right|),
\end{split}
\label{methods:pzTBsk}
\end{equation}
where $\gamma_0$ and $\gamma_1$ are 2.7 eV and 0.48 eV, respectively, $b=1.42$ {\AA} is the nearest intralayer carbon-carbon distance, and $F_c$ is a smooth function
\begin{equation}
F_c(r) = (1+e^{(r-0.265)/5})^{-1}.
\end{equation}
In our calculations, a large cutoff carbon-carbon distance for the hopping energy is adopted upto 5 {\AA}. In this $p_z$ orbital based TB model, the van der Waals interaction has been included in the interlayer hopping.
The $p_z$ orbital TB model has been widely used to predict the electronic structures and exotic states \cite{TBG_TB_Num,science_QC,TBG_TB_Ele,Moon_QC,TBG_TB_zhen,Yu_TDBG_QC,kerelsky2019maximized,liang2020effect,
do2019time,wang2019one} in good accordance with experimental results\cite{science_QC,TBG_TB_Ele,TBG_TB_zhen,kerelsky2019maximized} in twisted bilayer graphene systems. Using the $p_z$ orbital based TB model in Eqs.~\eqref{methods:pzTB} and ~\eqref{methods:pzTBsk}, we calculate the hybridization matrix elements within the basis functions of $C_6$ in size-$2$ and size-$8$ quantum dot structures, as shown in Fig. \ref{fig:size2pz} and Fig. \ref{fig:size8pz}, respectively, including the $D_{6h}$ untwisted BG structures with $\theta_t=0^\circ$, $D_{6}$ twisted BG structures with $\theta_t=10^\circ$ and $\theta_t=20^\circ$, and $D_{6d}$ graphene quasicrystal structures with $\theta_t=30^\circ$. The mapped distributions of these nonzero hybridization matrix elements manifest the corresponding hybridization selection rules of the three BG systems.
\begin{figure}[!htbp]
\centering
\includegraphics[width=14 cm]{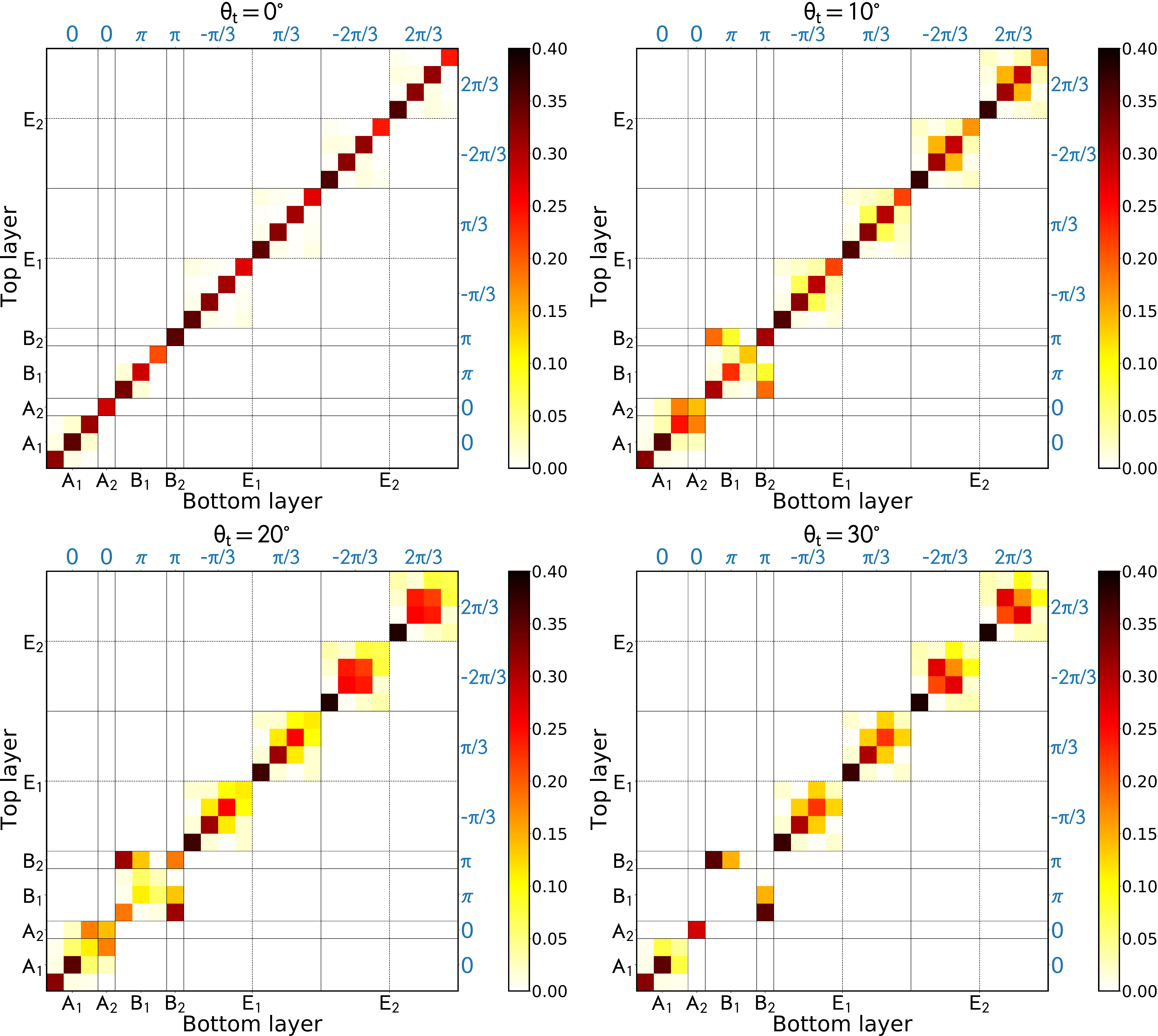}
\caption{Interlayer hybridization matrix elements with their absolute values $| \braket{\varphi_{ir,\theta}^b|U|\varphi_{ir^{\prime},\theta^{\prime}}^{t}}|$ of size-$2$ quantum dot structures in unit of eV from Wannier orbital TB model, for untwisted BG with $\theta_t=0^{\circ}$, twisted BG with $\theta_t=10^{\circ}$, twisted BG with $\theta_t=20^{\circ}$, and graphene quasicrystal with $\theta_t=30^{\circ}$. These eigenstates of the two layers $\varphi_{ir,\theta}^{b}$ and $\varphi_{ir^{\prime},\theta^{\prime}}^{t}$ are classified by the irreducible representation $(ir)$ of $C_{6v}$ and $\theta$ in the eigenvalue $e^{i\theta}$ of rotation operation $C_6$.}
\label{fig:size2wannier}
\end{figure}

\subsection{S4.2. Wannier orbital TB model}
\begin{figure}[!htbp]
\centering
\includegraphics[width=14 cm]{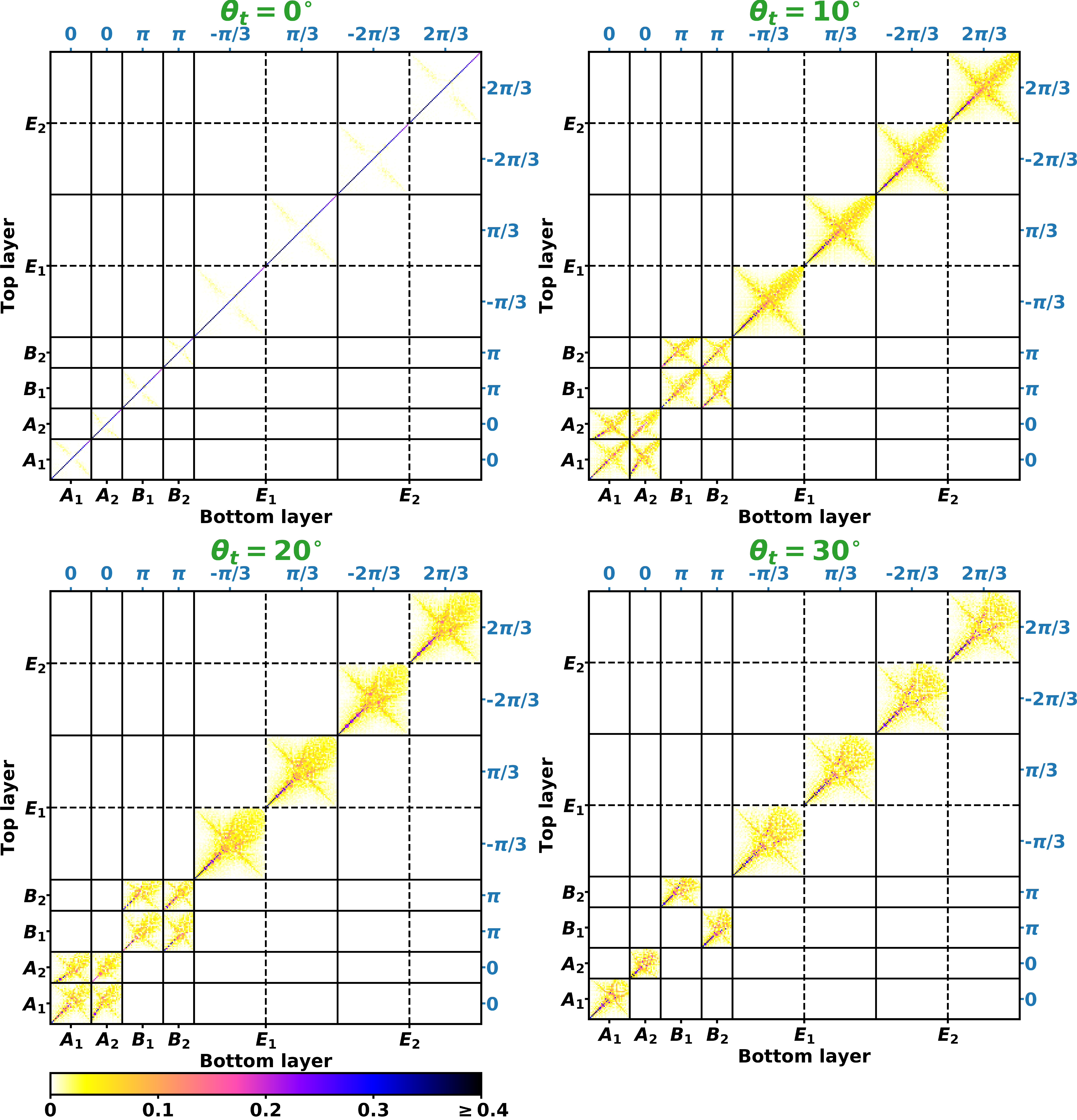}
\caption{Interlayer hybridization matrix elements with their absolute values $| \braket{\varphi_{ir,\theta}^b|U|\varphi_{ir^{\prime},\theta^{\prime}}^{t}}|$ of size-$8$ quantum dot structures in unit of eV from Wannier orbital TB model, for untwisted BG with $\theta_t=0^{\circ}$, twisted BG with $\theta_t=10^{\circ}$, twisted BG with $\theta_t=20^{\circ}$, and graphene quasicrystal with $\theta_t=30^{\circ}$. These eigenstates of the two layers $\varphi_{ir,\theta}^{b}$ and $\varphi_{ir^{\prime},\theta^{\prime}}^{t}$ are classified by the irreducible representation $(ir)$ of $C_{6v}$ and $\theta$ in the eigenvalue $e^{i\theta}$ of rotation operation $C_6$.}
\label{fig:size8wannier}
\end{figure}

The Wannier orbital TB model in twisted BG is proposed by Fang and Kaxiras\cite{fang2016electronic}. Compared with the $p_z$-orbital TB model, the Wannier orbital TB model can reproduce the electronic structure of DFT calculations on twisted bilayer graphene accurately even in higher energy region without increasing the computational cost. In this model, the intralayer hopping energies up to the eighth nearest neighbors, which are $-2.8922$ eV, $0.2425$ eV, $-0.2656$ eV, $0.0235$ eV, $0.0524$ eV, $-0.0209$ eV, $-0.0148$ eV and $-0.0211$ eV from the first to eighth nearest neighbors, respectively, are used to describe the Hamiltonian of graphene monolayer. The interlayer hopping in a functional form depending on both distance and orientation reads \cite{fang2016electronic}
\begin{equation}
\begin{split}
  t(\bm{r}) = V_0(r) +  V_3(r)[cos(3\theta_{12}) + cos(3\theta_{21})] + V_6(r)[cos(6\theta_{12}) + cos(6\theta_{21})],
\end{split}
\label{methods:wannierTB}
\end{equation}
where $\bm{r}$ is the in-plane part of the vector connecting two sites, $r = |\bm{r}|$ descries the projected distance between two Wannier functions, and $\theta_{12}$ and $\theta_{21}$ are the angles between the projected interlayer bond and the in-plane nearest-neighbor bonds, describing the relative orientation of the two Wannier functions. The three radial functions in Eq.~\eqref{methods:wannierTB} depend on ten hopping parameters ($\bar{r}=r/a$), as follows:
\begin{equation}
\begin{split}
V_0(r) = \lambda_0e^{-\xi_0\bar{r}^2}cos(\kappa_0\bar{r}),\\
V_3(r) = \lambda_3\bar{r}^2 e^{-\xi_3 (\bar{r}-x_3)^2},\\
V_6(r) = \lambda_6 e^{-\xi_6 (\bar{r}-x_6)^2} sin(\kappa_6\bar{r}),
\end{split}
\label{methods:wannierTBV}
\end{equation}
where related parameters are listed in Tab. \ref{table:hopping_ci}.
\begin{table}[htp!]
\caption{The ten interlayer hopping parameters in units of eV\cite{fang2016electronic}.}
\centering
\setlength{\tabcolsep}{4.5mm}{
\begin{tabular}{c c c c c c c c c c}
\hline\hline
 $\lambda_0$ & $\xi_0$ & $\kappa_0$ & $\lambda_3$ & $\xi_3$ & $x_3$ & $\lambda_6$ & $\xi_6$ & $x_6$ & $\kappa_6$\\
 \hline
 0.310 & 1.750 & 1.990 & $-0.068$ & 3.286 & 0.500 & $-0.008$ & 2.727 & 1.217 & 1.562\\
 \hline\hline
  \end{tabular}}
 \label{table:hopping_ci}
 \end{table}
Using the Wannier orbital TB model in Eqs.~\eqref{methods:wannierTB} and \eqref{methods:wannierTBV}, we calculate the hybridization matrix elements within the basis functions of $C_6$ in size-$2$ and size-$8$ quantum dot structures, as shown in Fig. \ref{fig:size2wannier} and Fig. \ref{fig:size8wannier}, respectively, including the $D_{6h}$ untwisted BG structures with $\theta_t=0^\circ$, $D_{6}$ twisted BG structures with $\theta_t=10^\circ$ and $\theta_t=20^\circ$, and $D_{6d}$ graphene quasicrystal structures with $\theta_t=30^\circ$. The mapped distributions of these nonzero hybridization matrix elements manifest the corresponding hybridization selection rules of the three BG systems and hence also agree well with the results from the $p_z$ orbital TB model.

\subsection{S4.3. Density functional theory}
\begin{figure}[!htbp]
\centering
\includegraphics[width=17 cm]{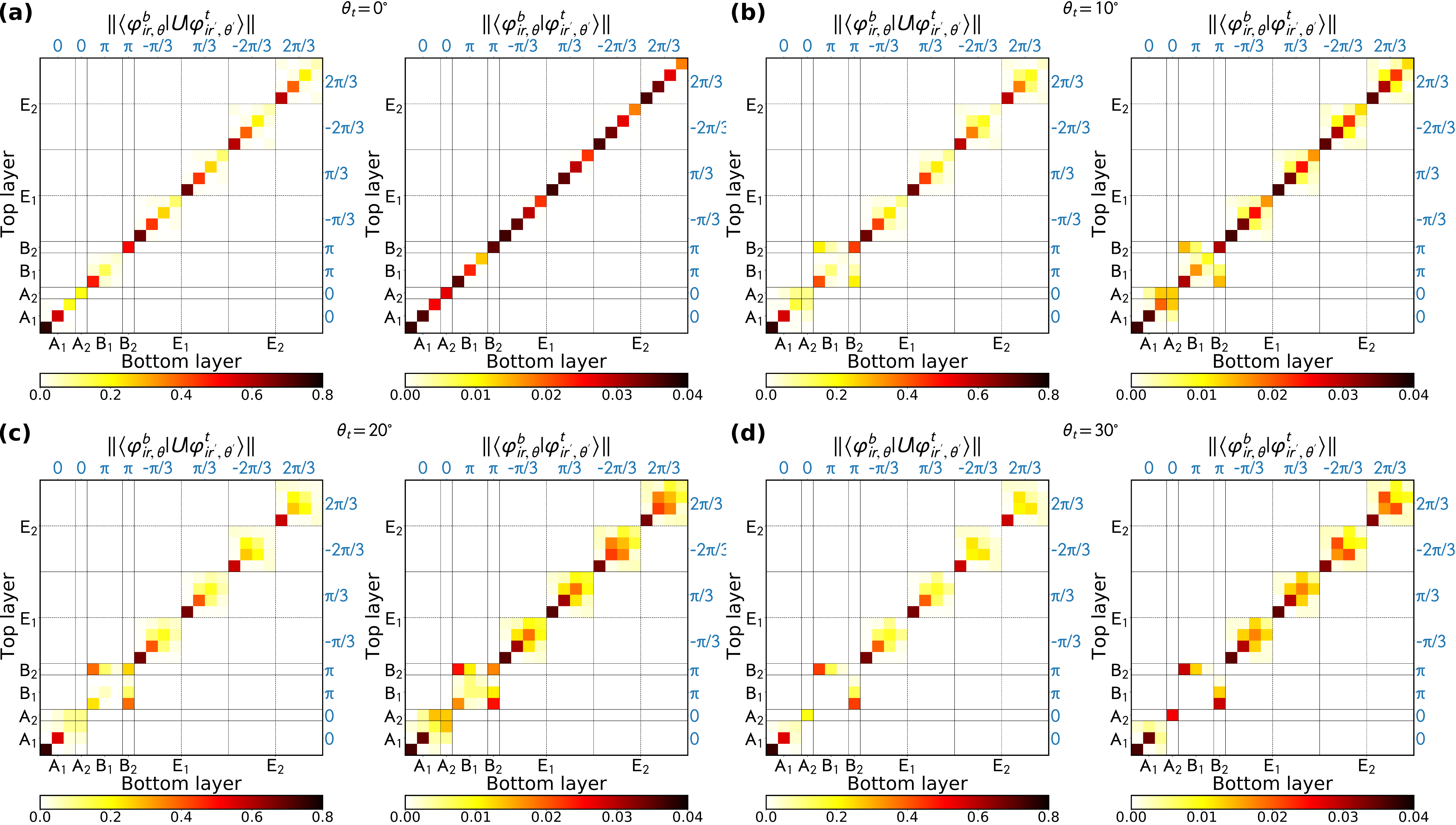}
\caption{Interlayer hybridization and overlap matrix elements with their absolute values $| \braket{\varphi_{ir,\theta}^b|U|\varphi_{ir^{\prime},\theta^{\prime}}^{t}}|$ and $| \braket{\varphi_{ir,\theta}^b|\varphi_{ir^{\prime},\theta^{\prime}}^{t}}|$ for size-$2$ quantum dot structures in unit of eV from DFT calculations, for (a) untwisted BG with $\theta_t=0^{\circ}$, (b) twisted BG with $\theta_t=10^{\circ}$, (c) twisted BG with $\theta_t=20^{\circ}$, and (d) graphene quasicrystal with $\theta_t=30^{\circ}$. These eigenstates of the two layers $\varphi_{ir,\theta}^{b}$ and $\varphi_{ir^{\prime},\theta^{\prime}}^{t}$ are classified by the irreducible representation $(ir)$ of $C_{6v}$ and $\theta$ in the eigenvalue $e^{i\theta}$ of rotation operation $C_6$.}
\label{fig:dft_size2}
\end{figure}

The density functional theory (DFT) calculations are implemented in SIESTA code\cite{soler2002siesta}. Firstly, graphene monolayer quantum dot, inside a 30 {\AA} $\times$ 30 {\AA} $\times$ 15 {\AA} box avoiding the interaction between adjacent images, is relaxed till the maximal force less than 0.04 eV/{\AA} under the GGA-PBE functional and double zeta polarization (DZP) basis sets. Secondly, we perform the self-consistent calculations on the size-2 untwisted and twisted BG quantum dots inside a 30 {\AA} $\times$ 30 {\AA} $\times$ 30 {\AA} box, where two relaxed monolayers are overlaid in $z$ direction with the same twist angle $\theta_t$ and the interlayer distance $h=3.35$ {\AA} as the value used in $p_z$ orbital and Wannier orbital TB models. The optB88-vdW functional and SZ bases are adopted during the self-consistent calculation. Note that for all DFT calculations, the edge carbon atom are saturated by hydrogen atoms. Finally, the Hamiltonian ($H$) and overlap ($S$) matrices are obtained with the help of sisl tool\cite{zerothisisl}. Because the energy states inside low-energy area have mainly the $p_z$ orbital character, the components with $p_z$ orbital involved are picked up. The eigenstates are obtained by solving the generalized eigen equation $H\phi=\varepsilon S \phi$. After separating the eigen states of each layer according to the irreducible representations of $C_{6v}$ point group and the rotation $C_6$, we obtain the interlayer hybridization matrix elements and overlap matrices. Fig. \ref{fig:dft_size2} shows the calculated results on a size-$2$ quantum dot structure, including the $D_{6h}$ untwisted BG structures with $\theta_t=0^\circ$, $D_{6}$ twisted BG structures with $\theta_t=10^\circ$ and $\theta_t=20^\circ$, and $D_{6d}$ graphene quasicrystal structures with $\theta_t=30^\circ$. The mapped distributions of these nonzero hybridization matrix elements manifest the corresponding hybridization selection rules of the three BG systems and hence also agree well with the results from the $p_z$ orbital TB model as well as the Wannier orbital TB model.

\section{S5. Energy dependence of hybridization strength}
\subsection{S5.1. Hamiltonian in momentum space}
Within the $p_z$ orbital TB model for both untwisted BG and twisted BG, the Bloch basis functions of the two monolayers are defined as
\begin{equation}
\begin{split}
\left| \bm{k}^b,X^b \right\rangle &= {{1}\over{\sqrt{N}}} \sum_{\bm{L}^b}  e^{i\bm{k}^b\cdot(\bm{L}^b+\bm{\tau}^b_{X^b})} \left| \bm{L}^b+\bm{\tau}^b_{X^b} \right\rangle \quad \textrm{(Bottom layer)}, \\
\left| \bm{k}^t,X^t \right\rangle &= {{1}\over{\sqrt{N}}} \sum_{\bm{L}^t}  e^{i\bm{k}^t\cdot(\bm{L}^t+\bm{\tau}^t_{X^t})} \left| \bm{L}^t+\bm{\tau}^t_{X^t} \right\rangle \quad \textrm{(Top layer)},
\end{split}
\end{equation}
where $N$ is the normalization factor, $\bm{L}^b=n_1^b\bm{a}_1^b+n_2^b\bm{a}_2^b$ ($\bm{L}^t=n_1^t\bm{a}_1^t+n_2^t\bm{a}_2^t$) is the unit cell vector, and $|\bm{L}^b +\bm{\tau}^b_{X^b}\rangle$ $(|\bm{L}^t + \bm{\tau}^t_{X^t}\rangle)$ denotes the $p_z$ orbital located at sublattice $X^b$ ($X^t$) in unit cell $\bm{L}^b$ ($\bm{L}^t$). The intralayer matrix elements inside the two graphene monolayers respectively read
\begin{equation}
\begin{split}
\braket{\bm{k}^b,X^b|H_0^b|\bm{k}^{b,\prime},X^{b,\prime}}&=\delta_{\bm{k}^b\bm{k}^{b,\prime}} \sum_{\bm{L}^b}t(\bm{L}^b+\bm{\tau}^b_{X^{b,\prime}X^b})e^{i\bm{k}^b\cdot(\bm{L}^b+\bm{\tau}^b_{X^{b,\prime}X^b})},\\
\braket{\bm{k}^t,X^t|H_0^t|\bm{k}^{t,\prime},X^{t,\prime}}&=\delta_{\bm{k}^t\bm{k}^{t,\prime}}
\sum_{\bm{L}^t}t(\bm{L}^t+\bm{\tau}^t_{X^{t,\prime}X^t})e^{i\bm{k}^t\cdot(\bm{L}^t+\bm{\tau}^t_{X^{t,\prime}X^t})},\\
\end{split}
\end{equation}
where $\bm{\tau}^b_{X^{b,\prime}X^b} =\bm{\tau}^b_{X^{b,\prime}}-\bm{\tau}^b_{X^b}$ and $\bm{\tau}^t_{X^{t,\prime}X^t} =\bm{\tau}^t_{X^{t,\prime}}-\bm{\tau}^t_{X^t}$. The interlayer matrix element reads\cite{Mele_twist,Moon_QC,Yu_TDBG_QC}
\begin{equation}
\braket{\bm{k}^b, X^b|U|\bm{k}^t, X^t} = \sum_{\bm{G}^b \bm{G}^t} T(|\bm{k}^b+\bm{G}^b|) e^{i\bm{G}^b\cdot \bm{\tau}^b_{X^b}} e^{-i\bm{G}^t\cdot \bm{\tau}^t_{X^t}} \delta_{\bm{k}^b+\bm{G}^b, \bm{k}^t+\bm{G}^t},
\label{eq:Umatrix}
\end{equation}
where $T(|\bm{k}^b+\bm{G}^b|)$ is the $xy$-plane Fourier transforms of interlayer hopping function $t(\bm{r_{xy}}+h\hat{\bm{e}_z})$ at $\bm{k}^b+\bm{G}^b$, $\bm{G}^b=m^b\bm{b}_1^b+n^b\bm{b}_2^b$, and $\bm{G}^t=m^t\bm{b}_1^t+n^t\bm{b}_2^t$, with two reciprocal lattice vectors $\bm{b}_1^{b/t}$ and $\bm{b}_2^{b/t}$ for two monolayers and the integers $m^{b/t}$ and $n^{b/t}$. The in-plane Fourier transform $T(\bm{q})$ of the interlayer hopping $t(\bm{r})$ at $\bm{q}$ is defined as
\begin{equation}
T(\bm{q})={1\over S}\int t(\bm{r}_{xy}+h\bm{\hat{e}}_z) e^{-i\bm{q}\cdot\bm{r}_{xy}}d\bm{r}_{xy},
\end{equation}
where $S$ is the area of the unit cell in graphene, and $\bm{\hat{e}}_z$ is the unit vector along $z$ axis. Eq.~\eqref{eq:Umatrix} indicates that only the Bloch bases of the two monolayers with the wave vectors satisfying $\bm{k}^b+\bm{G}^b=\bm{k}^t+\bm{G}^t$ are nonzero. This wave-vector-dependent interlayer coupling condition $\bm{k}^b+\bm{G}^b=\bm{k}^t+\bm{G}^t$ can be rewritten as\cite{shallcross2010,shallcross2013,weckbecker2016}
\begin{equation}
\bm{k}^b-\bm{k}^t=m\bm{g}_1+n\bm{g}_2,
\label{eq:coupk}
\end{equation}
where $m$ and $n$ are two arbitrary integer values, and $\bm{g}_1$ and $\bm{g}_2$ take the forms as
\begin{equation}
\begin{split}
\bm{g}_1 = \bm{b}_1^b - \bm{b}_1^t,\\
\bm{g}_2 = \bm{b}_2^b - \bm{b}_2^t.
\end{split}
\label{eq:coupg}
\end{equation}

\subsection{S5.2. Hybridization strength near the Fermi level}
\begin{figure}[!htbp]
\centering
\includegraphics[width=8 cm]{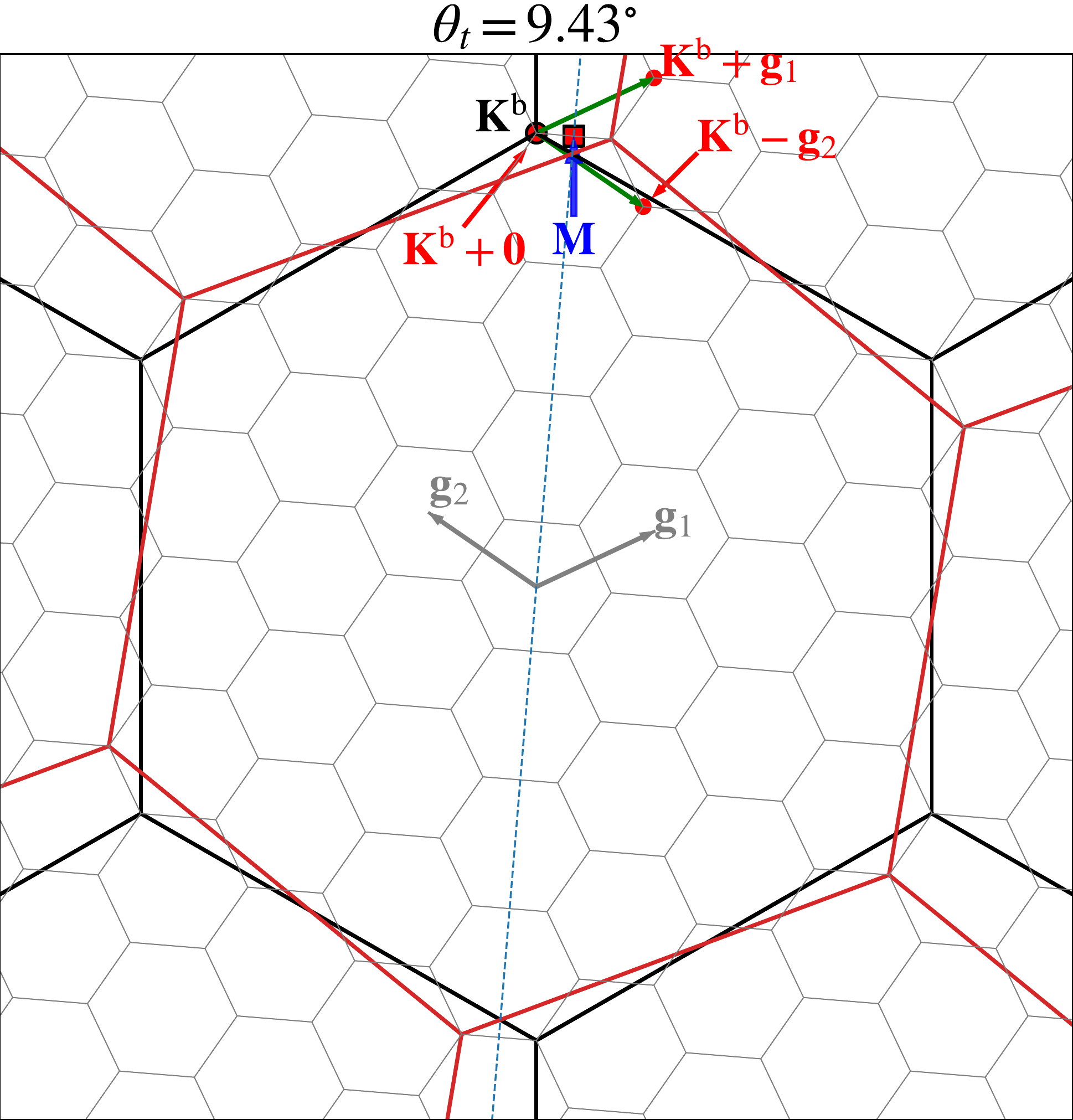}
\caption{Reciprocal lattice vectors and Brillouin zones in the twisted BG with $\theta_t=9.43^{\circ}$. Black and red big hexagons are the Brillouin zones of the bottom and top graphene, respectively. Gray hexagon network is the Brillouin zone of twisted BG with $\bm{g}_1$ and $\bm{g}_2$ as its reciprocal lattice vectors. The wave-vector-dependent interlayer coupling condition in Eqs.~\eqref{eq:coupk} and ~\eqref{eq:coupg} requires that the Bloch function from the bottom layer at $\bm{k}^b=\bm{K}^b$ mainly couple with three Bloch functions from the top layer at $\bm{k}^t=\bm{K}^b+\bm{0}$, $\bm{K}^b+\bm{g}_1$, and $\bm{K}^b-\bm{g}_2$. If the equal wave vectors $\bm{k}^b$ and $\bm{k}^t$ are on the blue dashed line across $M$ point, the Bloch functions of the two monolayers have the same energy. On the blue dashed line, $M$ point is closest to the Dirac points $\bm{K}^b$ and $\bm{K}^t$.}
\label{fig:BZs_coupling}
\end{figure}

\begin{figure}[!htbp]
\centering
\includegraphics[width=14 cm]{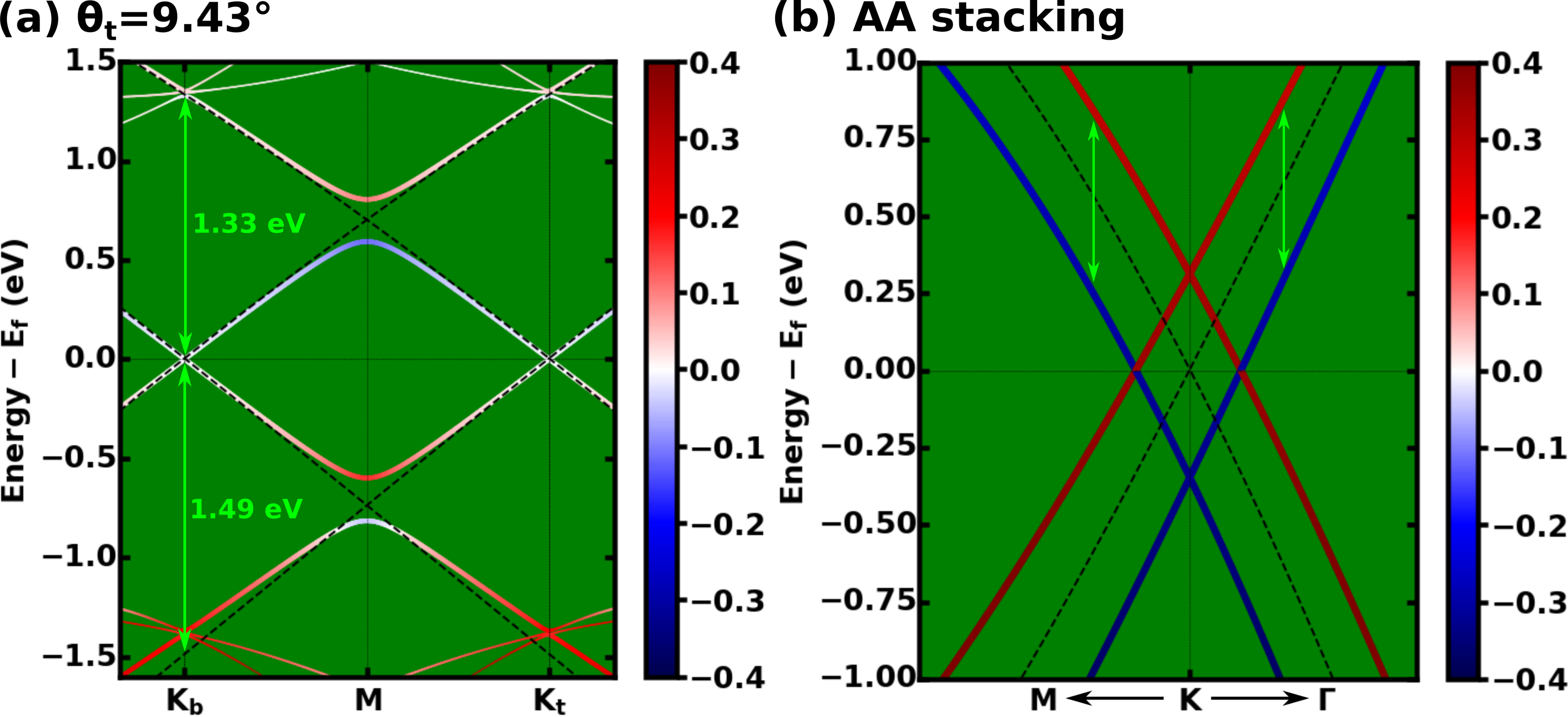}
\caption{(a) The calculated unfolded band structures for twisted BG with $\theta_t=9.43^{\circ}$ and (b) the calculated band structures for AA stacking BG, where the color denotes the interlayer hybridization strength $\triangle\varepsilon$, and black dashed lines stand for the band structures of the bottom and top graphene monolayers.}
\label{fig:dirac_coupling}
\end{figure}

As shown in Fig. 2(c) in main text, the energy-dependent hybridization strength shows two characteristics: (i) the hybridization strengths in twisted BG are weak near the Fermi level except a large strength in AA-stacking BG, and (ii) an electron-hole asymmetry of hybridization strength with stronger interlayer coupling for holes appears in high-energy areas. In this subsection, we discuss the physics supporting the first characteristic. We consider a commensurate twisted BG with $\theta_t=9.43^{\circ}$ as an example. Fig. \ref{fig:BZs_coupling} shows the reciprocal lattice vectors and Brillouin zones of two graphene monolayers in the twisted BG with $\theta_t=9.43^{\circ}$. According to the wave-vector-dependent interlayer coupling condition in Eqs.~\eqref{eq:coupk} and ~\eqref{eq:coupg}, the Dirac point $\bm{K}^b$ of the bottom layer can couple with the $\bm{k}^t$ points of the top layer mainly at $\bm{k}^t=\bm{K}^b+\bm{0}$, $\bm{K}^b+\bm{g}_1$ and $\bm{K}^b-\bm{g}_2$, and the coupling strengths between $\bm{K}^b$ and the above three $\bm{k}^t$ points are the same, i.e., $T(|\bm{K}^b|)\sim 0.11$ eV\cite{Bistritzer_Moir,coupling_condition}. The corresponding hybridization-generated band structures together with hybridization strengths of the twisted BG with $\theta_t=9.43^{\circ}$ are plotted in Fig. \ref{fig:dirac_coupling}(a), where the color represents the interlayer hybridization strength $\Delta \varepsilon $ in Eq. (6) of main text, and the black dash lines represent the Dirac band structures of the Hamiltonians $H_0^b$ and $H_0^t$ of graphene monolayers. As we can see, the hybridization strengths around $\bm{K}^b$ and $\bm{K}^t$ (near the Fermi level) are indeed weak and the hybridization strengths around $M$ inside relatively high energy areas are obviously stronger in Fig. \ref{fig:dirac_coupling}(a). The weak strength near the Fermi level can be well explained by the second-order non-degenerate perturbation theory, where the first-order energy correction is $E_n^{(1)}=\langle \varphi_n^{(0)}|U|\varphi_n^{(0)}\rangle$ with the wave function of ground state $|\varphi_n^{(0)}\rangle$ of unperturbed Hamiltonian $H_0$, and the second-order energy correction $E_n^{(2)}$ reads
\begin{equation}
E_n^{(2)} = \sum_{m\neq n}  {{\left| \braket{\varphi_m^{(0)}|U|\varphi_n^{(0)}} \right|^2}\over{E_n^{(0)}-E_m^{(0)}}}.
\label{eq:nondpertub2}
\end{equation}
The wave-vector-dependent interlayer coupling condition in Eqs.~\eqref{eq:coupk} and \eqref{eq:coupg} indicates the direct interlayer coupling between $\bm{k}^b=\bm{K}^b$ and $\bm{k}^t=\bm{K}^t$ points is not allowed, i.e., the first-order energy correction $E_n^{(1)}$ is zero. Around $\bm{K}^b$ and $\bm{K}^t$, the energy differences between the two Dirac linear bands (denoted by dash lines) are about $1.33$ eV and $1.49$ eV in Fig. \ref{fig:dirac_coupling}(a), and the coupling strength $|\braket{\varphi_m^{(0)}|U|\varphi_n^{(0)}}|$ is $0.11$ eV\cite{Bistritzer_Moir,coupling_condition}, and hence the second-order energy correction $E_n^{(2)}$ will be very small in Eq.~\eqref{eq:nondpertub2}. Consequently, the hybridization strength is very weak near the Fermi level.

For these k-points on the blue dashed line in Fig. \ref{fig:BZs_coupling}, the states from the two monolayers have the same energy, and the first-order degenerate perturbation theory can be applied to illustrate the relatively large hybridization strength near the blue dashed line. According to the first-order degenerate perturbation theory, the original degenerate energy bands are split, and hence the hybridization strength is relatively large, such as the hybridization strength (denoted by the color) around $M$ point in Fig. \ref{fig:dirac_coupling}(a). For AA stacking BG, the energy bands (dashed lines) for the top and bottom monolayers are degenerate at all k-points, and hence the degenerate energy bands are split with the energy difference denoted by the green double arrow and the relatively large hybridization strength at each k-point, as shown in Fig. \ref{fig:dirac_coupling}(b), as a result of the first-order degenerate perturbation theory.

\subsection{S5.3. Electron-hole asymmetrical hybridization strength inside high energy area}
\subsubsection{S5.3.1. A toy model of two six-carbon rings as a starting point}
\begin{figure}[!htbp]
\centering
\includegraphics[width=6 cm]{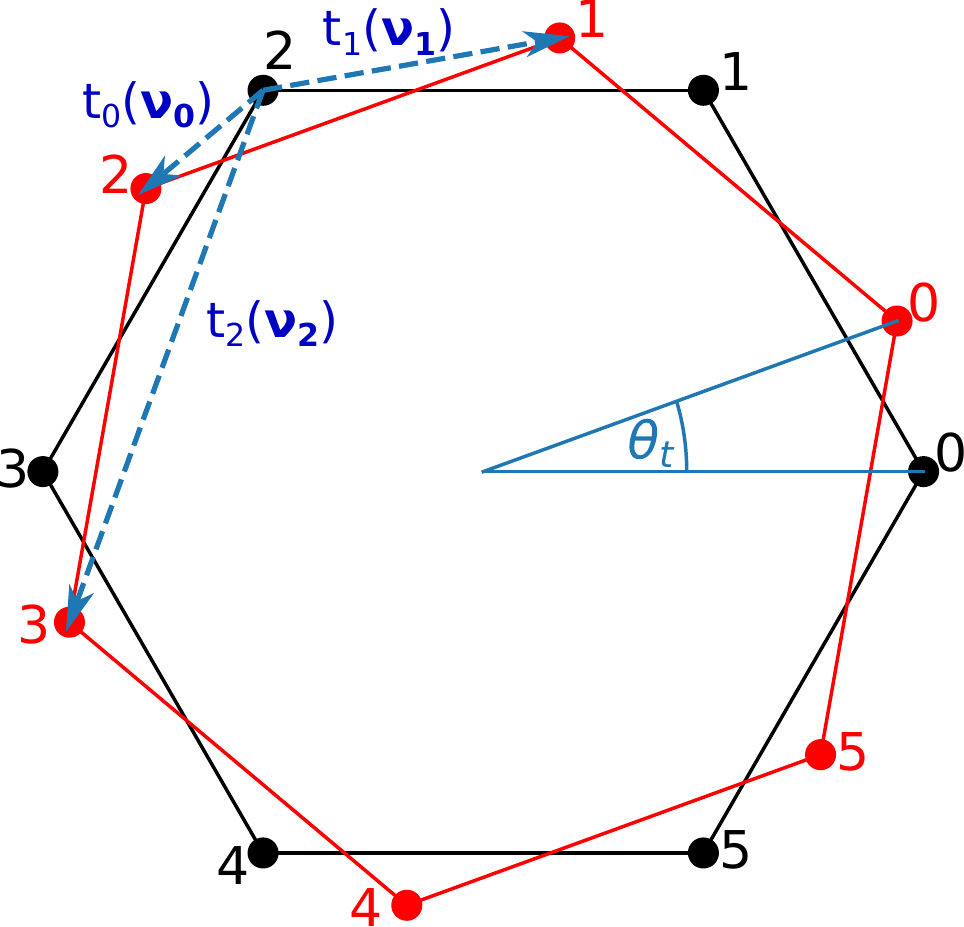}
\caption{Two six-carbon rings with a twist angle of $\theta_t$, where the bottom and top rings are denoted by black and red colors respectively, $t_0$, $t_1$ and $t_2$ are the interlayer NN, NNN and TNN hoppings, respectively, and $\bm{\nu}_0$, $\bm{\nu}_1$, and $\bm{\nu}_2$ are the interlayer NN, NNN and TNN displacements, respectively.}
\label{fig:t_rings}
\end{figure}

Fig. \ref{fig:t_rings} shows the structure of two six-carbon rings. The Hamiltonian of the twisted two six-carbon rings (see Fig. \ref{fig:t_rings}) reads
\begin{equation}
H = \left[
\begin{matrix}
H_0^b & U \\
U^{\dagger} & H_0^t
\end{matrix}
\right],
\end{equation}
where $H_0^b$ and $H_0^t$ are the Hamiltonians of the bottom and top rings, respectively, and $U$ is the interlayer coupling. For simplicity we consider the nearest-neighbour approximation within the monolayers and write $H_0^b$ and $H_0^t$ as
\begin{equation}
H_0^b = H_0^t = \left[
\begin{matrix}
0 & -t & 0 & 0 & 0 & -t \\
-t & 0 & -t & 0 & 0 & 0 \\
0 & -t & 0 & -t & 0 & 0 \\
0 & 0 & -t & 0 & -t & 0 \\
0 & 0 & 0 & -t & 0 & -t \\
-t & 0 & 0 & 0 & -t & 0
\end{matrix}
\right],
\label{eq:Hmono}
\end{equation}
where $-t$ is the nearest-neighbour hopping energy. The interlayer coupling $U$ within the third-nearest-neighbor approximation of the interlayer hoppings is written as
\begin{equation}
U= \left[
\begin{matrix}
t_0 & t_2 & 0 & 0 & 0 & t_1 \\
t_1 & t_0 & t_2 & 0 & 0 & 0 \\
0 & t_1 & t_0 & t_2 & 0 & 0 \\
0 & 0 & t_1 & t_0 & t_2 & 0 \\
0 & 0 & 0 & t_1 & t_0 & t_2 \\
t_2 & 0 & 0 & 0 & t_1 & t_0
\end{matrix}
\right],
\label{eq:BG_U}
\end{equation}
where $t_0$, $t_1$ and $t_2$ are the nearest-neighbor (NN), next-nearest-neighbor (NNN) and third-nearest-neighbor (TNN) interlayer hoppings, respectively, as shown in Fig. \ref{fig:t_rings}. From Eqs.~\eqref{methods:pzTB} and ~\eqref{methods:pzTBsk}, one can find that all $t_0$, $t_1$ and $t_2$ are positive. In addition, for untwisted case with $\theta_t=0^{\circ}$, $t_0>t_1=t_2>0$, and for $\theta_t=30^{\circ}$, $t_0=t_1\gg t_2\sim0$. The real eigenvalues $\epsilon$ and eigenvectors $\psi$ for $H_0^b$ and $H_0^t$ in Eq.~\eqref{eq:Hmono} read
\begin{equation}
\begin{split}
A_1: \quad \psi_0^{b/t} &= {{1}\over{\sqrt{6}}}(1, 1, 1, 1, 1, 1)^T \quad (\epsilon = -2t),\\
E_1: \quad \psi_1^{b/t} &= {{1}\over{2}}(-1, -1, 0, 1, 1, 0)^T \quad (\epsilon = -t), \\
E_1: \quad \psi_2^{b/t} &= {{1}\over{2}}(1, 0,  -1, -1, 0, 1)^T \quad (\epsilon = -t), \\
E_2: \quad \psi_3^{b/t} &= {{1}\over{2}}(-1, 1, 0,  -1, 1, 0)^T \quad (\epsilon = t), \\
E_2: \quad \psi_4^{b/t} &= {{1}\over{2}}(-1, 0, 1, -1, 0, 1)^T \quad (\epsilon = t), \\
B_1: \quad \psi_5^{b/t} &= {{1}\over{\sqrt{6}}}(-1, 1, -1, 1, -1, 1)^T \quad (\epsilon = 2t).
\end{split}
\end{equation}
\begin{figure}[!htbp]
\centering
\includegraphics[width=14 cm]{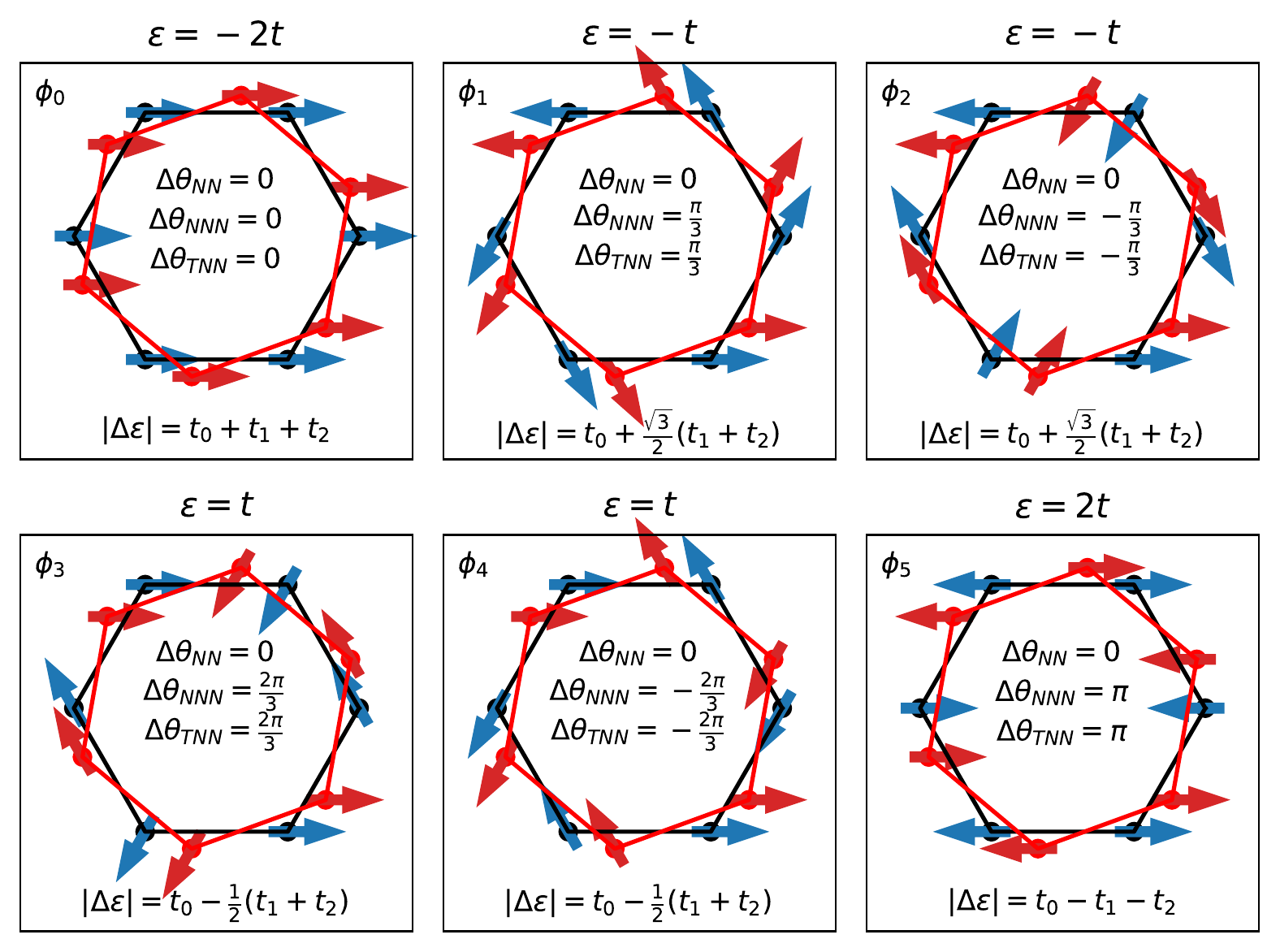}
\caption{Phases of $12$ eigen states $\phi_i^{b/t}$ (with $i=0,...,5$, and $\phi_i^{b/t}=\phi_i$) in Eq.~\eqref{eq:phi_i} for the two six-carbon rings with a twist angle of $\theta_t$. The wave-function phase difference of hybridization paring states $\phi_i^{b/t}$ for interlayer NN, NNN and TNN displacements are denoted by $\Delta\theta_{NN}$, $\Delta\theta_{NNN}$ and $\Delta\theta_{TNN}$, respectively. The calculated interlayer hybridization strengths $\Delta\varepsilon_{anti}$ (here labeled as $\Delta\varepsilon$) and $\Delta\varepsilon_{bond}$ (i.e., $-\Delta\varepsilon$) in Eq.~\eqref{eq:hybDelta} decreases with the increasing energy $\varepsilon$.}
\label{fig:phase_map}
\end{figure}
Within the basis function of rotation operation $C_{6}$, we rewrite the eigenvectors as
\begin{equation}
\begin{split}
A_1, \theta=0: \quad \phi_0^{{b/t}} &= \psi_0^{b/t} = {{1}\over{\sqrt{6}}}(1, 1, 1, 1, 1, 1)^T \quad (\epsilon = -2t),\\
E_1, \theta=-{{\pi}\over{3}}: \quad \phi_1^{b/t} &= {{2}\over{\sqrt{6}}}(e^{-i{{\pi}\over{3}}}\psi_1^{b/t} + \psi_2^{b/t}) \quad (\epsilon = -t) \\
      &= {{1}\over{\sqrt{6}}}(e^{i{{\pi}\over{3}}}, e^{i{{2\pi}\over{3}}}, e^{i{\pi}}, e^{-i{{2\pi}\over{3}}}, e^{-i{{\pi}\over{3}}}, e^{i0})^T , \\
E_1, \theta={{\pi}\over{3}}: \quad \phi_2^{b/t} &= {{2}\over{\sqrt{6}}}(e^{i{{\pi}\over{3}}}\psi_1^{b/t} + \psi_2^{b/t}) \quad (\epsilon = -t), \\
     &= {{1}\over{\sqrt{6}}}(e^{-i{{\pi}\over{3}}}, e^{-i{{2\pi}\over{3}}}, e^{i{\pi}}, e^{i{{2\pi}\over{3}}}, e^{i{{\pi}\over{3}}}, e^{i0})^T , \\
E_2, \theta=-{{2\pi}\over{3}}: \quad \phi_3^{b/t} &= {{2}\over{\sqrt{6}}}(e^{-i{{2\pi}\over{3}}}\psi_3^{b/t} + \psi_4^{b/t}) \quad (\epsilon = t), \\
     &= {{1}\over{\sqrt{6}}}(e^{i{{2\pi}\over{3}}}, e^{-i{{2\pi}\over{3}}}, e^{i{0}}, e^{i{{2\pi}\over{3}}}, e^{-i{{2\pi}\over{3}}}, e^{i0})^T , \\
E_2, \theta={{2\pi}\over{3}}: \quad \phi_4^{b/t} &= {{2}\over{\sqrt{6}}}(e^{i{{2\pi}\over{3}}}\psi_3^{b/t} + \psi_4^{b/t}) \quad (\epsilon = t), \\
     &= {{1}\over{\sqrt{6}}}(e^{-i{{2\pi}\over{3}}}, e^{i{{2\pi}\over{3}}}, e^{i{0}}, e^{-i{{2\pi}\over{3}}}, e^{i{{2\pi}\over{3}}}, e^{i0})^T , \\
B_1, \theta=\pi: \quad \phi_5^{b/t} &= \psi_5^{b/t} ={{1}\over{\sqrt{6}}}(-1, 1, -1, 1, -1, 1)^T \quad (\epsilon = 2t),
\end{split}
\label{eq:phi_i}
\end{equation}
where $\theta$ is related to the eigenvalue of $C_6$, i.e., $C_6\phi_i^{b/t} = e^{i\theta}\phi_i^{b/t}$. Using Eq.~\eqref{eq:phi_i} and the hybridization selection rule in Eq. (5) of main text, we can write a uniform formula for an intralayer state as $\phi^{b/t} = (a_0^{b/t}, a_1^{b/t}, a_2^{b/t}, a_3^{b/t}, a_4^{b/t}, a_5^{b/t})^T$ with $\sum_{i=0}^5 \left|a_i^{b/t}\right|^2 = 1$. The bonding $(-)$ and anti-bonding $(+)$ states for the intralayer state read
\begin{equation}
\begin{split}
\varphi_- &= {{1}\over{\sqrt{2}}}\begin{pmatrix}
\phi^b \\
- \phi^t
\end{pmatrix}
 ={{1}\over{\sqrt{2}}}(a_0^b, a_1^b, a_2^b, a_3^b, a_4^b, a_5^b, -a_0^t, -a_1^t, -a_2^t, -a_3^t, -a_4^t, -a_5^t)^T, \\
\varphi_+ &= {{1}\over{\sqrt{2}}}\begin{pmatrix}
\phi^b \\
\phi^t
\end{pmatrix} ={{1}\over{\sqrt{2}}}(a_0^b, a_1^b, a_2^b, a_3^b, a_4^b, a_5^b, a_0^t, a_1^t, a_2^t, a_3^t, a_4^t, a_5^t)^T.
\end{split}
\label{eq:BG_phi}
\end{equation}
The energy for bonding $(-)$ and anti-bonding $(+)$ states read
\begin{equation}
\varepsilon_{\pm}=\varphi_{\pm}^{\dagger}H\varphi_{\pm}={{1}\over{2}}{[\phi^b]}^{\dagger}H_0^b\phi^b + {{1}\over{2}}{[\phi^t]}^{\dagger}H_0^t\phi^t \pm {{1}\over{2}}{[\phi^b]}^{\dagger}U\phi^t \pm {{1}\over{2}}{[\phi^t]}^{\dagger}U^{\dagger}\phi^b ={{1}\over{2}} \varepsilon^b + {{1}\over{2}}\varepsilon^t \pm {{1}\over{2}}{[\phi^b]}^{\dagger}U\phi^t \pm {{1}\over{2}}{[\phi^t]}^{\dagger}U^{\dagger}\phi^b .
\end{equation}
Then, the corresponding interlayer hybridization strengths for the two states are
\begin{equation}
\begin{split}
\Delta \varepsilon_{bond} &= - {{1}\over{2}}{[\phi^b]}^{\dagger}U\phi^t - {{1}\over{2}}{[\phi^t]}^{\dagger}U^{\dagger}\phi^b,\\
\Delta \varepsilon_{anti} &= {{1}\over{2}}{[\phi^b]}^{\dagger}U\phi^t + {{1}\over{2}}{[\phi^t]}^{\dagger}U^{\dagger}\phi^b,
\end{split}
\label{eq:hybDelta}
\end{equation}
where ${[\phi^b]}^{\dagger}U\phi^t$ and ${[\phi^t]}^{\dagger}U^{\dagger}\phi^b$ takes the form as
\begin{equation}
\begin{split}
{[\phi^b]}^{\dagger}U\phi^t =  \sum_{i=0}^5t_0(a_i^{b,*}a_i^t)
+ t_1(a_0^{b,*}a_5^t+a_1^{b,*}a_0^t+a_2^{b,*}a_1^t +a_3^{b,*}a_2^t +a_4^{b,*}a_3^t +a_5^{b,*}a_4^t) \\
+ t_2(a_0^{b,*} a_1^t +a_1^{b,*}a_2^t+a_2^{b,*}a_3^t +a_3^{b,*}a_4^t +a_4^{b,*} a_5^t+a_5^{b,*} a_0^t),\\
{[\phi^t]}^{\dagger}U^{\dagger}\phi^b =  \sum_{i=0}^5t_0(a_i^ba_i^{t,*})
+ t_1(a_0^ba_5^{t,*}+a_1^ba_0^{t,*}+a_2^ba_1^{t,*} +a_3^ba_2^{t,*} +a_4^ba_3^{t,*} +a_5^ba_4^{t,*}) \\
+ t_2(a_0^ba_1^{t,*} +a_1^ba_2^{t,*}+a_2^ba_3^{t,*} +a_3^ba_4^{t,*} +a_4^ba_5^{t,*}+a_5^ba_0^{t,*}).
\end{split}
\label{eq:all_terms}
\end{equation}
In Eq.~\eqref{eq:all_terms}, Eqs.~\eqref{eq:BG_U} and ~\eqref{eq:BG_phi} are used. Eq.~\eqref{eq:all_terms} is further rewritten as
\begin{equation}
\begin{split}
{[\phi^b]}^{\dagger}U\phi^t =  \sum_{j=0}^5t_0(a_{\bm{r}_j}^{b,*}a_{\bm{r}_j+\bm{\nu}_0}^t)
+ \sum_{j=0}^5t_1(a_{\bm{r}_j}^{b,*}a_{\bm{r}_j+\bm{\nu}_1}^t)
+ \sum_{j=0}^5t_2(a_{\bm{r}_j}^{b,*}a_{\bm{r}_j+\bm{\nu}_2}^t),\\
{[\phi^t]}^{\dagger}U^{\dagger}\phi^b =  \sum_{j=0}^5t_0(a_{\bm{r}_j}^ba_{\bm{r}_j+\bm{\nu}_0}^{t,*})
+ \sum_{j=0}^5t_1(a_{\bm{r}_j}^ba_{\bm{r}_j+\bm{\nu}_1}^{t,*})
+ \sum_{j=0}^5t_2(a_{\bm{r}_j}^ba_{\bm{r}_j+\bm{\nu}_2}^{t,*}),
\end{split}
\label{eq:all_termsr}
\end{equation}
where $\bm{\nu}_0$, $\bm{\nu}_1$, and $\bm{\nu}_2$ are the interlayer NN, NNN and TNN displacements. Eqs.~\eqref{eq:hybDelta} and ~\eqref{eq:all_termsr} indicate that the interlayer hybridization strength $\Delta \varepsilon_{bond}$ and $\Delta \varepsilon_{anti}$ are determined by these interlayer hoppings $t_0$, $t_1$ and $t_2$ and the wave-function phase difference of hybridization paring states for these NN, NNN and TNN displacements. As shown in Fig. \ref{fig:phase_map} and Eq.~\eqref{eq:phi_i}, with the increasing energy, i.e., $\phi_i$ from $\phi_0$ to $\phi_5$, the absolute values $|\Delta\theta_{NN}^j|$ (with $e^{i\Delta\theta_{NN}^j}=a_{\bm{r}_j}^{b,*}a_{\bm{r}_j+\bm{\nu}_0}^t$) of phase differences $\Delta\theta_{NN}^j$ of each hybridization paring states $\phi_i^{b}$ and $\phi_i^{t}$ for the interlayer NN displacements are always equal to $0$, namely $\Delta\theta_{NN}=0$. For $\phi_i$ from $\phi_0$ to $\phi_5$, both the absolute values $|\Delta\theta_{NNN}^j|$ (with $e^{i\Delta\theta_{NNN}^j}=a_{\bm{r}_j}^{b,*}a_{\bm{r}_j+\bm{\nu}_1}^t$) and $|\Delta\theta_{TNN}^j|$ (with $e^{i\Delta\theta_{TNN}^j}=a_{\bm{r}_j}^{b,*}a_{\bm{r}_j+\bm{\nu}_2}^t$) of phase differences $\Delta\theta_{NNN}^j$ and $\Delta\theta_{TNN}^j$ for the interlayer NNN and TNN displacements vary in a form of
\begin{equation}
|\Delta\theta_{NNN}^j|=|\Delta\theta_{TNN}^j|=|\Delta\theta_{NNN}|=|\Delta\theta_{TNN}|=0\rightarrow \pi/3 \rightarrow \pi/3 \rightarrow 2\pi/3 \rightarrow 2\pi/3 \rightarrow \pi.
\end{equation}
Therefore, with the increasing energy from negative to positive values, the size $\Delta\varepsilon$ of the interlayer hybridization strength for the bonding and anti-bonding states (i.e., $\Delta\varepsilon=\Delta\varepsilon_{anti}=-\Delta\varepsilon_{bond}$) vary with
\begin{equation}
\Delta\varepsilon = t_0+t_1+t_2 \rightarrow t_0+\sqrt{3}(t_1+t_2)/2 \rightarrow t_0+\sqrt{3}(t_1+t_2)/2 \rightarrow t_0-(t_1+t_2)/2 \rightarrow t_0-(t_1+t_2)/2 \rightarrow t_0-t_1-t_2.
\label{eq:ehring}
\end{equation}
Eq.~\eqref{eq:ehring} indicates $t_1$, $t_2$, $\Delta\theta_{NNN}^j$ and $\Delta\theta_{TNN}^j$ are reponsible for the electron-hole asymmetry of the interlayer hybridization with the stronger hybridization strength for holes in the structure of two six-carbon rings.

\subsubsection{S5.3.2. Generalized formula for hybridization strength in an arbitrary BG system}
For an arbitrary BG system including $2N$ atoms with $N$ as the number of atoms in one monolayer and a given twist angle $\theta_t\neq 0^{\circ}$ (twisted) or $\theta_t= 0^{\circ}$ (untwisted), the interlayer NN hopping $(t_0)$, NNN hopping $(t_1)$ and TNN hopping $(t_2)$ are determined by Eqs.~\eqref{methods:pzTB} and ~\eqref{methods:pzTBsk}. The hybridization paring states $\phi^{b/t}$ are written as
\begin{equation}
\phi^{b/t} = (a_0^{b/t}, a_1^{b/t}, \cdots, a_{N-1}^{b/t})^T,
\end{equation}
where $\sum_{i=0}^{N-1} \left|a_i^{b/t}\right|^2 = 1$. The bonding $(-)$ and anti-bonding $(+)$ states for the intralayer state read
\begin{equation}
\begin{split}
\varphi_- &= {{1}\over{\sqrt{2}}}\begin{pmatrix}
\phi^b \\
- \phi^t
\end{pmatrix}
 ={{1}\over{\sqrt{2}}}(a_0^b, a_1^b, \cdots, a_{N-1}^b, -a_0^t, -a_1^t, \cdots, -a_{N-1}^t)^T, \\
\varphi_+ &= {{1}\over{\sqrt{2}}}\begin{pmatrix}
\phi^b \\
\phi^t
\end{pmatrix}
 ={{1}\over{\sqrt{2}}}(a_0^b, a_1^b, \cdots, a_{N-1}^b, a_0^t, a_1^t,\cdots, a_{N-1}^t)^T.
\end{split}
\label{eq:BG_phiN}
\end{equation}
The interlayer hybridization strengths of bonding and anti-bonding states have the same forms as Eq.~\eqref{eq:hybDelta}, where ${[\phi^b]}^{\dagger}U\phi^t$ and ${[\phi^t]}^{\dagger}U^{\dagger}\phi^b$ for the arbitrary BG system including $2N$ atoms can be written as
\begin{equation}
\begin{split}
{[\phi^b]}^{\dagger}U\phi^t = \sum_{j=0}^{N-1}\sum_{\bm{\nu}_0}t_0(a_{\bm{r}_j}^{b,*}a_{\bm{r}_j+\bm{\nu}_0}^t)
+ \sum_{j=0}^{N-1}\sum_{\bm{\nu}_1}t_1(a_{\bm{r}_j}^{b,*}a_{\bm{r}_j+\bm{\nu}_1}^t)
+ \sum_{j=0}^{N-1}\sum_{\bm{\nu}_2}t_2(a_{\bm{r}_j}^{b,*}a_{\bm{r}_j+\bm{\nu}_2}^t),\\
{[\phi^t]}^{\dagger}U^{\dagger}\phi^b =  \sum_{j=0}^{N-1}\sum_{\bm{\nu}_0}t_0(a_{\bm{r}_j}^ba_{\bm{r}_j+\bm{\nu}_0}^{t,*})
+ \sum_{j=0}^{N-1}\sum_{\bm{\nu}_1}t_1(a_{\bm{r}_j}^ba_{\bm{r}_j+\bm{\nu}_1}^{t,*})
+ \sum_{j=0}^{N-1}\sum_{\bm{\nu}_2}t_2(a_{\bm{r}_j}^ba_{\bm{r}_j+\bm{\nu}_2}^{t,*}).
\end{split}
\label{eq:all_termsrNeq}
\end{equation}
The hybridization paring states $\phi^{b/t}$ are the same for equivalent and mixed hybridizations according to the hybridization selection rules in Table I of main text, and hence the interlayer NN phase difference $\Delta\theta_{NN}^j$ is always $0$, i.e., $a_{\bm{r}_j}^{b,*}a_{\bm{r}_j+\bm{\nu}_0}^t=1$, in Eq.~\eqref{eq:all_termsrNeq}. As a result, with the increasing energy, the hybridization strength $\Delta\varepsilon$ is changed by the interlayer non-nearest-neighbor hoppings $t_1$ and $t_2$ and phase differences $\Delta\theta_{NNN}^j$ (with $e^{i\Delta\theta_{NNN}^j}=a_{\bm{r}_j}^{b,*}a_{\bm{r}_j+\bm{\nu}_1}^t$) and $\Delta\theta_{TNN}^j$ (with $e^{i\Delta\theta_{TNN}^j}=a_{\bm{r}_j}^{b,*}a_{\bm{r}_j+\bm{\nu}_2}^t$).

For non-equivalent hybridizations in $D_{6d}$ graphene quasicrystal systems in Eq.~\eqref{hyb:D6d_nonequi}, $\phi_{B_i}^b$ and $\phi_{B_i}^t$  with irrep $B_i$ have the same energy $\varepsilon_{B_i}^b=\varepsilon_{B_i}^t$ with $i=1,2$, after the local reflection plane is chosen, i.e., $\sigma_{v,0}^b=\sigma_x$ and $\sigma_{v,i}^{t}=\sigma_{d,i}^{b}$. Consequently, two non-equivalent hybridizations ${B_1}^b+{B_2}^t$ and ${B_2}^b+{B_1}^t$ in Eq.~\eqref{hyb:D6d_nonequi} with sign $+$ denoting paring are actually the same, and hence we can perform a following transformation of the wave functions of two non-equivalent hybridizations to obtain a uniform expression of the hybridization paring states
\begin{equation}
\begin{pmatrix}
\phi_{+}^{b/t} \\
\phi_{-}^{b/t}
\end{pmatrix} = {{1}\over{\sqrt{2}}}\begin{pmatrix}
 1 & 1\\
 1 & -1\\
\end{pmatrix}\begin{pmatrix}
\phi_{B_1}^{b/t} \\
\phi_{B_2}^{b/t}
\end{pmatrix},
\end{equation}
where the hybridization paring states $\phi_\lambda^{b/t}$ with $\lambda = +$ or $-$ are written as
\begin{equation}
\phi_\lambda^{b/t} = (c_{\lambda,0}^{b/t}, c_{\lambda,1}^{b/t}, \cdots, c_{\lambda,N-1}^{b/t})^T,
\end{equation}
with $\sum_{i=0}^{N-1} \left|c_{\lambda,i}^{b/t}\right|^2 = 1$. The bonding $(-)$ and anti-bonding $(+)$ states for the intralayer state read
\begin{equation}
\begin{split}
\varphi_{\lambda,-} &= {{1}\over{\sqrt{2}}}\begin{pmatrix}
\phi_{\lambda}^b \\
- \phi_{\lambda}^t
\end{pmatrix}
 ={{1}\over{\sqrt{2}}}(c_{\lambda,0}^b, c_{\lambda,1}^b, \cdots, c_{\lambda,N-1}^b, -c_{\lambda,0}^t, -c_{\lambda,1}^t, \cdots, -c_{\lambda,N-1}^t)^T, \\
\varphi_{\lambda,+} &= {{1}\over{\sqrt{2}}}\begin{pmatrix}
\phi_{\lambda}^b \\
\phi_{\lambda}^t
\end{pmatrix}
 ={{1}\over{\sqrt{2}}}(c_{\lambda,0}^b, c_{\lambda,1}^b, \cdots, c_{\lambda,N-1}^b, c_{\lambda,0}^t, c_{\lambda,1}^t, \cdots, c_{\lambda,N-1}^t)^T.
\end{split}
\label{eq:BG_phiNc}
\end{equation}
The interlayer hybridization strengths of bonding and anti-bonding states have the same forms as Eq.~\eqref{eq:hybDelta}, where ${[\phi_{\lambda}^b]}^{\dagger}U\phi_{\lambda}^t$ and ${[\phi_{\lambda}^t]}^{\dagger}U^{\dagger}\phi_{\lambda}^b$ for the arbitrary BG system including $2N$ atoms can be written as
\begin{equation}
\begin{split}
{[\phi_{\lambda}^b]}^{\dagger}U\phi_{\lambda}^t = \sum_{j=0}^{N-1}\sum_{\bm{\nu}_0}t_0(c_{\lambda,\bm{r}_j}^{b,*}c_{\lambda,\bm{r}_j+\bm{\nu}_0}^t)
+ \sum_{j=0}^{N-1}\sum_{\bm{\nu}_1}t_1(c_{\lambda,\bm{r}_j}^{b,*}c_{\lambda,\bm{r}_j+\bm{\nu}_1}^t)
+ \sum_{j=0}^{N-1}\sum_{\bm{\nu}_2}t_2(c_{\lambda,\bm{r}_j}^{b,*}c_{\lambda,\bm{r}_j+\bm{\nu}_2}^t),\\
{[\phi_{\lambda}^t]}^{\dagger}U^{\dagger}\phi_{\lambda}^b =  \sum_{j=0}^{N-1}\sum_{\bm{\nu}_0}t_0(c_{\lambda,\bm{r}_j}^bc_{\lambda,\bm{r}_j+\bm{\nu}_0}^{t,*})
+ \sum_{j=0}^{N-1}\sum_{\bm{\nu}_1}t_1(c_{\lambda,\bm{r}_j}^bc_{\lambda,\bm{r}_j+\bm{\nu}_1}^{t,*})
+ \sum_{j=0}^{N-1}\sum_{\bm{\nu}_2}t_2(c_{\lambda,\bm{r}_j}^bc_{\lambda,\bm{r}_j+\bm{\nu}_2}^{t,*}).
\end{split}
\label{eq:all_termsrNeqc}
\end{equation}
In this respect, for the same $\lambda$, $\phi_{\lambda}^b$ and $\phi_{\lambda}^t$ are the same, and hence  the interlayer NN phase difference $\Delta\theta_{NN}^j$ is also $0$, i.e., $c_{\lambda,\bm{r}_j}^{b,*}c_{\lambda,\bm{r}_j+\bm{\nu}_0}^t=1$. As a result, with the increasing energy, the hybridization strength $\Delta\varepsilon$ is still changed by the interlayer non-nearest-neighbor hoppings $t_1$ and $t_2$ and the phase differences $\Delta\theta_{NNN}^j$ (with $e^{i\Delta\theta_{NNN}^j}=c_{\lambda,\bm{r}_j}^{b,*}c_{\lambda,\bm{r}_j+\bm{\nu}_1}^t)$ and $\Delta\theta_{TNN}^j$ (with $e^{i\Delta\theta_{TNN}^j}=c_{\lambda,\bm{r}_j}^{b,*}c_{\lambda,\bm{r}_j+\bm{\nu}_2}^t$) of the hybridization paring states $\phi_{\lambda}^{b/t}$. In a word, for three categories of hybridizations in Table I of main text, with the increasing energy, the electron-hole asymmetrical hybridization with the stronger hybridization strength for holes is a result of non-nearest-neighbor interlayer hoppings and phase differences of hybridization paring states.

\subsection{S5.4. Electric field effects on hybridization strength}
\begin{figure}[!htbp]
\centering
\includegraphics[width=14 cm]{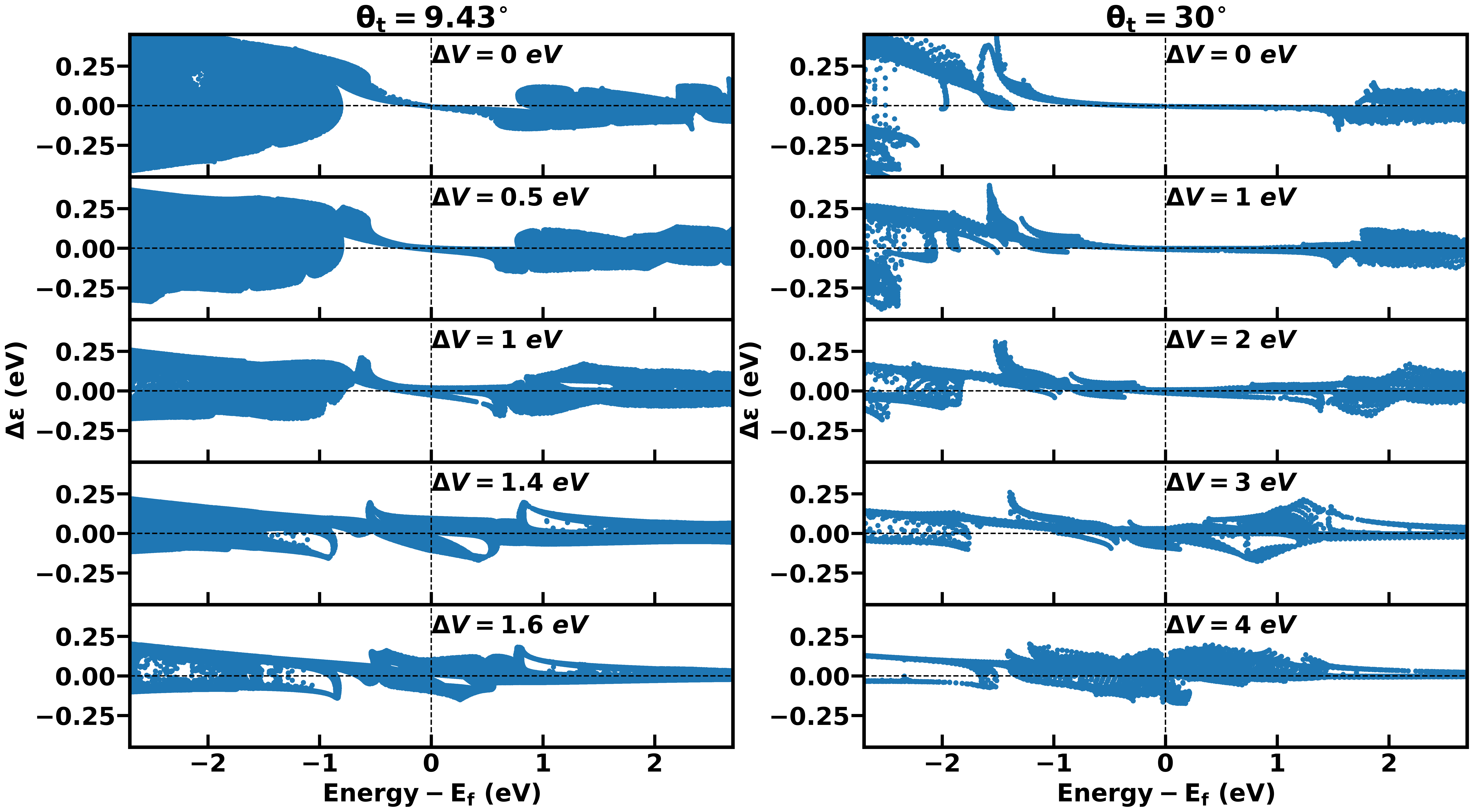}
\caption{The hybridization strengths under different interlayer potential difference $\triangle V$ as a function of energy for twisted BG with $\theta_t=9.43^{\circ}$ (left panel) and graphene quasicrystal simulated by a periodic 15/26 approximant (right panel).}
\label{fig:hybrid_strength_Ele}
\end{figure}

\begin{figure}[!htbp]
\centering
\includegraphics[width=16 cm]{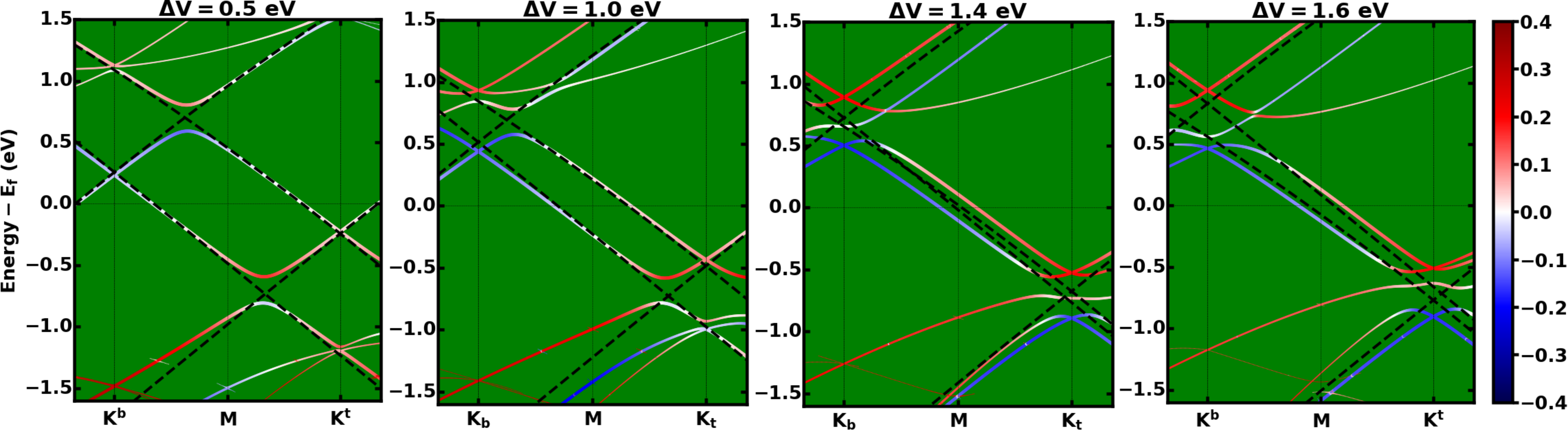}
\caption{The calculated unfolded band structures for twisted BG with $\theta_t=9.43^{\circ}$ under different interlayer potential difference $\triangle V$, where the color map denotes the interlayer hybridization strength $\triangle\varepsilon$, and black dashed lines stand for the band structures of the bottom and top graphene monolayers.}
\label{fig:coupling_Ele}
\end{figure}

The interlayer potential difference $\triangle V$ induced by a vertical electric field is usually applied to modify the electronic structures and transport properties of twisted BG to achieve gate-controllable devices. Here we reveal the electric field effects on the hybridization strength in twisted BG. Fig. \ref{fig:hybrid_strength_Ele} shows the hybridization strengths under different interlayer potential difference $\triangle V$ in twisted BG with $\theta_t=9.43^{\circ}$ and graphene quasicrystal within a periodic 15/26 approximant. With the increasing interlayer potential difference $\triangle V$, the hybridization strengths near the Fermi level are enhanced. The results can be still explained by the second-order non-degenerate perturbation theory for non-degenerate energy positions and the first-order degenerate perturbation theory for degenerate energy positions. For example, for the unfolded band structures of twisted BG with $\theta_t=9.43^{\circ}$ in Fig. \ref{fig:coupling_Ele}, the interlayer potential difference $\triangle V$ lifts the Dirac linear dispersion around $K^b$ of the bottom graphene and moves down the Dirac linear dispersion around $K^t$ of the top graphene. Consequently, the energy difference between the two  linear bands is reduced near the Fermi level, and hence the hybridization strength is increased near the Fermi level according to the second-order non-degenerate perturbation theory. For the degenerate energy positions, the first-order degenerate perturbation theory induces an obvious hybridization strength as well.

\section{S6. Symmetry and electric-field effects on resonant quasicrystalline states}
\subsection{S6.1. Hamiltonian for resonant quasicyrstalline states}
\begin{figure}[!htbp]
\centering
\includegraphics[width=4 cm]{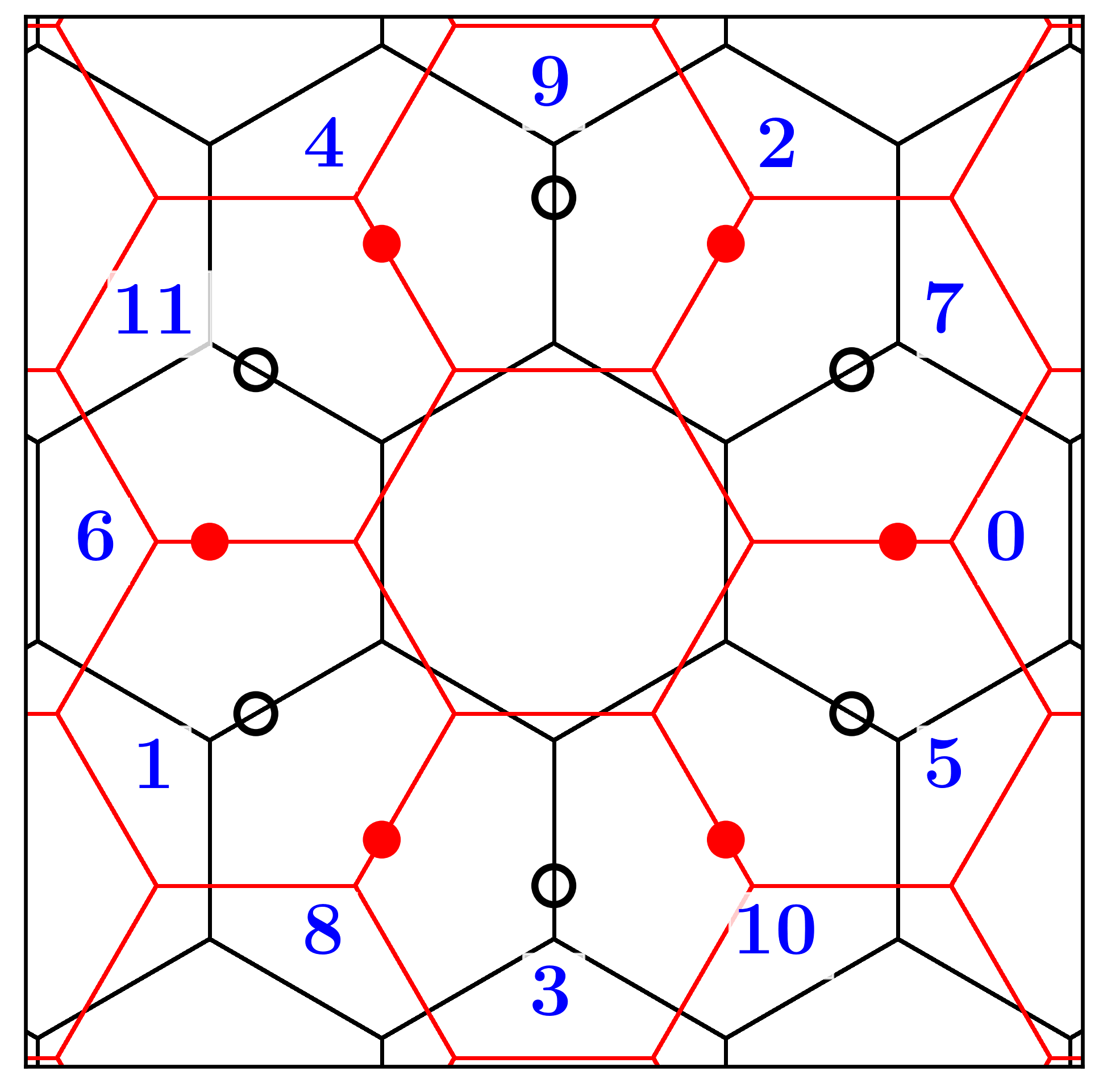}
\caption{The 12 k-points for constructing the 24$\times$24 Hamiltonian of the resonant states in reciprocal space. The black and red hexagon networks respectively denote the Brillouin zones for the bottom and top layers. The 12 k-points are labelled by black empty and red solid dots with corresponding numbers from the two $C_{6v}$ subsystems.}
\label{fig:GQQD_recip}
\end{figure}

Graphene quasicyrstalline has various special electronic states, such as the multiple Dirac cones together with 12-fold rotational symmetry. Another one unique electronic state in $\bm{k}$ space is the 12-fold symmetric resonant quasicyrstalline states, which are these states inside the $\bm{k}_0$-based subspace with $\bm{k}^b = \bm{k}_0 + \bm{G}^t$ and $\bm{k}^t = \bm{k}_0 + \bm{G}^b$, where $\bm{k}_0=0$ is focused because they have the same energy. According to the wave-vector-dependent interlayer coupling condition in Eqs.~\eqref{eq:coupk} and \eqref{eq:coupg}, the 12 k-points in Fig. \ref{fig:GQQD_recip}, i.e., these vectors $\bm{G}^b$ and $\bm{G}^t$ with their lengths less than ${4\pi}/{\sqrt{3}a}$, mainly contribute the resonant quasicyrstalline states. The 12 k-points are made up of two $C_{6v}$ subsystems of six k-points denoted by black empty dots from the bottom layer and six k-points denoted by red solid dots from the top layer in Fig. \ref{fig:GQQD_recip} and hence keeps the $D_{6d}$ point group symmetry, i.e.,
\begin{equation}
\begin{split}
E & = I(12) \otimes \sigma_0, \\
S_{12}^{2i+1} &= C_{12}^{2i+1}(\bm{k})\otimes \sigma_0, \quad (i=0,1,...,5), \\
C_{12}^{2i} &= C_{12}^{2i}(\bm{k})\otimes \sigma_0, \quad (i=1,2,...,5), \\
C_{2,i}^{\prime} &= C_{2,i}^{\prime}(\bm{k})\otimes\sigma_1,  \quad (i = 0,1,...,5),\\
\sigma_{d,i} &= \sigma_{d,i}(\bm{k})\otimes \sigma_1, \quad (i = 0,1,...,5),\\
\end{split}
\end{equation}
where $I(12)$, $C_{12}^{2i+1}(\bm{k})$, $C_{12}^{2i}(\bm{k})$, $C_{2,i}^{\prime}(\bm{k})$ and $\sigma_{d,i}(\bm{k})$ are the symmetry operations for the 12 k-points, and the sublattice-dependent $\sigma_0$ and $\sigma_1$ are the identity matrix $I(2)$ and the $x$ component of Pauli matrices, respectively. In the direct product space of the 12 k-points and the two sublattices in reciprocal space, one can construct a 24$\times$24 tight-binding Hamiltonian at $\bm{k}_0=\bm{0}$ for the resonant quasicyrstalline states\cite{Moon_QC}, as follows:
\begin{equation}
H =
\begin{pmatrix}
H^{(0)}     &    W_{0,1}        &         &               &              & W_{11,0} \\
W_{0,1}^{\dagger} & H^{(1)}     &  W_{1,2}      &               &              &  \\
            & W_{1,2}^{\dagger} & H^{(2)} &        \ddots       &              &  \\
            &             & \ddots  &  \ddots       & W_{9,10}       &  \\
            &             &         &  W_{9,10}^{\dagger}  & H^{(10)}     & W_{10,11} \\
W_{11,0}^{\dagger}           &             &         &               & W_{10,11}^{\dagger} & H^{(11)}
\end{pmatrix}.\\
\label{Hmat}
\end{equation}
Here, $H^{(n)}$ is the Hamiltonian at $nth$ k-point with the form
\begin{equation}
\begin{split}
H^{(n)} =
\begin{pmatrix}
\Delta V/ 2 &    0.682\gamma_0      \\
0.682\gamma_0      & \Delta V/2  \\
\end{pmatrix} \quad \textrm{(odd $n$)}, \\
H^{(n)} =
\begin{pmatrix}
-\Delta V/2 &    0.682\gamma_0      \\
0.682\gamma_0      & -\Delta V/2  \\
\end{pmatrix}\quad \textrm{(even $n$)},
\end{split}
\label{Hmat}
\end{equation}
where $\Delta V$ is the interlayer potential difference induced by the vertical electric field. $W_{n,n+1}$ is the interlayer interaction between $nth$ and $(n+1)th$ k-points. If the sublattice order is rearranged as $(A^b,B^b)$ or ($A^t,B^t$) when $n\mod 4=2,3$ and $(B^b,A^b)$ or ($B^t,A^t$) when $n\mod 4=0,1$, all of the interlayer interaction matrices $W_{n,n+1}$'s have the same form
\begin{equation}
W_{n,n+1} =T_0
\begin{pmatrix}
e^{{{2\pi}\over{3}}i} &    1      \\
1      & e^{-{{2\pi}\over{3}}i} \\
\end{pmatrix},
\label{Hmat}
\end{equation}
where $T_0$ is 0.157 eV according to the tight-binding parameters we adopted. The rearrangement of the sublattice order has no influence on $H^{(n)}$.

For each isolated $C_{6v}$ subsystem, the energy gap is 3.68 eV because of the intralayer sublattice interaction, as denoted by the energy difference between the positive and negative energy levels with black lines in Fig. \ref{fig:hybrid_picture_recip}(a). Because of the interlayer coupling, the hybridization-induced quasicrystalline states are generated, as denoted by these antibonding and bonding states with red lines also shown in Fig. \ref{fig:hybrid_picture_recip}(a). All these states from the monolayers follow the interlayer hybridization rules in Eq.~\eqref{D6_AB_rule} and Eq.~\eqref{D6_E_rule}, and the hybridization strength $\Delta \varepsilon$ for the valence band is stronger than that for the conduction band for equivalent hybridizations with $A_1$, $A_2$, $E_1$ and $E_2$ irreps. As shown by the insets in Fig. \ref{fig:hybrid_picture_recip}(a), the antibonding and bonding states generated by the equivalent hybridizations are 12-fold symmetric. The non-equivalent hybridizations $B_1+B_2$ and $B_2+B_1$ are quite weak because of the large energy difference about 3.68 eV. Therefore, the four 6-fold symmetric $E_3$ states are less stable than other states.

\subsection{S6.2. Electric-field effects on the resonant quasicrystalline states}

\begin{figure*}[!htbp]
\centering
\includegraphics[width=16 cm]{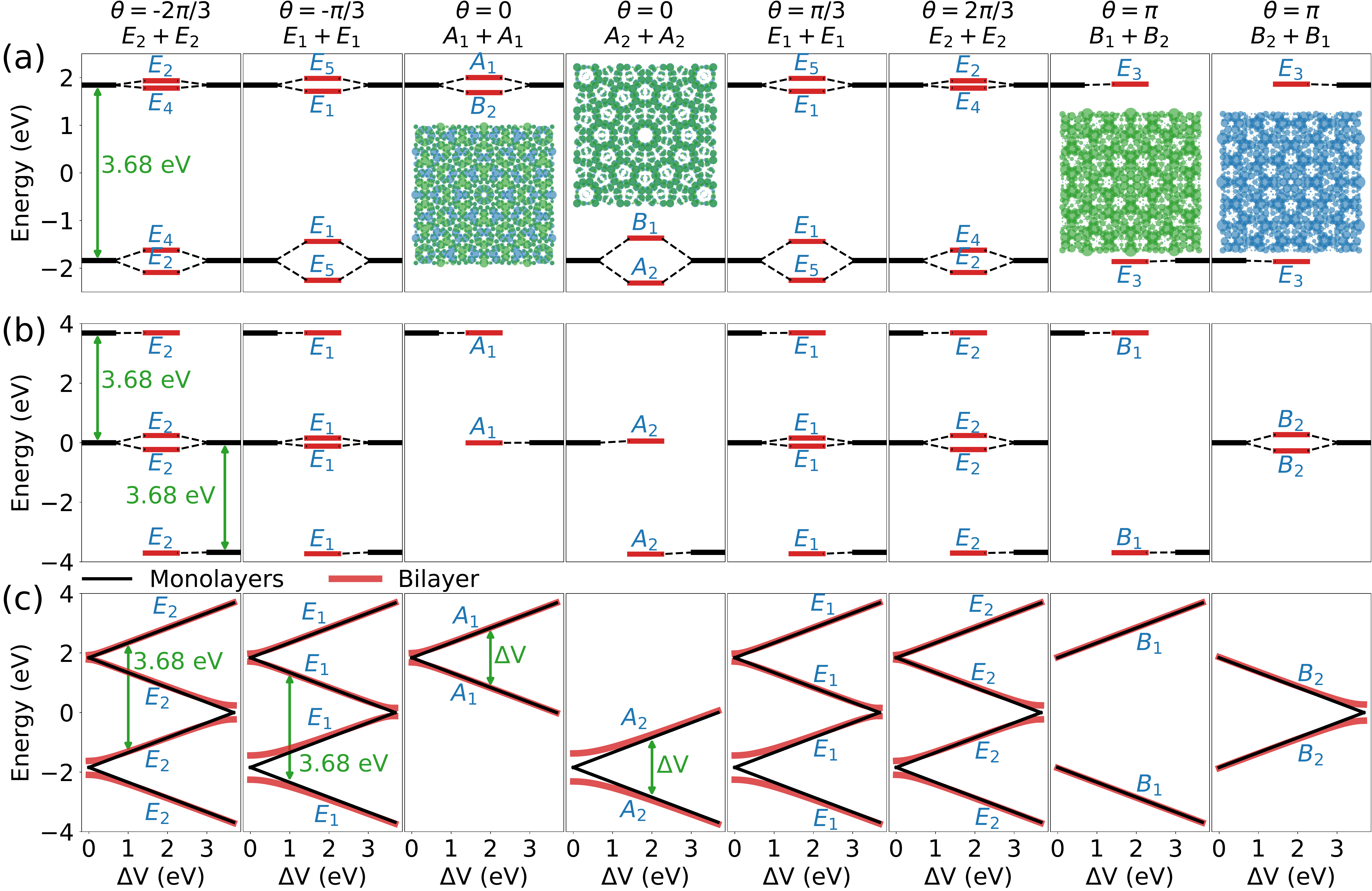}
\caption{The eigen energy spectrum and their irreps of the 12 k-points model in reciprocal space with respect to $C_6$ for (a) $\Delta V =0$ eV, and (b) $\Delta V =3.68$ eV. The left and right black lines in each subplot denote the hybridization paring states from two $C_{6v}$ subsystems, and the middle red lines denote the bonding and antibonding paring states for $D_{6d}$ bilayer. The insets show the real-space electron density for the bonding $B_2$ state, the antibonding $B_1$ state and two bonding $E_3$ states in corresponding subplots. (c) The hybridization paring energy levels and the bonding and antibonding paring energy levels as a function of $\Delta V$.}
\label{fig:hybrid_picture_recip}
\end{figure*}

If a vertical electric field is applied, an additional interlayer potential difference $\Delta V$ should be taken into account. In this respect, the $D_{6d}$ symmetry of the bilayer system is broken into $C_{6v}$. The previous $E_4$ and $E_5$ states formed by the equivalent hybridizations become $E_2$ and $E_1$ states, and the $E_3$ states from the non-equivalent hybridizations become $B_1$ and $B_2$ states, as shown in Figs. \ref{fig:hybrid_picture_recip}(b) and~\ref{fig:hybrid_picture_recip}(c). The hybridization-generated states in Eq.~\eqref{D6_AB_rule} will also lose the 12-fold symmetry. With the increasing $\Delta V$ from $0$ eV to $3.68$ eV, the hybridizations of $A_1+A_1$, $A_2+A_2$ and $B_1+B_2$ weaken much because of the enlarged energy difference, and the hybridizations of $B_2+B_1$, $E_1+E_1$ and $E_2+E_2$ weaken slightly at first and then strengthen, as shown in Fig. \ref{fig:hybrid_picture_recip}(c). Therefore, the electric field acts as a polarizer, which filters the hybridizations of $A_1+A_1$, $A_2+A_2$ and $B_1+B_2$ and allows the hybridizations of $B_2+B_1$, $E_1+E_1$ and $E_2+E_2$.

\section{S7. Interband transitions in graphene quasicrystal}
\begin{figure}[!htbp]
\centering
\includegraphics[width=16 cm]{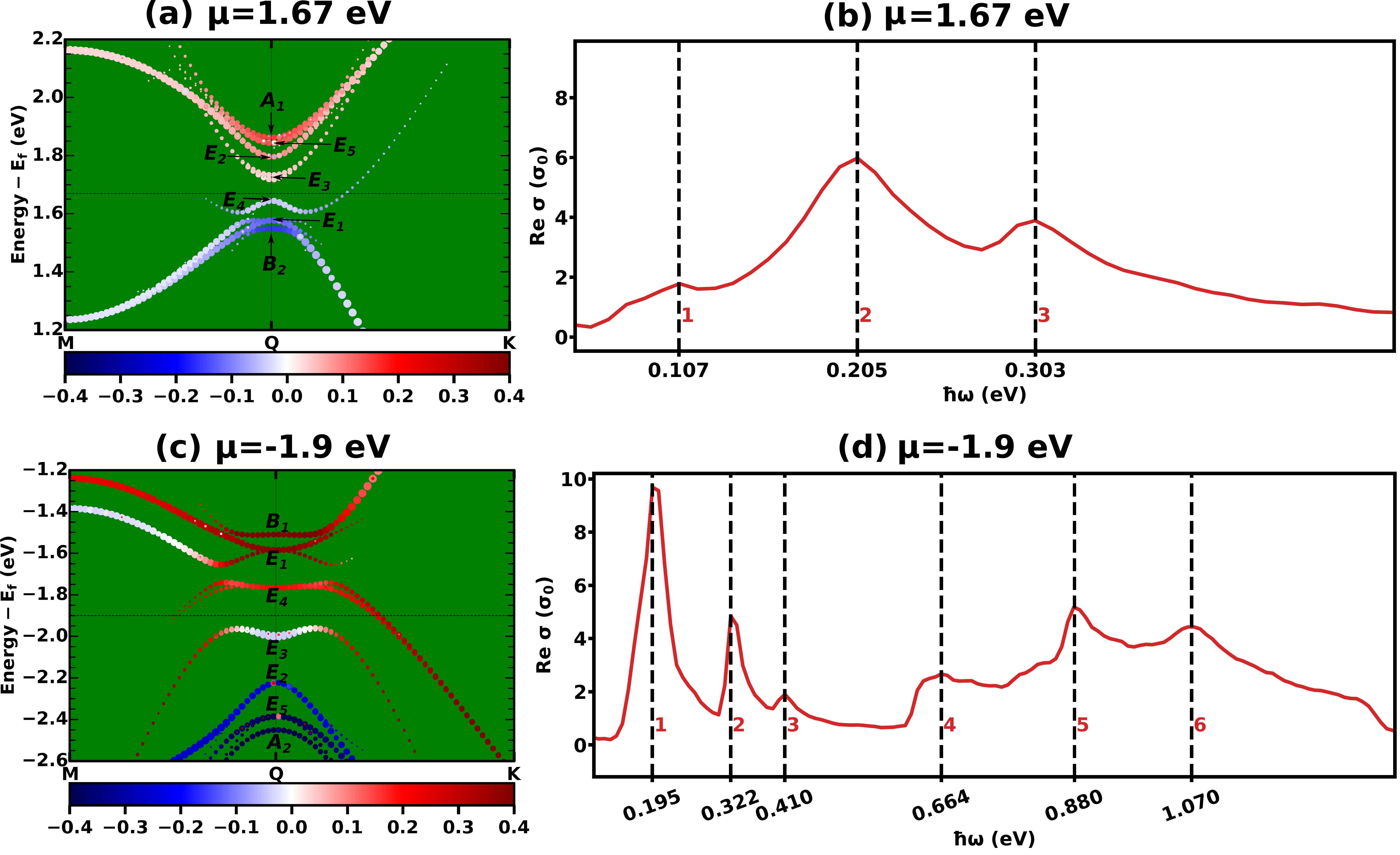}
\caption{The zoomed-in images of unfolded band structures of Fig. 3(a) of main text in (a) near $\mu=1.67$ eV and in (c) near $\mu=-1.9$ eV and corresponding zoomed-in optical conductivity spectra here in (b) and (d), respectively.}
\label{fig:optical_transition}
\end{figure}

The zoomed-in images of unfolded band structures and corresponding optical conductivity spectra in Fig. 3 of main text are plotted in Fig. \ref{fig:optical_transition}. The irreducible representations of bands at $Q$ point are obtained and labeled near $\mu=-1.9$ eV and $\mu=1.67$ eV, respectively. The optical selection rules of $D_{6d}$ point group read\cite{wang2021polarization}: $A_1\leftrightarrow E_1$, $A_2\leftrightarrow E_1$, $B_1\leftrightarrow E_5$, $B_2\leftrightarrow E_5$, $E_1\leftrightarrow E_2$, $E_2\leftrightarrow E_3$, $E_3\leftrightarrow E_4$, and $E_4\leftrightarrow E_5$. Following the optical selection rules, the allowed transitions at $\mu=1.67$ eV in Fig. \ref{fig:optical_transition}(a) corresponding to the absorption peaks are $E_4\leftrightarrow E_3$ for peak 1, $E_4\leftrightarrow E_5$ for peak 2, and both $E_1\leftrightarrow A_1$ and $B_2\leftrightarrow E_5$ for peak 3 in Fig. \ref{fig:optical_transition}(b). The allowed transitions at $\mu=-1.9$ eV in Fig. \ref{fig:optical_transition}(c) corresponding to the absorption peaks are $E_3\leftrightarrow E_4$ for peaks 1, 2 and 3 because of the special profiles of $E_3$ and $E_4$ bands and the splitting of $E_4$ band, both $E_2\leftrightarrow E_1$  and $E_5\leftrightarrow E_4$ for peak 4, and both $E_5\leftrightarrow B_1$ and $A_2\leftrightarrow E_1$ for peak 5, and $A_2\leftrightarrow E_1$ for peak 6 because of the special profiles and splitting of $E_1$ band in Fig. \ref{fig:optical_transition}(d).

\bibliography{supReferences}